\documentclass[usenatbib]{mn2e}
\usepackage[dvips]{graphicx}
\usepackage{amssymb}
\usepackage{natbib}
\usepackage{mn2e-breakabs}
\usepackage{lscape}

\def\gtsim {>\kern-1.2em\lower1.1ex\hbox{$\sim$}~}   
\def\ltsim {<\kern-1.2em\lower1.1ex\hbox{$\sim$}~}   
\newcommand{\aap  }{A\&A}

\newcommand{\apjs }{ApJS}
\newcommand{\apjl }{ApJL}

\title[The Aquila Comparison Project]{The Aquila comparison Project: The Effects of Feedback and Numerical Methods on Simulations of Galaxy Formation}
\author[Scannapieco, Wadepuhl, Parry, Navarro et al. ]{C. Scannapieco,$^{1}$
        M. Wadepuhl,$^{2}$
        O.H. Parry,$^{3,4}$ 
        J.F. Navarro,$^{5}$
        A. Jenkins,$^{3}$
        \newauthor
        {V. Springel,$^{6,7}$
         R. Teyssier,$^{8,9}$
        E. Carlson,$^{10}$
        H.M.P. Couchman,$^{11}$}
        \newauthor
        {R.A. Crain,$^{12,13}$
        C. Dalla Vecchia,$^{14}$
        C.S. Frenk,$^{3}$
        C. Kobayashi,$^{15,16}$}
        \newauthor
        {P. Monaco,$^{17,18}$
        G. Murante,$^{17,19}$
        T. Okamoto,$^{20}$
        T. Quinn,$^{10}$
        J. Schaye,$^{13}$}
        \newauthor
        {G. S. Stinson,$^{21}$
        T. Theuns,$^{3,22}$
        J. Wadsley,$^{11}$
        S.D.M. White,$^{2}$
        R. Woods$^{11}$}\\
$^1$ Leibniz-Institute for Astrophysics Potsdam (AIP), An
  der Sternwarte 16, 14482 Potsdam, Germany\\ 
$^2$ Max-Planck
  Institute for Astrophysics, Karl-Schwarzschild-Str. 1,
85741 Garching, German\\
$^3$ Institute for Computational Cosmology, Department of Physics, Durham
  University, Durham DH1 3LE, United Kingdom\\
$^4$ Department of Astronomy, University of Maryland, College Park, MD 20745, USA\\
$^5$ Department of Physics and Astronomy, University of
  Victoria, Victoria, BC, V8P 5C2, Canada\\
$^6$ Heidelberg Institute for Theoretical Studies,
  Schloss-Wolfsbrunnenweg 35, 69118 Heidelberg, Germany\\
$^7$ Zentrum f\"{u}r Astronomie der Universit\"{a}t Heidelberg, 
 ARI, M\"{o}nchhofstr. 12-14, 69120 Heidelberg, Germany\\
$^8$ CEA, IRFU, SAp, 91191 Gif-sur-Yvette, France\\
$^9$ Institute for Theoretical Physics, University of Z\'urich, CH-8057 Z\"urich, Switzerland\\
$^{10}$ Department of Astronomy, University of Washington, Box 351580, Seattle, WA 98195-1580, USA\\
$^{11}$ Department of Physics \& Astronomy, McMaster University, Hamilton, ON L8S, 4M1, Canada\\
$^{12}$ Centre for Astrophysics \& Supercomputing, Swinburne University of Technology, Hawthorn, Victoria 3122, Australia\\
$^{13}$ Leiden Observatory, Leiden University, PO Box 9513, 2300 RA Leiden, The Netherlands\\
$^{14}$ Max Planck Institute for Extraterrestrial Physics, Gissenbachstra\ss{}e 1, 85748 Garching, Germany\\
$^{15}$ School of Physics, Astronomy and Mathematics, University of Hertfordshire, Hatfield  AL10 9AB, UK\\
$^{16}$ Research School of Astronomy \& Astrophysics, The Australian National University, Cotter Road, Weston, ACT 2611, Australia\\
$^{17}$ Dipartimento di Fisica - Sezione di Astronomia, Universit\`a di Trieste, via Tiepolo 11, I- 34131 Trieste, Italy\\
$^{18}$ INAF, Osservatorio Astronomico di Trieste, Via Tiepolo 11, I-34131 Trieste, Italy\\
$^{19}$ INAF, Osservatorio Astronomico di Torino, Strada Osservatorio 20, I-10025 Pino Torinese, Italy\\
$^{20}$ Center for Computational Sciences, University of Tsukuba, 1-1-1 Tennodai, Tsukuba 305-8577 Ibaraki, Japan\\
$^{21}$ Max-Planck-Institut f\"ur Astronomie, K\"onigstuhl 17, 69117, Heidelberg, Germany\\
$^{22}$ Universiteit Antwerpen, Campus Groenenborger, Groenenborgerlaan
171, B-2020 Antwerpen, Belgium}

\def\LaTeX{L\kern-.36em\raise.3ex\hbox{a}\kern-.15em
    T\kern-.1667em\lower.7ex\hbox{E}\kern-.125emX}

\begin{document}

   \maketitle

\begin{abstract}
  We compare the results of various cosmological gas-dynamical codes
  used to simulate the formation of a galaxy in the $\Lambda$CDM
  structure formation paradigm. The various runs (thirteen in total)
  differ in their numerical hydrodynamical treatment (SPH, moving-mesh
  and AMR) but share the {\it same} initial conditions and adopt in
  each case their latest published model of gas cooling, star
  formation and feedback. Despite the common halo assembly history, we
  find large code-to-code variations in the stellar mass, size,
  morphology and gas content of the galaxy at $z=0$, due mainly to the
  different implementations of star formation and feedback. Compared
  with observation, most codes tend to produce an overly massive
  galaxy, smaller and less gas-rich than typical spirals, with a
  massive bulge and a declining rotation curve. A stellar disk is
  discernible in most simulations, although its prominence varies
  widely from code to code. There is a well-defined trend between the
  effects of feedback and the severity of the disagreement with
  observed spirals. In general, models that are more effective at
  limiting the baryonic mass of the galaxy come closer to matching
  observed galaxy scaling laws, but often to the detriment of the disk
  component.  Although numerical convergence is not particularly
    good for any of the codes, our conclusions hold at two different
  numerical resolutions.  Some differences can also be traced to the
  different numerical techniques; for example, more gas seems able to
  cool and become available for star formation in grid-based codes
  than in SPH. However, this effect is small compared to the
  variations induced by different feedback prescriptions. We conclude
  that state-of-the-art simulations cannot yet uniquely predict the
  properties of the baryonic component of a galaxy, even when the
  assembly history of its host halo is fully specified. Developing
  feedback algorithms that can effectively regulate the mass of a
  galaxy without hindering the formation of high-angular momentum
  stellar disks remains a challenge.
\end{abstract}

\begin{keywords} cosmology: theory -- methods: numerical -- galaxies: formation -- galaxies: evolution 
\end{keywords}

\section{Introduction}
\label{SecIntro}

Numerical simulations play a central role in studies of cosmic
structure formation.  Collisionless N-body simulations have now become
the main theoretical tool to predict the non-linear evolution of
dark-matter dominated structures once initial conditions are
specified. Their high accuracy and huge dynamic range have allowed a
detailed comparison of their outcome with observations of the
large-scale structure of the universe.  The impressive agreement
between these observations and the predictions of the $\Lambda$ Cold
Dark Matter ($\Lambda$CDM) model has helped to establish it as the
current paradigm of structure formation \citep{Springel2006}.

Simulating the evolution of the visible Universe is much
more complex, because it requires understanding the many astrophysical
processes which drive the evolution of the baryonic component
under the gravitational influence of the dark matter.
Numerical hydrodynamics in cosmological
simulations has traditionally used either the Lagrangian Smoothed
Particle Hydrodynamics technique (SPH; \citealt{Lucy77, Gingold77,
  Monaghan92, Katz1996,Springel2010_SPH}) or Eulerian grid-based
solvers sometimes aided
by Adaptive Mesh Refinement (AMR) techniques
(\citealt{Cen1992,Bryan95,Kravtsov1999,Fryxell2000,Teyssier2002,Quilis2004}).

Both approaches have advantages and disadvantages. It is widely
appreciated that SPH is not able to capture shocks with high accuracy
and that in certain situations fluid instabilities can be suppressed,
at least for standard implementations of SPH
\citep{Agertz07,Creasey2011}.  On the other hand, mesh-based codes are
not Galilean invariant and may in some cases generate entropy
spuriously through artificial mixing \citep{Wadsley08}. As a result,
even for some simple  non-radiative problems, Lagrangian and Eulerian
codes do not converge to the same solution (e.g., \citealt{Okamoto03,Agertz07,
  Tasker08, Mitchell2009}). Novel techniques, such as the Lagrangian,
moving-mesh {\sc arepo} code introduced by \citet{Springel2010},
hold the promise of improving this state of affairs, but their 
application is still in its infancy.

An even more uncertain ingredient of galaxy formation simulations 
are the descriptions of poorly understood physical
processes such as star formation and feedback. The huge dynamic range
between the super-Megaparsec scales needed to follow the hierarchical
growth of a galaxy and the sub-parsec scales that govern the
transformation of its gas into stars implies that direct simulation of
all relevant physical processes is still out of reach of even the most
powerful computers and best available algorithms. Star formation and
feedback are therefore introduced in cosmological simulations as
``sub-grid'' parameterized prescriptions of limited physical content
and lacking numerical rigor.

These difficulties have hampered the progress of simulations of galaxy
formation within the $\Lambda$CDM paradigm, but a few results of
general applicability have nevertheless emerged.  For example, the
formation of {\it realistic disk galaxies} in dark matter halos formed
hierarchically, as expected in $\Lambda$CDM, was recognized as a
challenge even in early simulations (see,
e.g., \citealt{Navarro1995,NavarroSteinmetz1997}).  Simulated galaxies
suffered from ``overcooling'' and from a dearth of angular momentum
due to the transfer of angular momentum from the baryonic component to the
dark matter during the many merger events that characterize
hierarchical assembly
(\citealt{Navarro91,SL99,Navarro00}).
Feedback, as a general heating mechanism that can prevent overcooling
and regulate the assembly of a galaxy whilst avoiding catastrophic
angular momentum losses, emerged as a crucial ingredient of any
successful galaxy formation simulation \citep{W98,NavarroSteinmetz1997,SL99,CS2008,Zavala08}.

Considerable progress has been made since this time: recent simulations have
shown that the angular momentum problem can indeed be alleviated when
feedback from evolving stars, in particular supernovae (SNe), is included
(e.g.,
\citealt{Okamoto05,Governato2007,CS2008,Ceverino2009,CS2009,Joung2009,Colin2010,Sales2010,Stinson2010}).
Some authors have also investigated alternative feedback mechanisms,
such as energy liberated during the formation of supermassive black
holes as well as that carried by cosmic rays (e.g.,
\citealt{DiMatteo2003, SpringelDiMatteo2005,Jubelgas2008,Booth2009,Fabjan2010,
Wadepuhl11}).

Despite progress, difficulties remain. For example, simulations tend
to allow far too many baryons to accrete into galaxies to be
consistent with the observed stellar mass function of galaxies  
(see, e.g., \citealt{Guo2010}). In
addition, although stellar disks do form, they are often too
concentrated, with steeply-declining rotation curves at odds with
observation (e.g., \citealt{Abadi2003a,Abadi2003b,Stinson2010}).

The difficulties in simulating disk galaxies highlighted above are
compounded by the limited guidance afforded by analytic and
semi-analytic models of galaxy formation. In such modeling, the
properties of disk galaxies are typically envisioned to reflect those
of their surrounding halos
\citep{FE80,KWG93,Dalcanton97,Mo98,Bower06,deLucia07}: for example,
halos with high net spin and quiet recent merger histories are
commonly assumed to be likely sites of disk galaxy formation. However,
there are growing indications that these assumptions might be too
simplistic. \citet{CS2009} and \citet{Stinson2010}, for example, find
no clear relation between the presence of disks and the spin parameter
of the halo: disks form in halos with low and high spin parameters.
Moreover, \citet{CS2009} report that disk formation is not assured by
a quiescent assembly history, since they can be destroyed not only by
major or intermediate-mass mergers, but also by secular processes such
as a misalignment between the gaseous and stellar disks.

These difficulties have led to little consensus on what determines
the morphology of a galaxy; what the main feedback mechanisms are; and
what role they play on different mass scales and at different
times. Indeed, there is even debate about whether the difficulties in
reproducing realistic disks are predominantly a consequence of
insufficient numerical resolution (e.g., \citealt{Governato2004}),
inappropriate modeling of the relevant physics
(e.g., \citealt{Mayer2008,Piontek2011}), or a failure of the cosmological model
(e.g., \citealt{Sommer-Larsen2001}).

Progress in this unsettled field requires at least a careful
evaluation of the different numerical techniques, resolution, and
choice of sub-grid physics adopted by various authors. Most groups do
carry out and publish resolution tests and convergence studies of
their own numerical setup. However, because of the complexity of the
problem, the lack of telling test cases with known solution, and the
absence of clear theoretical predictions for individual systems, each
group chooses to tune the relevant numerical parameters according to
different priorities and/or prejudice, and there has so far
been little effort invested in comparing the results of different
techniques and codes. Would they give similar results if they followed
the formation of a galaxy in the {\it same} dark matter halo?

The main goal of the Aquila Project is to address this question by
comparing the predictions of different codes using {\it common initial
  conditions} and a {\it homogeneous set of analysis tools}. Rather
than focusing on whether individual codes perform better or worse
than others, we contrast their predictions for the stellar mass,
angular momentum content, star formation rates and galaxy size, with
observation.

This paper is organized as follows. \S~\ref{sec:simulations}
describes the initial conditions and the simulation setup.
\S~\ref{sec:results} compares the galaxy morphology, star
formation history, size, angular momentum, and gas content of the $13$
simulated galaxies, together with a brief discussion of the effects of
numerical resolution on the results.  Finally, \S~\ref{SecConc}
summarizes our main findings and lists our main conclusions.

\begin{figure*}
\includegraphics[width=10cm]{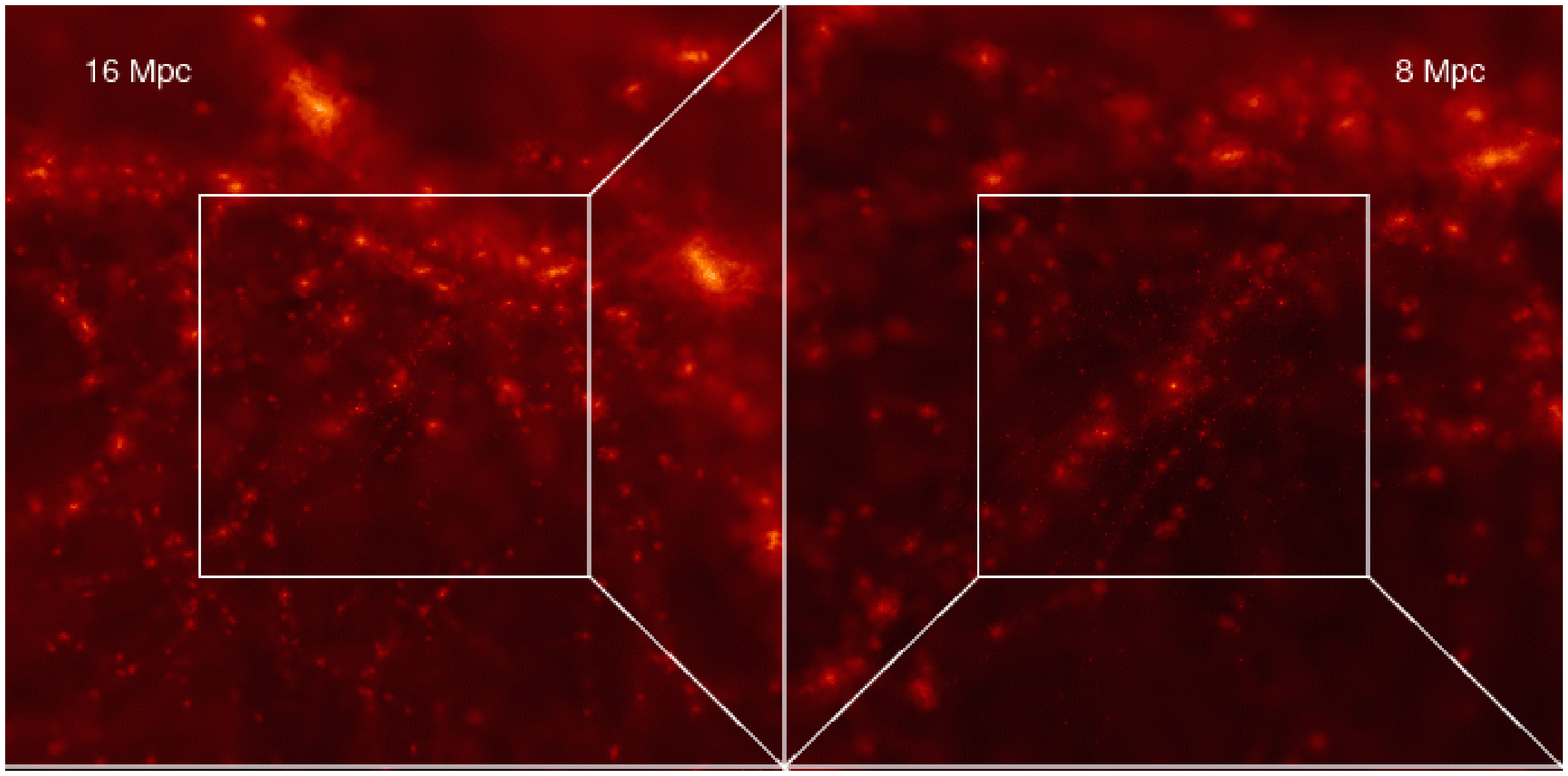}

\vspace{-0.1cm}

\includegraphics[width=10cm]{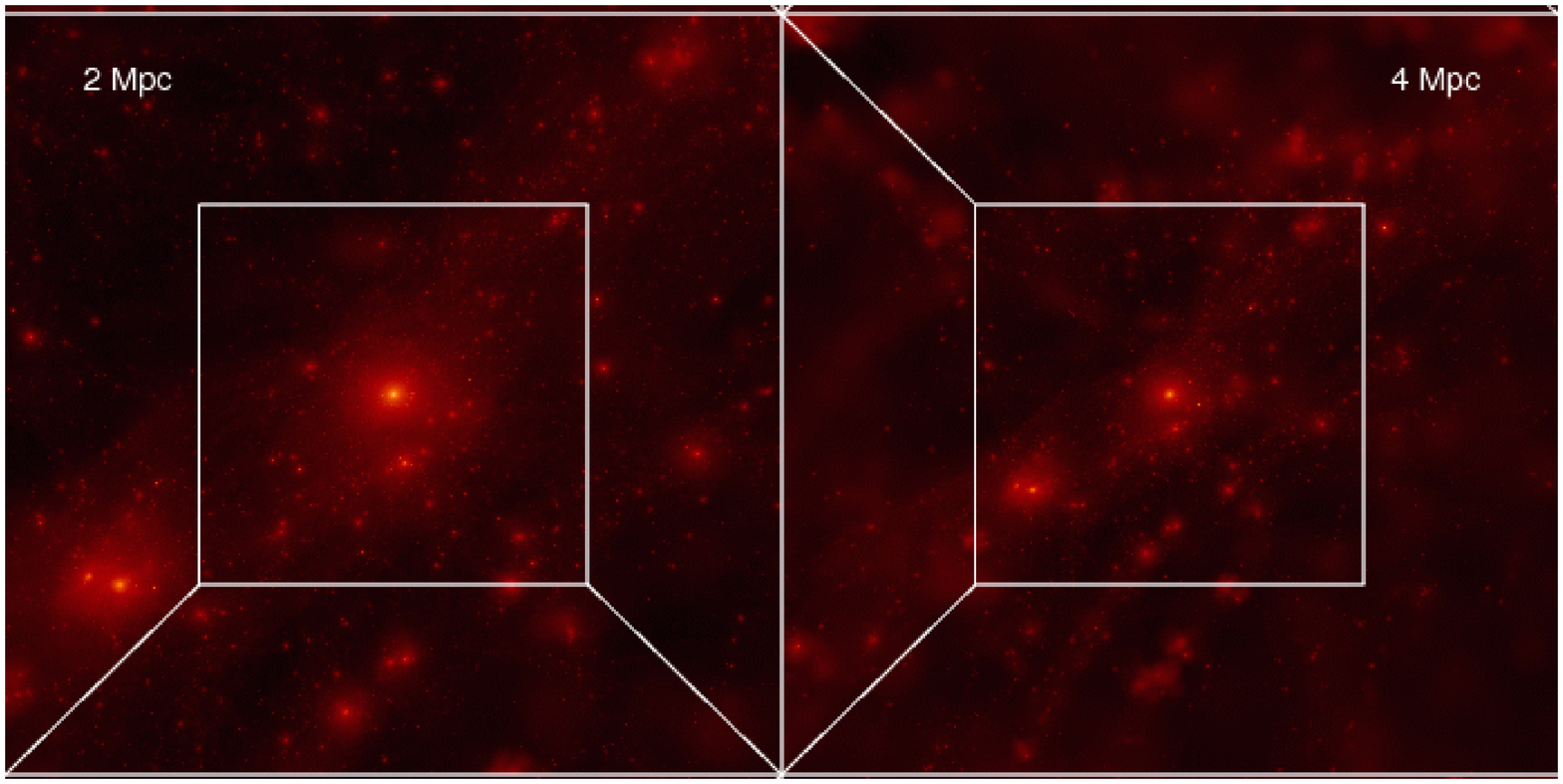}

\vspace{-0.1cm}

\includegraphics[width=10cm]{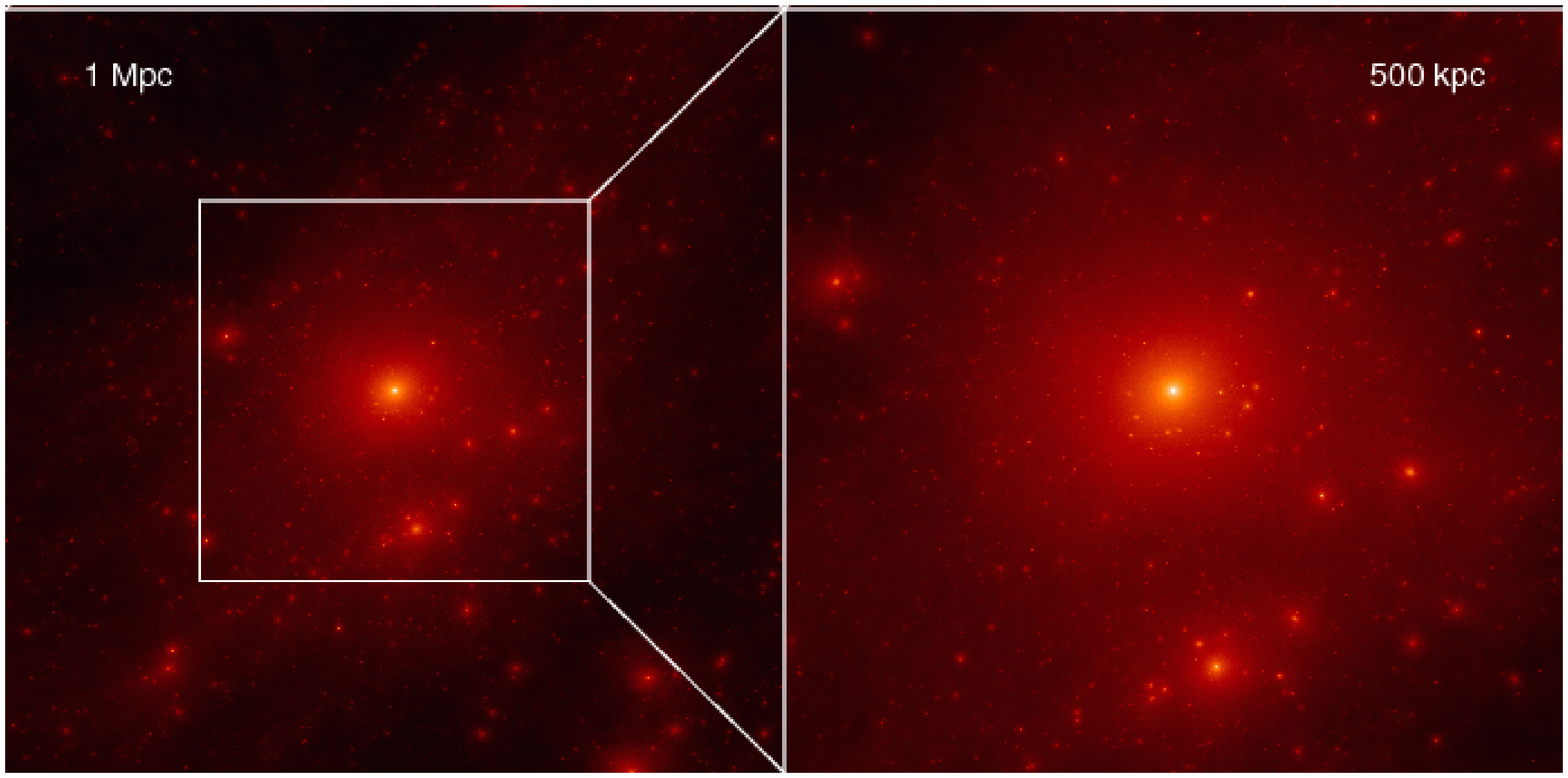}
\caption{Maps of the dark matter distribution in the region
  surrounding the Aquila halo at $z=0$.  The dark matter is
  projected in cubic volumes of side length as indicated in each
  panel.  Pixel brightness corresponds to the dark matter
  density using a logarithmic scale. }
\vspace{1cm}
  \label{DMoverview}
\end{figure*}

\begin{table*}
\caption{Summary of code characteristics and sub-grid physics}
\label{table_codes}
\begin{tabular}{lcccccc}
\hline\hline
Code & Reference & Type & \multicolumn{2}{c}{UV background} &Cooling & Feedback \\ 
& & & ($z_{\rm UV}$) &  (spectrum) & & \\
\hline
 {\small G3} ({\sc gadget3})& [1] & SPH & 6 & [10] & primordial [13]& SN (thermal)
\\ \\

{\small G3-BH} & [1] & SPH & 6 & [10] & primordial [13]& SN (thermal),
BH \\ \\ 

{\small G3-CR} & [1] & SPH & 6 & [10] & primordial [13]& SN (thermal), BH,
CR \\ \\ 

{\small G3-CS} & [2] & SPH & 6 & [10] & metal-dependent [14] & SN
(thermal) \\ \\

{\small G3-TO} & [3] & SPH & 9 & [11] & element-by-element [15] & SN
(thermal+kinetic)\\ \\ 

{\small G3-GIMIC} & [4] & SPH & 9 & [11] & element-by-element [15]
& SN (kinetic)\\ \\ 

{\small G3-MM} & [5] & SPH & 6 & [10] & primordial [13] & SN
(thermal)\\ \\ 

{\small G3-CK} & [6] & SPH & 6 & [10] & metal-dependent [14] & SN (thermal)\\ \\

{\small GAS} ({\sc gasoline})& [7] & SPH & 10 & [12] & metal-dependent
[16] & SN (thermal)\\ \\ 

{\small R} ({\sc ramses}) & [8] & AMR & 12 & [10] & metal-dependent
[14] & SN (thermal)\\ \\

{\small R-LSFE} & [8] & AMR & 12 & [10] & metal-dependent [14] & SN (thermal)\\ \\ 

{\small R-AGN} & [8] & AMR & 12 & [10] & metal-dependent [14] & SN (thermal), BH\\ \\ 

{\small AREPO} & [9] & Moving Mesh & 6 & [10] & primordial [13] & SN (thermal)\\
\hline
\end{tabular}

{\sc note:} [1] \citet{Springel08}; [2] \citet{CS2005};
  \citet{CS2006}; [3] \citet{Okamoto2010}; [4] \citet{Crain2009}; [5]
  \citet{M10}; [6] \citet{kob07}; [7]
  \citet{Stinson2006}; [8] \citet{Teyssier2002}; \citet{Rasera2006};
  \citet{Dubois2008}; [9] \citet{Springel2010}; [10]
  \citet{HaardtMadau1996}; [11] \citet{HaardtMadau2001}; [12] Haardt
  \& Madau (2005, private communication); [13] \citet{Katz1996}; [14]
  \citet{SutherlandDopita1993}; [15] \citet{Wiersma2009a}; [16]
  \citet{Shen2010}. 
\end{table*}

\section{The Simulations}
\label{sec:simulations}

\subsection{The Codes}
\label{SecCodes}

The Aquila Project consists of $13$ different gas-dynamical simulations
of the formation of a galaxy in a $\Lambda$CDM halo of similar mass to that 
of the Milky Way. Nine different codes were used for the project (two codes were
run three times each with different sub-grid physics modules).  The
various codes differ in their hydrodynamical technique (SPH, AMR,
moving-mesh): seven codes use the SPH technique, six of which are
based on {\sc gadget} \citep[hereafter {\small G3} for short,
see][]{Springel2005} and one on {\sc gasoline} (hereafter referred to as
{\sc gas}; \citealt{Wadsley2004}). 

The {\small G3}-based codes share the same numerical
gravity/hydrodynamical treatment but differ in their cooling/star
formation/feedback modules. {\small G3} refers to the standard
\citet{SpringelHernquist2003} implementation;  {\small G3-CS} refers to the code
presented in \citet{CS2005,CS2006}; {\small G3-TO} to that developed
by \citet{Okamoto05,Okamoto2008,Okamoto2010}; {\small G3-GIMIC}
is described in \citet{Crain2009};
 {\small G3-MM} is introduced by
\citet{M10}; and {\small G3-CK} by \citet{kob07}.  Of
the two codes that do not use SPH one is the {\sc ramses} AMR code
\citep[hereafter {\small R}, for short, see][]{Teyssier2002}, and the
other is the moving-mesh code {\sc arepo} \citep{Springel2010}.

Each code was run by the group responsible for its development,
adopting (independently from the choices made by other Aquila
participants) their latest published model of cooling, star formation,
and feedback. These differ from code to code. Regarding radiative
cooling, for example, some codes assume primordial abundances to
compute cooling rates; others use metal-dependent cooling rate tables;
and in some cases cooling is implemented on an element-by-element
basis. Star formation also varies from code to code, although in
nearly all cases the efficiency of transformation of gas into stars is
set by attempting to reproduce the Kennicutt-Schmidt empirical
relation \citep{Kennicutt98} in simulations of isolated disks (see also
Fig.~\ref{fig:Kennicutt}).

The numerical treatment of feedback also varies from code to code. In
most cases, {\it thermal feedback} is used, where supernova energy is
injected into the interstellar medium (ISM) as thermal energy. The dispersal
of the input energy is usually delayed artificially, in order to
promote the pressurization of the ISM, the onset of winds, and
the effective regulation of subsequent star formation. A few codes adopt
{\it kinetic feedback}, where kinetic energy is dumped directly into
the gas.  In the G3-TO code, the outflowing gas is temporarily {\it
  decoupled} hydrodynamically from the rest of the ISM to ensure that
the specified mass loading (wind mass per unit mass of stars formed)
and velocity are not modified by viscous drag until gas escapes the
star forming region.

Two further simulations were run with the {\small G3} code: one
({\small G3-BH}) that included, in addition to SN, the feedback energy
associated with the assembly of supermassive black holes and a third
({\small G3-CR}) where another form of feedback, that associated with
energy deposition by cosmic rays, was also included.

Two further {\sc ramses} runs are also part of our series, one
({\small R-LSFE}) where the star formation timescale is much longer than
the fiducial choice, delaying substantially the transformation of gas
into stars, and another ({\small R-AGN}) where the feedback energy was
augmented by the contribution of a putative AGN associated with a
central supermassive black hole.

The main characteristics of each code are summarized in
Table~\ref{table_codes}. Appendix~\ref{code_description} presents a
more detailed description of each of the codes and their numerical
choices.

\begin{figure*}
\begin{center}
{\includegraphics[width=35mm]{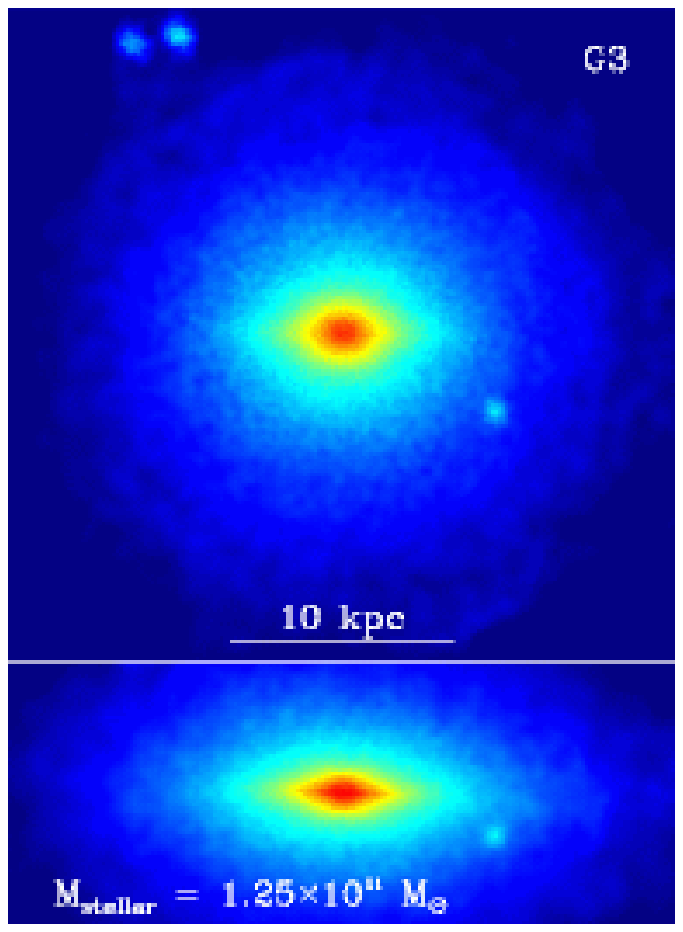}\includegraphics[width=35mm]{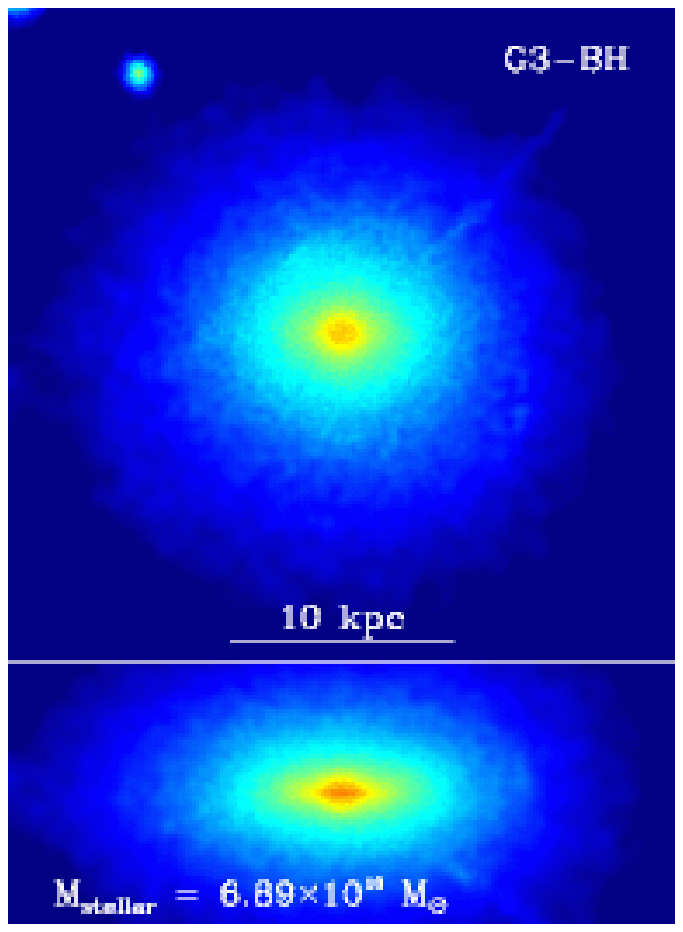}\includegraphics[width=35mm]{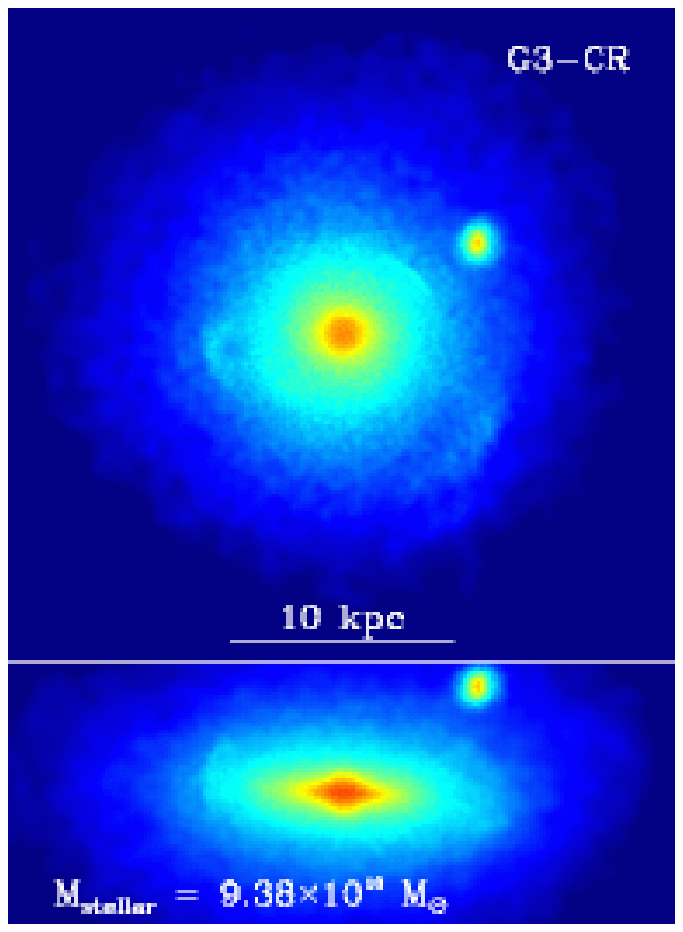}\includegraphics[width=35mm]{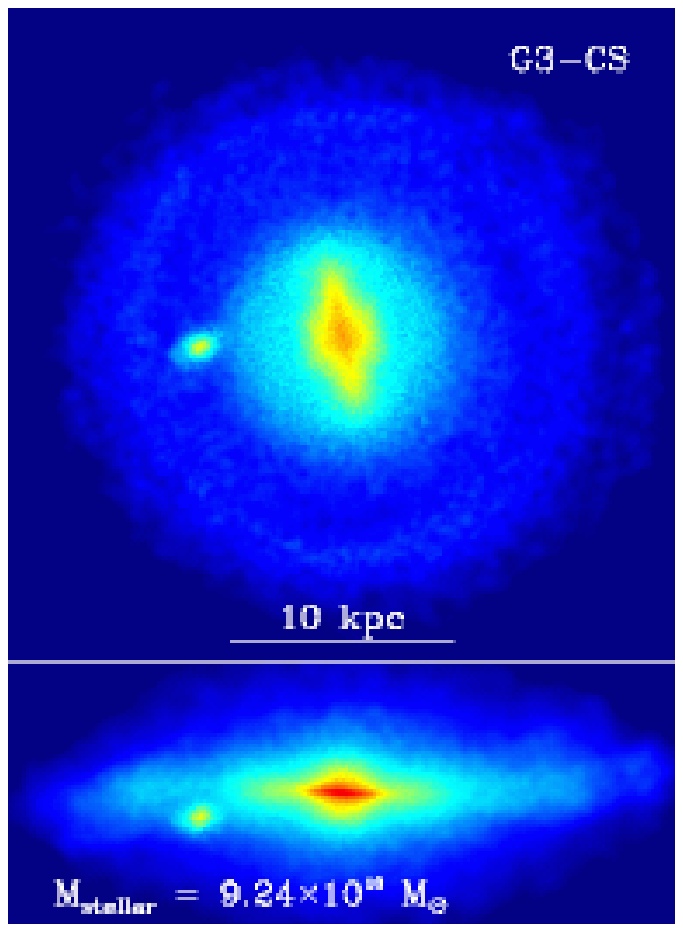}}

\vspace{0.5cm}

{\includegraphics[width=35mm]{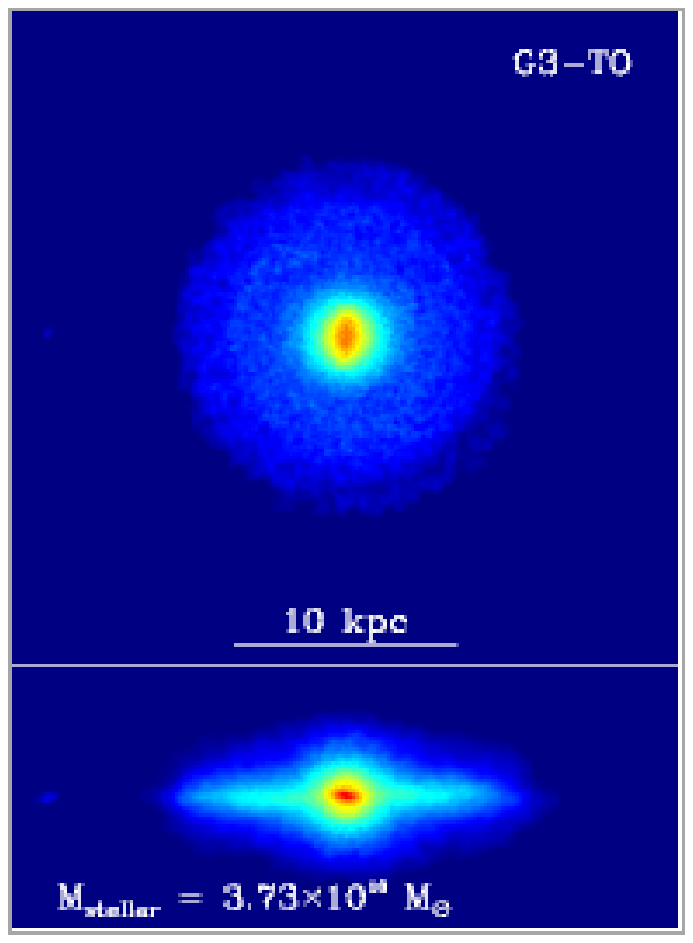}\includegraphics[width=35mm]{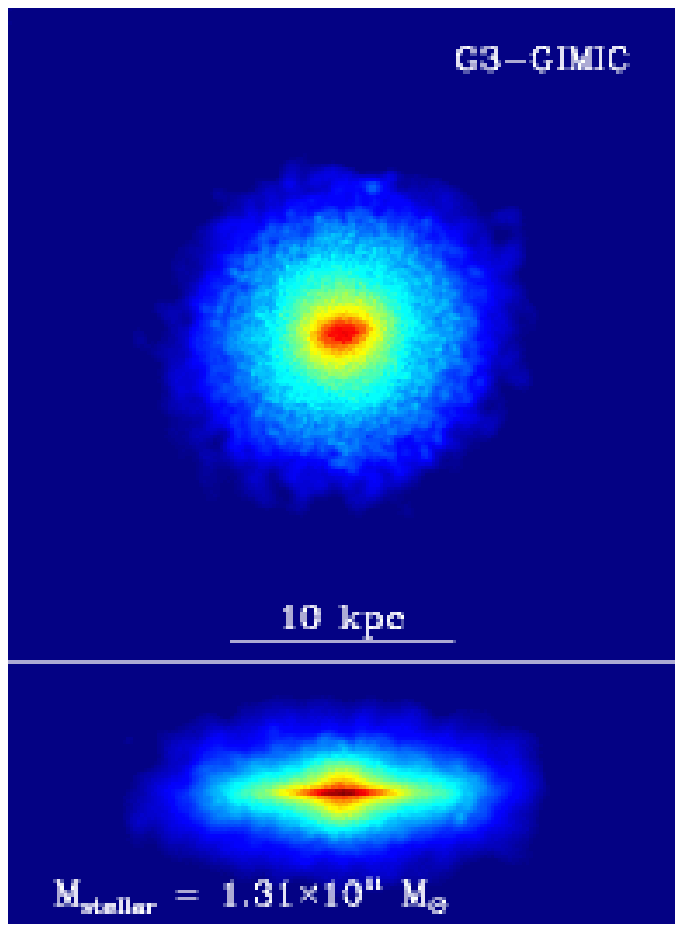}\includegraphics[width=35mm]{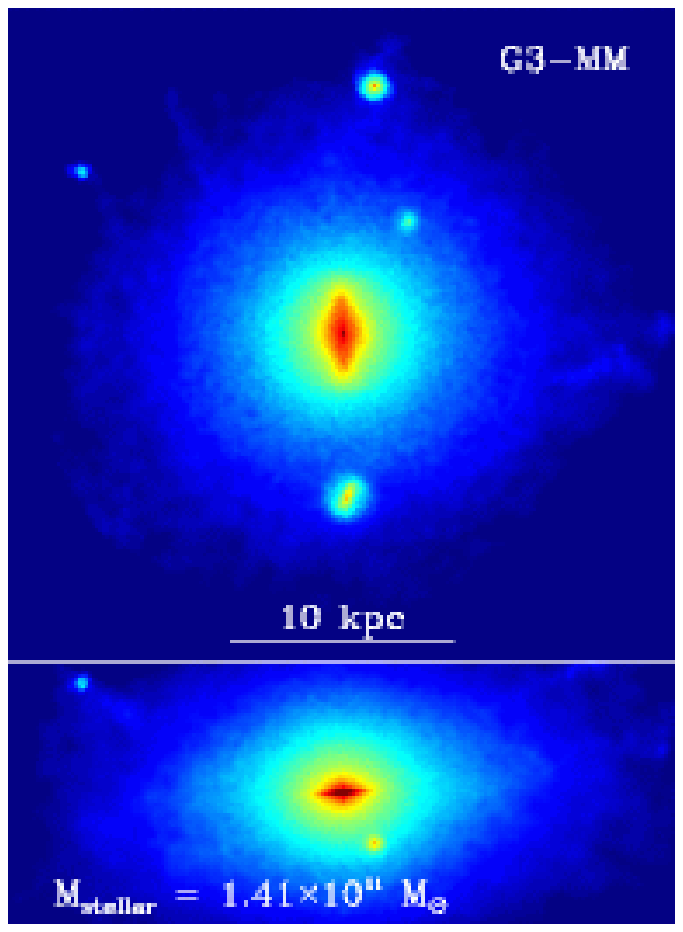}\includegraphics[width=35mm]{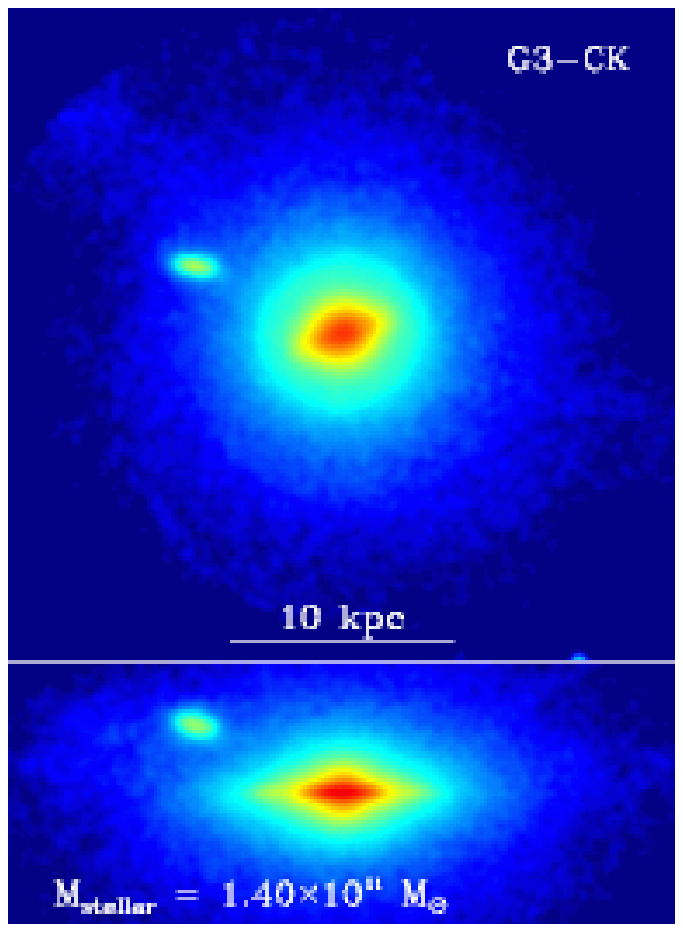}\includegraphics[width=35mm]{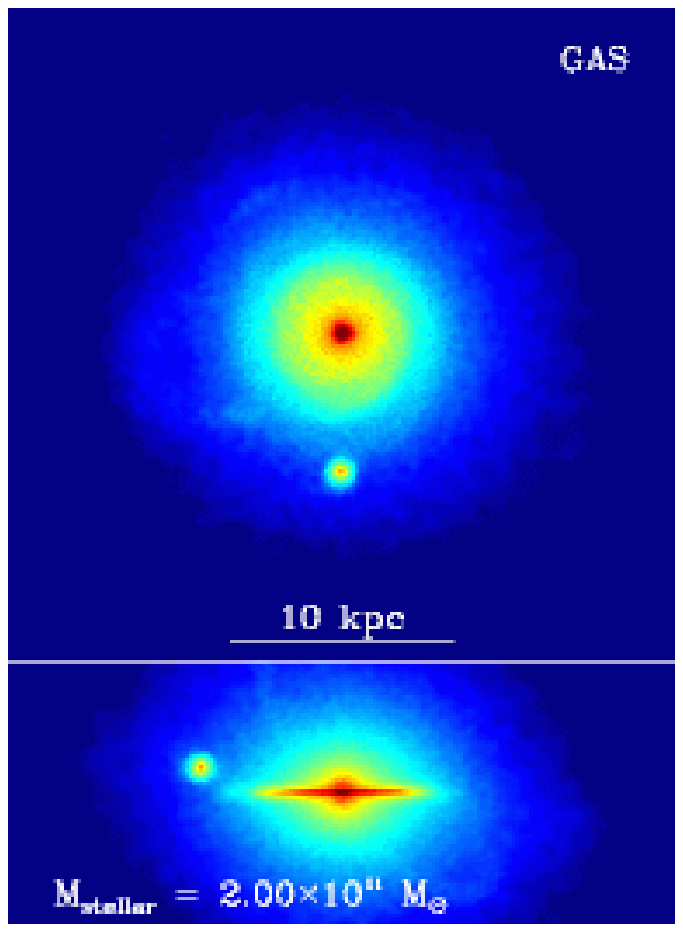}}

\vspace{0.5cm}

       {\includegraphics[width=35mm]{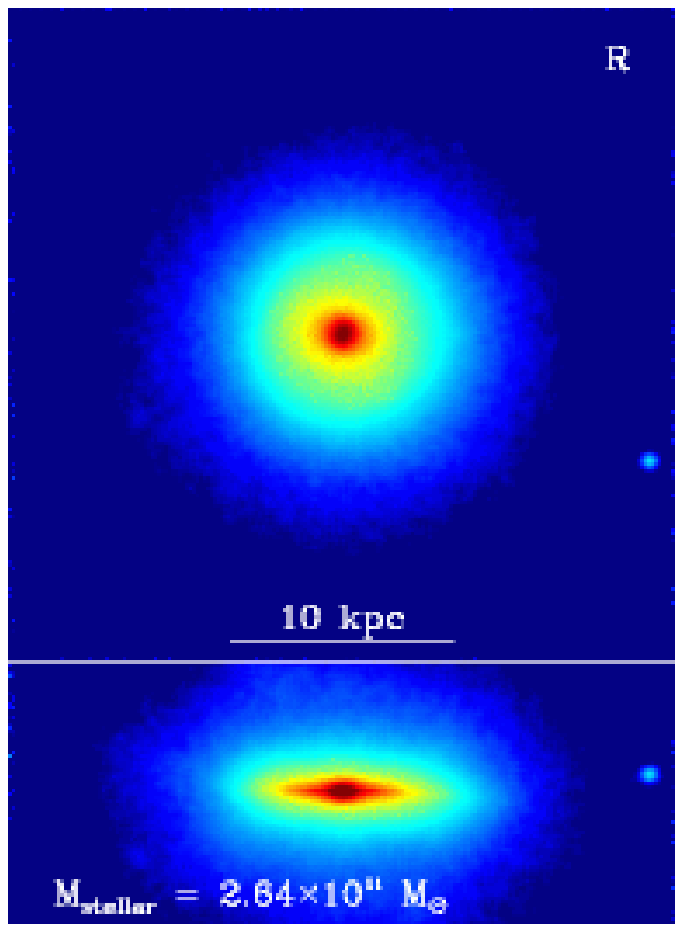}\includegraphics[width=35mm]{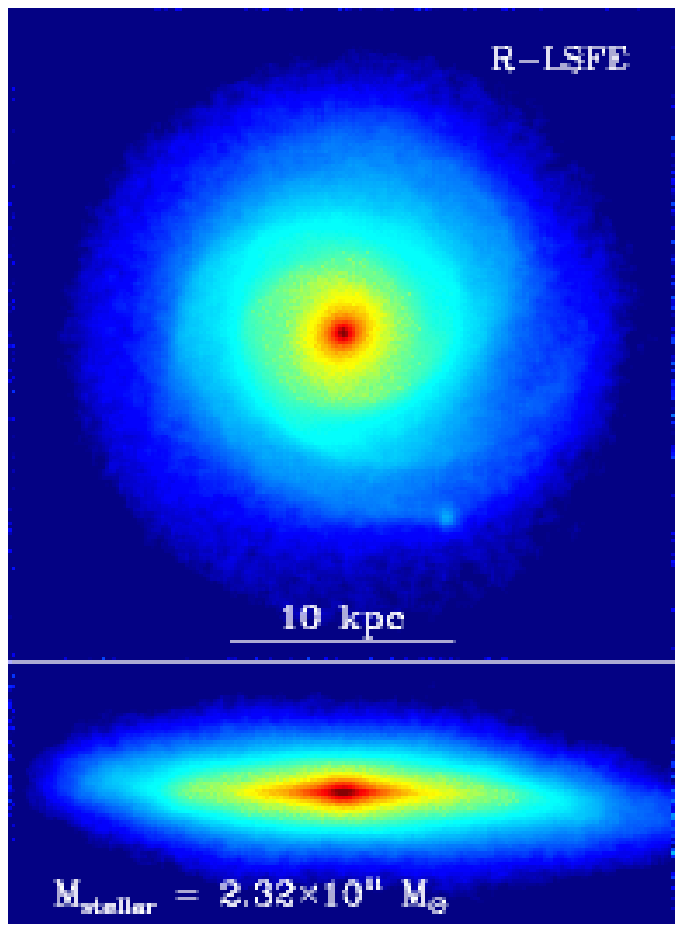}\includegraphics[width=35mm]{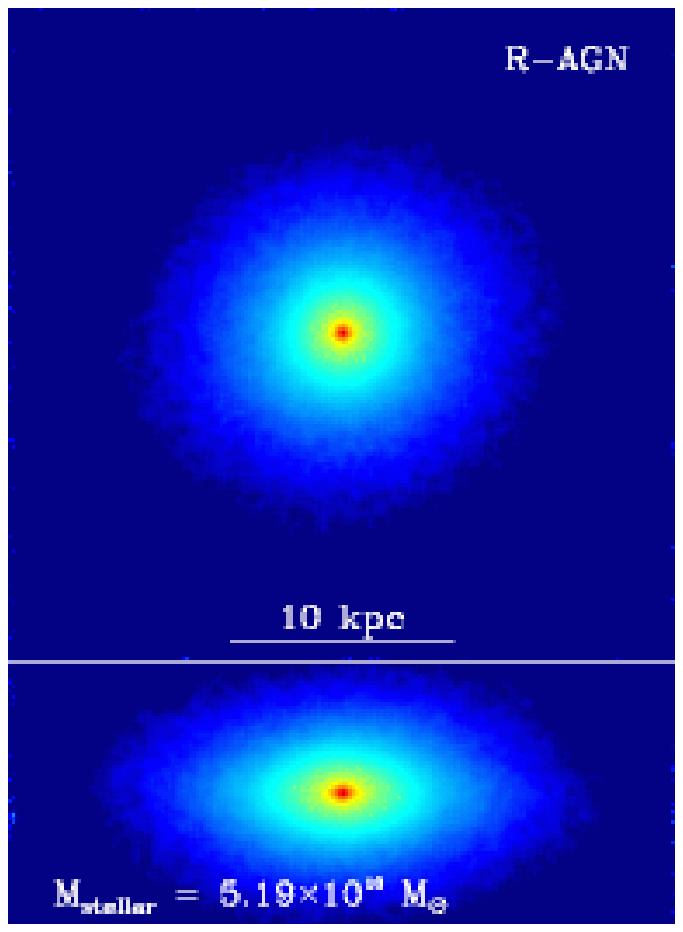}\includegraphics[width=35mm]{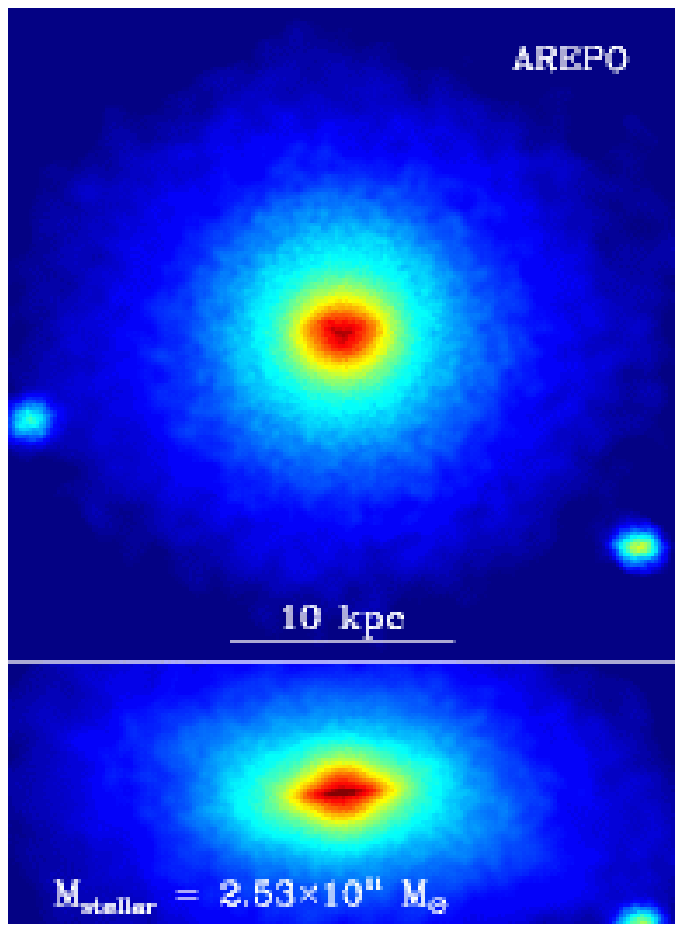}}

\vspace{0.5cm}
\includegraphics[width=80mm]{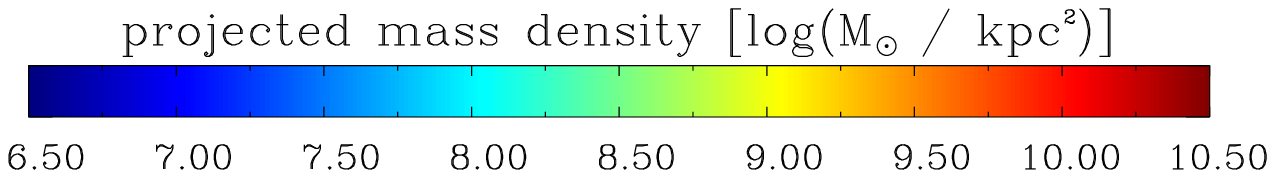}
\end{center}

\caption{Face-on and edge-on maps of projected stellar mass
  density. The face-on projection is along the direction of the
  angular momentum vector of galaxy stars.  The face-on and edge-on
  maps are $30\times 30$ kpc$^2$ and $30\times 12$ kpc$^2$,
  respectively. The size of each pixel is $58.6$ pc on a side and its
  color is drawn from a logarithmic color map of the surface stellar
  mass density.  The total stellar mass within the galaxy radius
  ($r_{\rm gal}=0.1\, r_{\rm 200}\sim 25$ kpc) is listed in the legend
  of each panel.}
\label{fig:stellar_maps}
\end{figure*}

\begin{figure*}
\begin{center}
 \includegraphics[width=175mm]{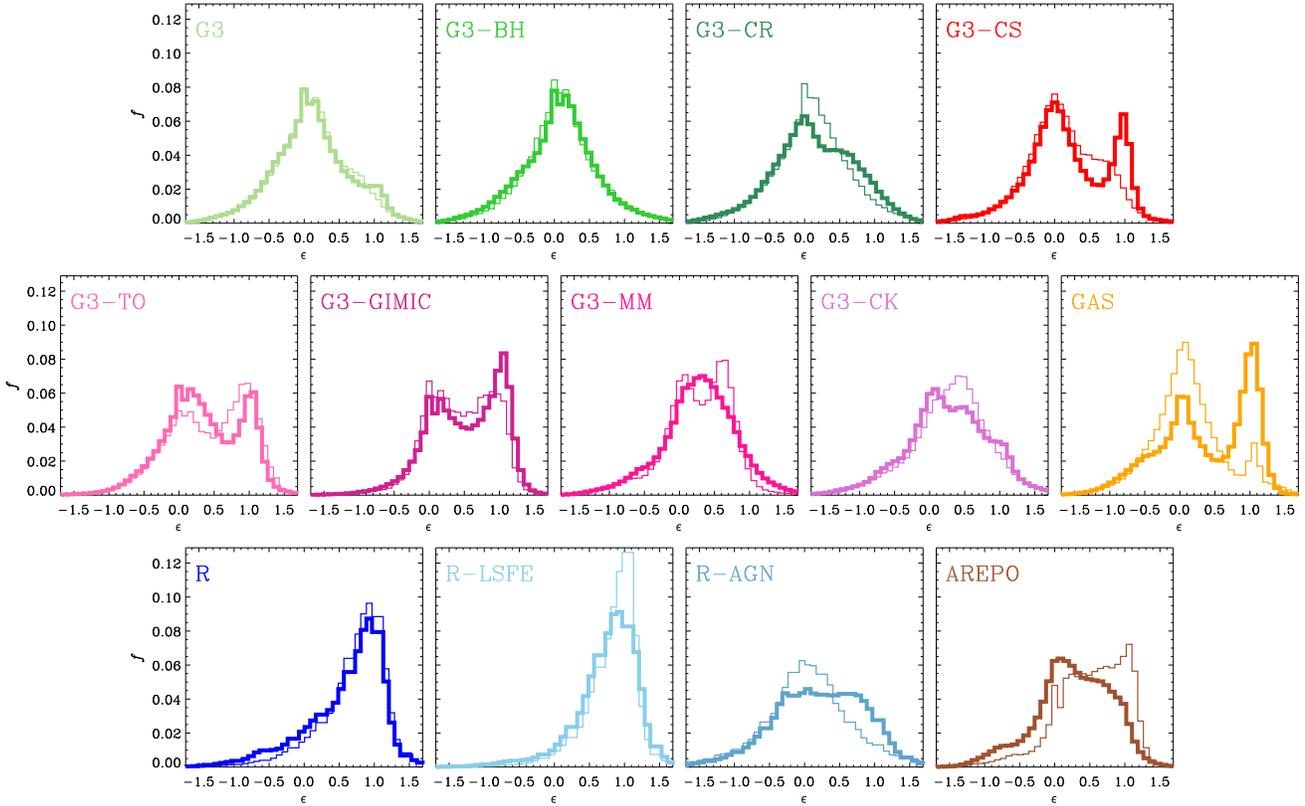}
\caption{Distribution of stellar circularities, $\epsilon = j_z/j_{\rm
    c}$, for the different models.  The circularity
  parameter is the z-component of the specific angular momentum of a
  star particle, $j_z$, expressed in units of the circular orbit value,
  $j_{\rm c}$, {\it at that radius}.  Stars with $\epsilon\approx 1$
  typically belong to a rotationally-supported disk component.  Thick and
  thin lines correspond to level-5 and level-6 resolution runs, respectively.
  \label{fig:circularities}}
\end{center}
\end{figure*}

\begin{table*}
\begin{minipage}{90mm}
\caption{Code parameters for
  simulations at level-5 resolution (level-6 parameters are given between
  parentheses)}
\label{table_params} 
 
\begin{tabular}{lccccc}
\hline\hline

 Code &$f_{\rm b}$&$m_{\rm DM}$  & $m_{\rm gas}$ &\multicolumn{2}{c}{Softening} \\ 
    &($\Omega_{\rm  b}/\Omega_{\rm m}$) & [10$^6$M$_\odot$]  &[10$^6$M$_\odot$] & $\epsilon_{\rm g}^{z=0}$ [kpc] &$z_{\rm fix}$\\
\hline

 G3 &&&&&\\ G3-BH  &&&&&\\G3-CR & 0.16 & $2.2$ & $0.4$ & 0.7 & 0 \\ G3-CS & &
($17$) & ($3.3$) & (1.4) & (0) \\ G3-CK &&&&&\\Arepo&&&&&\\  \hline

  G3-TO &0.18 & $2.1$ & $0.5$ & 0.5 & 3 \\G3-GIMIC& &
  ($17$)&($3.7$)& (1) & (3) \\  \hline

G3-MM & 0.16 & 2.2 & 0.4 & 0.7 & 2 \\
      &      & (17) & (3.3) & (1.4) & (2) \\

 \hline 

  GAS &0.18 &$2.1$ &$0.5$ &0.46 & 8 \\ 
 & &($17$) &($3.7$)&(0.9) &(8) \\

 \hline 

 R &0.16 & $1.4$ & $0.2$ & $0.26$ & $9$\\ R-LSFE & & ($11$)&
 ($1.8$)& ($0.5$)& ($9$)\\ R-AGN &&&&&\\ 
\hline 

\end{tabular}

{\sc note:}
   $f_{\rm b}$: baryon fraction; $m_{\rm DM}$: mass of dark matter
   particles in the high resolution region; $m_{\rm gas}$: initial mass of gas particles;
   $\epsilon_{\rm g}^{z=0}$: gravitational softening at $z=0$; $z_{\rm
     fix}$: redshift after which the gravitational softening is kept fixed in
   physical coordinates. The softening is fixed in comoving
   coordinates at $z>z_{\rm fix}$ (see Appendix~\ref{app:softening}). 

\end{minipage}
\end{table*}

\subsection{Initial Conditions}
\label{SecICs}

All simulations share {\it the same initial conditions} (ICs), a
zoomed-in resimulation of one of the halos of the Aquarius
Project \citep[halo ``Aq-C'', in the notation of][]{Springel08}.  The
ICs were generated using Fourier methods as described in
\citet{Springel08}, modified to include a gas component.  The
displacement field is first calculated on a set of grids and then
interpolated onto the nodes of the unperturbed particle positions, chosen from a
glass-like configuration. For SPH codes these particles represent the
matter distribution, while for the AMR initial conditions they
represent the dark matter only.

The displacement field is used to perturb the particle positions and
to assign them velocities consistent with the growing mode of the
density fluctuations. For AMR runs the density and velocity fields of
the gas are needed on a set of meshes. These quantities were calculated
in the same way as described above except that the displacement and
density fields were interpolated onto the vertices of a pre-defined
set of nested hierarchical grids tailored to requirements of the AMR
code.

For SPH runs, each high-resolution dark matter particle is split into
two to create one dark matter and one gas particle, with relative
masses given by the assumed value of the universal baryon abundance
parameter $\Omega_{\rm b}$ (Table~\ref{table_params}). The positions
of the new particles are such that their center of mass position and
velocity are identical to those of the original particle.  

The selected halo, Aq-C, has a present-day mass similar to the Milky
Way ($\sim 1.6\times 10^{12} \, M_\odot$)
\citep[e.g.][]{Dehnen2006,LiWhite2008,Smith2007,Xue2008,Watkins2010} and has a
relatively quiet formation history. It is also mildly isolated at
$z=0$, with no neighboring halo more massive than half its mass within
a radius of $1\ h^{-1}$ Mpc.  Maps of the dark matter distribution in
boxes of various sizes are shown in Fig.~\ref{DMoverview}.

\subsection{Cosmology}
\label{SecCosm}

We assume a $\Lambda$CDM cosmology with the following parameters:
$\Omega_{\rm m} = 0.25$, $\Omega_{\rm \Lambda}=0.75$, $\sigma_8=0.9$,
$n_{\rm s}=1$, and a Hubble constant of $H_0 =100\,h\,{\rm
  km\,s^{-1}\,Mpc^{-1}} = 73\,{\rm km\,s^{-1}\,Mpc^{-1}}$.  These
parameters are consistent with the WMAP 1- and 5-year results at the
$3\sigma$ level and are identical to the parameters used for the
Millennium and Millennium-II simulations
(\citealt{Millenium,MilleniumII}).  The value of $\Omega_{\rm b}$ used
in each simulation is given in Table~\ref{table_params}.

The Millennium-II is, in fact, a resimulation of the cosmological
volume from which the Aquarius halos were originally selected\footnote{ 
In the convention of the Aquarius Project (lower numbers indicate
higher resolution), the Millennium-II simulation
has a resolution intermediate between levels 5 and 6 (the two resolutions
used in our set of simulations).}.  Aq-C
is thus present both in this simulation and in the semi-analytic model
of \citet{Guo2011} which was tuned to fit the luminosity, stellar
mass, size and gas content functions measured for galaxies in the
Sloan Digital Sky Survey (SDSS).  Similarly, \citet{Cooper2010} used
the {\sc galform} code to model star formation in all six Aquarius
halos\footnote{ In this case, the level-$4$ resolution simulations were used.}, 
using parameters very similar to those of \citet{Bower06}, which
reproduce local galaxy luminosity functions. The properties found for
the central galaxy of Aq-C in these two models thus give an indication
of what direct simulations should produce if implementation on a
cosmological volume is to reproduce observed galaxy abundances.

\subsection{The Runs}
\label{SecRuns}

All $13$ simulations were run at two different numerical resolutions:
level 5 and level 6, respectively, following the naming convention of
the Aquarius project (lower numbers indicate higher resolution).
Table~\ref{table_params} gives the dark matter and (initial) gas
particle masses for the two resolutions.  The gravitational softening
is kept fixed in comoving coordinates until redshift $z_{\rm fix}$,
after which its value is fixed in physical coordinates. The simulations vary
in their choice of $z_{\rm fix}$ and therefore the gravitational
softening has slightly different values at $z=0$, listed as
$\epsilon_{\rm g}^{z=0}$ in Table~\ref{table_params} (see also
Fig.~\ref{fig:softenings}).

\subsection{Analysis}

We describe here some of the conventions and definitions used in the
analysis of the simulated galaxies.  The center of the galaxy is
defined to coincide with the position of the baryonic particle
with minimum gravitational potential.  The \emph{virial radius},
$r_{200}$, is the radius of a sphere, centered on the galaxy, with
mean density equal to $200\, \rho_{\rm crit}$, where $\rho_{\rm crit} =
3H^2/8\pi G$ is the critical density for closure.  We use the term
\emph{halo} to refer to all the mass within $r_{\rm 200}$ and
\emph{galaxy} to the baryonic component within a radius $r_{\rm gal} =
0.1\times r_{200}$ from the center.

Where a distinction is drawn between ``hot'' and ``cold'' gas, we
adopt a temperature threshold of $10^{5}$K to separate the two phases.
Some codes (e.g. {\small G3}, {\small G3-BH}, {\small G3-CR}, {\small
  G3-GIMIC}, {\small G3-TO}, {\sc arepo}) adopt an
``effective'' equation of state to describe the ISM and
to circumvent numerical instabilities
in poorly-resolved regions. This may cause some fluid elements to have
nominal temperatures in excess of $10^{5}$K, but still be star-forming. 
In order to prevent assigning this gas to the hot phase, we
automatically assign all star-forming gas particles to the ``cold''
phase.

As we shall see below, different runs yield simulated galaxies of
widely-varying baryonic mass and angular momentum.  In particular,
  not only the specific angular momentum changes between simulations
  (as expected,  given the wide range in galaxy mass spanned by the
  various runs), but also its orientation, as we discuss in
  Appendix~\ref{app:disk_orientation}.  Because of this, for
orientation-dependent diagnostics, we rotate each simulated galaxy to
a new coordinate system where the angular momentum vector of its
stellar component coincides with the $z$ direction.

\section{Results}\label{sec:results}

We present here results concerning the stellar mass, morphology, size,
star formation history, and angular momentum content of the simulated
galaxy. Unless otherwise specified, all results correspond to $z=0$
and to the level-5 resolution runs. Numerical convergence between
level-5 and level-6 runs is discussed in \S~\ref{sec:resolution}.

\subsection{Galaxy morphology}
\label{SecMorph}

Fig.~\ref{fig:stellar_maps} shows face-on and edge-on maps of the
projected stellar mass density for the $13$
runs.  Labels in each panel
list the simulation name, as given in Table~\ref{table_codes}, as well
as the total stellar mass of the galaxy (i.e., within $r_{\rm gal}$).

These figures illustrate the complex morphology of the simulated
galaxies; bars, bulges, and extended disks are present, but their
relative prominence varies widely from run to run. The galaxy stellar
mass also shows large scatter, spanning about a decade from the least
({\small G3-TO}) to the most massive ({\small R}), respectively.

A quantitative measure of the importance of a rotationally-supported
component is provided by the distribution of stellar {\it
  circularities}, $\epsilon$, defined as the ratio between the
$z$-component of the specific angular momentum of a star and that of a
circular orbit at the same radius $r$:
\begin{equation}
\epsilon = {j_z\over{j_{\rm c}(r)}} = {j_z\over{r\ V_{\rm c}(r)}},
\end{equation}
where $V_{\rm c}(r)=\sqrt{GM(<r)/r}$ is the circular velocity. Stars
belonging to a disk are expected to have $\epsilon \sim 1$, whereas
stars belonging to a non-rotating spheroidal component should have an
$\epsilon$-distribution roughly symmetric around zero \citep[see,
e.g.,][]{Abadi2003b,CS2009}.

\begin{figure}
\begin{center}
{\includegraphics[width=8.5cm]{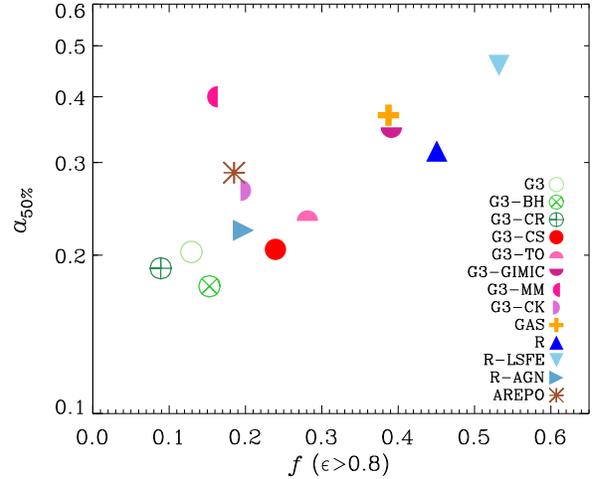}}
\caption{Median formation time of stars in the galaxy at $z=0$,
  expressed in terms of the expansion factor,
  $a_{50\%}=1/(1+z_{50\%})$, as a function of the fraction of stars
  with circularity exceeding $0.8$. }
\label{FigEcca50}
\end{center}
\end{figure}

We show the circularity distribution of all $13$ runs in
Fig.~\ref{fig:circularities}. Thick and thin lines correspond to
the level-5 and level-6 resolution simulations, respectively.  The
diversity in morphology seen in Fig.~\ref{fig:stellar_maps} 
is clearly reflected in the distribution of
circularities. Thin disks that appear prominently in the images show
up as well-defined peaks in the circularity distribution at $\epsilon
\sim 1$, a distinction that sharpens at higher numerical
resolution. In some cases, notably {\small G3}, {\small G3-MM},
{\small G3-CK}, and {\sc arepo}, the galaxy is noticeably flattened
and clearly rotating, but lacks a prominent thin disk.

The importance of a thin disk may be crudely estimated by the fraction
of stars with $\epsilon > 0.8$, $f(\epsilon>0.8)$\footnote{Note,
  however, that these fractions often compare poorly with photometric
  estimates of the disk-to-total ratios \citep{Abadi2003a,CS2010}.}. 
Only in four
simulated galaxies  do more than $\sim 40\%$ of stars satisfy this
condition, two SPH-based and two AMR-based: {\small R}, {\small
  R-LSFE}, {\small G3-GIMIC}, and {\sc gas}. The most extreme case,
{\small R-LSFE}, provides a clue to this behaviour. In this simulation
feedback is inefficient and star formation is deliberately delayed,
allowing gas to accrete into the galaxy and settle into a
centrifugally-supported structure before turning into stars.

Indeed, any mechanism that hinders the early transformation of gas
into stars without curtailing gas accretion later on is expected to
promote the formation of a disk \citep[see,
e.g.,][]{NavarroSteinmetz1997}. As a result, the galaxies with most
prominent disks are also the ones with the youngest stars \citep{Agertz11}.  This is
shown in Fig.~\ref{FigEcca50} , where we plot $f(\epsilon>0.8)$ versus the
median formation time of all stars in the galaxy (expressed in terms
of the expansion factor, $a_{50\%}$). A clear correlation emerges, with
disks increasing in prevalence in galaxies that make their stars
later. On the other hand, galaxies that make their stars early tend to
be spheroid-dominated.

An interesting outlier to this trend is {\small G3-MM}, which forms
stars as late as {\small R} but has a small fraction of stars in a
disk. Further investigation shows that the {\small G3-MM} galaxy did
harbour a disk, but it was severely impacted by a collision with a
massive satellite in recent times. 
The satellite is present in other
runs, but it has not yet collided with the main galaxy in the majority
of cases. This is due to the fact that even small differences in the
early evolution get amplified with time and can lead to large
discrepancies in the orbital phase of satellites later on. To the
extent that this can influence the morphology of the central galaxy, a
certain degree of stochasticity in the morphological evolution of a
simulated galaxy seems unavoidable.

Another interesting result to note is that neither {\small G3} nor
{\sc arepo} form prominent thin disks.  These two runs share the same
sub-grid physics, but use very different numerical hydrodynamical
techniques, which suggests that the morphology of the simulated
galaxies is indeed rooted mainly in how gas gets accreted and
transformed into stars and in the merger history of the particular
halo.  As discussed recently by, e.g., \citet{Torrey2011}, the
numerical scheme {\it does} make a difference when considering the
detailed properties of simulated disks, but it does not seem to be the
main reason why the {\small G3} and {\sc arepo} runs lack disks in
this halo. Rather, the failure of feedback to regulate effectively the
onset of early star formation and to allow for late gas accretion
seems the most likely culprit. Support for this interpretation is
provided by {\small G3-BH} and {\small G3-CR} which, despite the
increased feedback, fail to prevent most stars from forming quite
early. As a result, none of these models allows a sizeable thin disk
to develop.

Models with more efficient feedback schemes, such as those where
feedback regulates more effectively early star formation through
galactic winds (e.g., {\small G3-CS}, {\small G3-TO}, {\small
  G3-GIMIC}, {\sc gas}), yield galaxies with two well-defined
components: an old, non-rotating spheroid surrounded by a young
rotationally-supported disk. Still, even in this case the disk
component is subdominant in terms of total stellar mass, with
$f(\epsilon>0.8)$ around $\sim 30$-$40\%$.

Finally, it is worth noting that the morphology of a galaxy is
  often dissimilar at the two resolutions attempted here. In general,
  more prominent disks form at higher resolution but in some
  cases this trend is reversed. We further discuss
  resolution effects in Section~\ref{sec:resolution}.

\begin{figure*}
\begin{center}
\includegraphics[width=12cm]{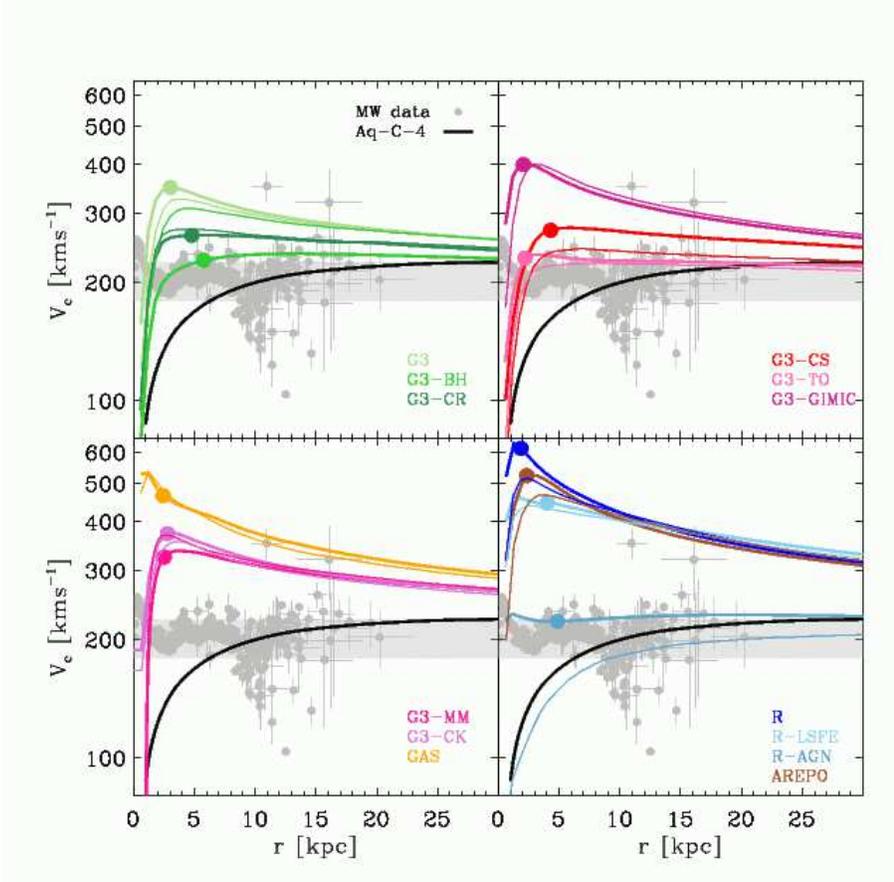}
\caption{Circular velocity curves of all galaxies in the level-5 runs, for
  the resolution level-5 runs.  The four panels group the results according
  to numerical technique.  The top-left panel corresponds to various
  feedback choices of the standard {\sc gadget} code; the top-right
  and bottom-left correspond to other, independent star
  formation/feedback modules developed for {\sc gadget}, as well as
  the SPH-based {\sc gasoline} code. The bottom-right panel groups the
  results of the AMR code {\sc ramses} and the moving-mesh code {\sc
    arepo}.    Thick and
  thin lines correspond to level-5 and level-6 resolution runs, respectively.
  The solid circles indicate, for the level-5 simulations, the position of the stellar
  half-mass radius of each simulated galaxy. 
  The thick black line
  shows the circular velocity of the dark-matter-only simulation of
  the same halo (Aq-C).  For reference, the region shaded in light
  grey is bounded by the peak and virial velocities of the Aquarius
  halo.  Dark grey points with error bars are observed data for the
  Milky Way's rotation curve, as compiled by \citet{Sofue2009}.
  \label{fig:vcirc}}
\end{center}
\end{figure*}

\subsection{Rotation curves}
\label{SecRotCur}

As discussed above, regulating star formation without hindering the
formation of a stellar disk is a challenging task for any feedback
implementation. One might argue that a solution might simply be to
delay star formation, such as in {\small R-LSFE}, but this comes at
the expense of unrealistic disk properties. A simple and convincing
diagnostic is the ``rotation curve'' of the disk which, for
simplicity, we represent by the circular velocity profile of the
galaxy, $V_c(r)$.

This is shown in Fig.~\ref{fig:vcirc}, where we group in four panels
the results of the $13$ level-5 Aquila runs, and compare them with the
circular velocity curve of the dark matter-only Aq-C run (dark solid
line) and, for reference, with the rotation curve of the Milky Way as
compiled by \citet{Sofue2009}\footnote{Note that \citet{Sofue2009}
  assume a Galacto-centric position and velocity of the sun of $8$ kpc
  and $200$ km s$^{-1}$, respectively.}.

This figure makes clear that the ``best disks'' in terms of morphology
(i.e., {\small R-LSFE}, {\small R}, {\sc gas}, and {\small G3-GIMIC})
all have steeply {\it declining} rotation curves, at odds with the flat
rotation curves characteristic of normal spirals. The extreme {\small
  R-LSFE} model again illustrates the problem: here feedback is
inefficient at removing baryons, allowing large amounts of gas to
collect in a central disk before being turned into stars. A large
fraction of these baryons have relatively low angular momentum,
however, leading to the formation of a disk that is unrealistically
concentrated and with a declining rotation curve. (A similar
consideration applies to {\small G3} and {\sc arepo}.) It seems
one could argue that a successful feedback mechanism must selectively
{\it remove} low-angular momentum material from the galaxy (see, e.g.,
\citealt{NavarroSteinmetz1997,vandenBosch2002,Brook11}).

Support for this view comes from inspection of the rotation curves of
galaxies where galactic winds play a more substantive
role, especially at high-redshift, when low-angular momentum baryons
are preferentially accreted: {\small G3-CS} and {\small
  G3-TO} show nearly flat rotation curves. A similar result is found
for {\small G3-BH}, {\small G3-CR} and {\small R-AGN}, but in these
cases the ``success'' must be qualified by noting that none of these
galaxies have a clearly-defined stellar disk: the flat $V_c$-curves
are just a reflection of the low baryonic mass of the galaxy that
results from adopting these extremely effective feedback models.

\begin{figure*}
\begin{center}
\includegraphics[width=12cm]{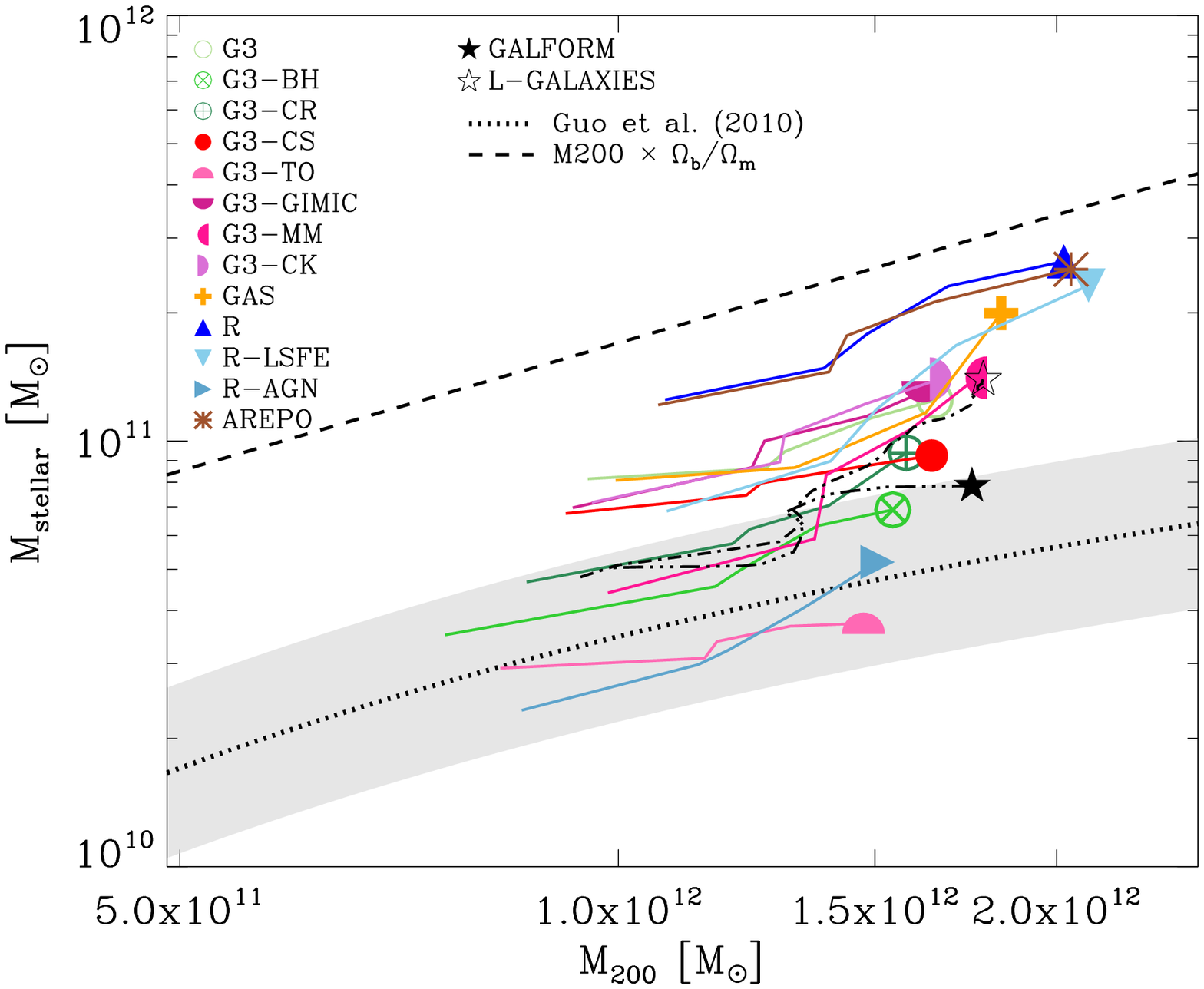}
\caption{The stellar mass of the central galaxy as a function of the
  virial mass of the surrounding halo. Curves of different color track
  the evolution of the galaxy in each simulation between $z=2$ and
  $z=0$. The dotted line indicates the stellar mass {\it expected} at
  $z=0$ from the abundance-matching analysis of \citet{Guo2010}; the
  shaded region corresponds to a $0.2$ dex uncertainty.  The dashed
  line indicates the mass of all baryons within the virial radius,
  $(\Omega_{\rm b}/\Omega_{\rm m}) \,M_{200}$.  The filled and open
  star symbols indicate 
  the predictions of the semi-analytic models {\sc galform}
  \citep{Cooper2010} and {\sc l-galaxies} \citep{Guo2011} for halo Aq-C, 
  respectively. The dot-dashed curves show the evolution since $z=2$
  according to these two
  models.
  \label{fig:mstar_mhalo}}
\end{center}
\end{figure*}

\begin{figure*}
\begin{center}
\includegraphics[width=12cm]{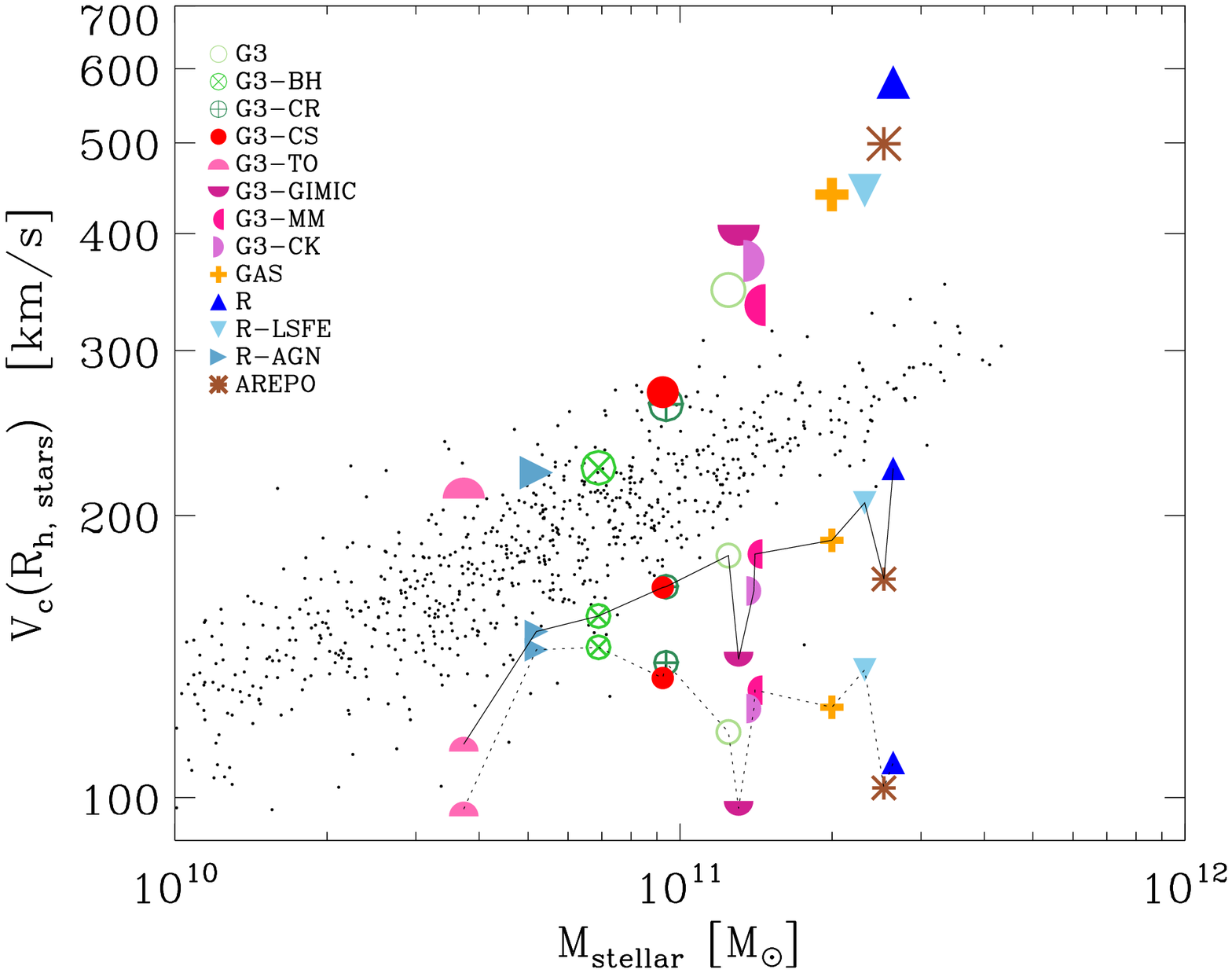}
\caption{ The Tully-Fisher relation.  The circular velocity at the
  stellar half-mass radius of each simulated galaxy is plotted as a
  function of stellar mass for all $13$ level-5 runs.
  Small black dots correspond to data for nearby spirals taken
  from \citet{Pizagno2007}, \citet{Verheijen2001} and
  \citet{Courteau2007}. The symbols connected by a solid line show the
  contribution of the dark matter to the circular velocity at $R_{\rm h, stars}$. Those
connected by the dotted line show the circular velocity of the dark
matter-only halo (Aq-C) at the same radii.
\vspace{0.4cm}
\label{fig:tully_fisher}}
\end{center}
\end{figure*}

\subsection{Stellar Mass}
\label{sec:mstar_mhalo}

The stellar mass of a galaxy is determined by the combined effects
of radiative cooling, the rate at which cold gas is transformed into
stars, and the ability of feedback to regulate the supply of
star-forming gas.  Fig.~\ref{fig:mstar_mhalo} shows how the various
implementations affect the stellar mass of the central galaxy, $M_{\rm
  stellar}$. This figure shows $M_{\rm stellar}$ as a function of 
$M_{200}(z)$, the
virial mass of the  main progenitor from $z=2$ to
$z=0$. (The symbols correspond to values at $z=0$.)

To guide the interpretation, we show with a dashed curve the total
baryonic mass within the virial radius corresponding to the universal
baryon fraction, $(\Omega_{\rm b}/\Omega_{\rm m}) M_{200}$, which sets
a hard upper limit to the stellar mass of the central galaxy.  The
shaded region surrounding the dotted curve corresponds to the stellar
masses predicted, at $z=0$, by requiring agreement between the halo
mass and galaxy stellar mass functions through simple abundance
matching \citep{Guo2010}. We also show the prediction
for Aq-C of the semi-analytic models {\sc galform}
\citep{Cooper2010} and {\sc l-galaxies} \citep{Guo2011}.

The most striking feature of Fig.~\ref{fig:mstar_mhalo} is the large
code-to-code scatter in the stellar mass of the galaxy, which varies
between $\sim 4\times 10^{10}\, M_\odot$ and $\sim3\times 10^{11} \,
M_\odot$ at $z=0$ (Table~\ref{TabFigParams}). The three largest
stellar masses are obtained with the mesh-based codes, {\small R},
{\small R-LSFE} and {\sc arepo}, and correspond to assembling nearly
all available baryons in the central galaxy. This illustrates the weak
efficiency of the feedback implementations chosen for these codes,
aided by the fact that, at comparable resolution, cooling efficiency
is enhanced in mesh-based codes relative to SPH
\citep{Vogelsberger2011,Sijacki2011,Keres2011}.

Indeed,  {\small G3} forms 
only about half as many stars as {\sc arepo}, despite sharing
exactly the same sub-grid physics. The galaxy formed by {\sc gas} also
has a large stellar mass, but this may be due to the fact that this
code has a more efficient cooling function through the addition of
metal-line cooling.  Although the numerical technique may effect some
changes in the stellar mass, these are small compared with the
variations introduced by the feedback implementation. This may be seen
by noting that, when including AGN feedback in {\sc ramses}, the
stellar mass decreases by a factor of $\sim 5$ and the disk component
is largely erased.

 Note that the large variations in the stellar mass
  of the galaxy formed in different runs imply that the dark halo will
  respond by contracting differently in each case, as we discuss in
  Section~\ref{sec:TF}.
Note also that the differences in stellar mass are dominated by differences 
prior to $z=2$; in fact, in some simulations the stellar mass at $z=2$ is already
above the $z=0$ stellar mass-halo mass relation.

It is also important to note that feedback {\it must} be roughly as
effective as that of {\small R-AGN} in order to obtain stellar masses
consistent (within the error) with the abundance-matching
predictions. Indeed, the only other codes to match this constraint,
and thus fall within the shaded area of Fig.~\ref{fig:mstar_mhalo} are
{\small G3-BH} and {\small G3-TO}; of these only the latter forms a
galaxy with a discernible disk (see Fig.~\ref{fig:stellar_maps}). All
other models give stellar masses well in excess of the
abundance-matching constraint, a shortcoming of most published galaxy
formation simulations to date \citep{Guo2010,Sawala11}. 

It is also worth noting that the abundance-matching models allow for
substantial scatter in the $M_{200}$-$M_{\rm stellar}$
relation. Indeed, the more sophisticated treatments of the {\sc
  l-galaxies} and {\sc galform} semianalytic codes indicate that Aq-C
might form a galaxy more massive than expected on average for a halo
of that mass (see open and filled starred symbols in
Fig.~\ref{fig:mstar_mhalo}). {\sc l-galaxies}, in particular, suggests
that Aq-C might be a 2$\sigma$ outlier from the relation, which would
alleviate, but not resolve, the disagreement between the results of
{\small R}, {\small R-LSFE}, {\sc arepo}, and {\sc gas} and the model
predictions. {\sc galform}, on the other hand, predicts 
that  Aq-C should be about $1\sigma$ above the mean abundance-matching
relation.

Taken altogether, these results illustrate the basic challenge faced
by disk galaxy formation models: feedback must be efficient enough
either to prevent the accretion, or to facilitate the removal, of most
baryons, whilst at the same time allowing enough high-angular material
to accrete and form an extended stellar disk.

\subsection{Tully-Fisher relation}\label{sec:TF}

The stellar mass and circular velocity of disk galaxies are strongly
linked by the Tully-Fisher relation, and it is therefore instructive
to compare the properties of simulated galaxies with those of observed
disks. This is done in Fig.~\ref{fig:tully_fisher}, where we compare
data compiled by \citet{Dutton2011} 
from \citet{Pizagno2007}, \citet{Verheijen2001}, and
\citet{Courteau2007} with the $13$ simulated galaxies. 

Because the rotation curves of simulated galaxies are not flat (see
Fig.~\ref{fig:vcirc}), we use velocities estimated at the stellar
half-mass radius in order to be as consistent as possible with the
rotation speeds estimated observationally from spatially-resolved
rotation curves \citep[see, e.g.,][]{Courteau2007}. The symbols
connected by a solid line show the contribution of the dark matter to
the circular velocity {\it at the same radius}. A dotted line shows
the same, but for the dark matter-only Aq-C halo; the difference
between solid and dotted curves indicates the degree of
``contraction'' of the dark halo.

There is a clear discrepancy between the observed Tully-Fisher
relation and simulated galaxies, which tend to have substantially
larger velocities at given $M_{\rm stellar}$. The disagreement worsens
for large stellar masses, emphasizing again the fact that too many
baryons are able to cool and form stars in these systems.
Interestingly, at low stellar mass simulated galaxies approach the
observed relation but still have, on average, higher rotation speeds
than typical disks. This suggests that, although these galaxies may
have stellar masses consistent with abundance-matching considerations
(see \S~\ref{sec:mstar_mhalo}), they must differ from typical spirals in
other respects, such as an excessive concentration of the dark
matter or luminous component. 

The dark matter {\it contribution} to the circular velocity (connected
symbols in Fig.~\ref{fig:tully_fisher}) lies well below the average
rotation speed expected from the Tully-Fisher relation. This suggests
that the concentration of dark matter is not the origin of the
disagreement; there should in principle be no problem matching the
observed relation provided that the luminous component of the galaxy
is extended enough. The offset from the observed Tully-Fisher relation
thus suggests that simulated galaxies are more concentrated than
normal spirals, resulting in disks that rotate too fast for their
stellar mass.  We analyze the size of simulated galaxies in
\S~\ref{sec:sizes}, after examining the importance of the
gaseous component of the galaxy next.

\begin{figure*}
\begin{center}
\includegraphics[width=12cm]{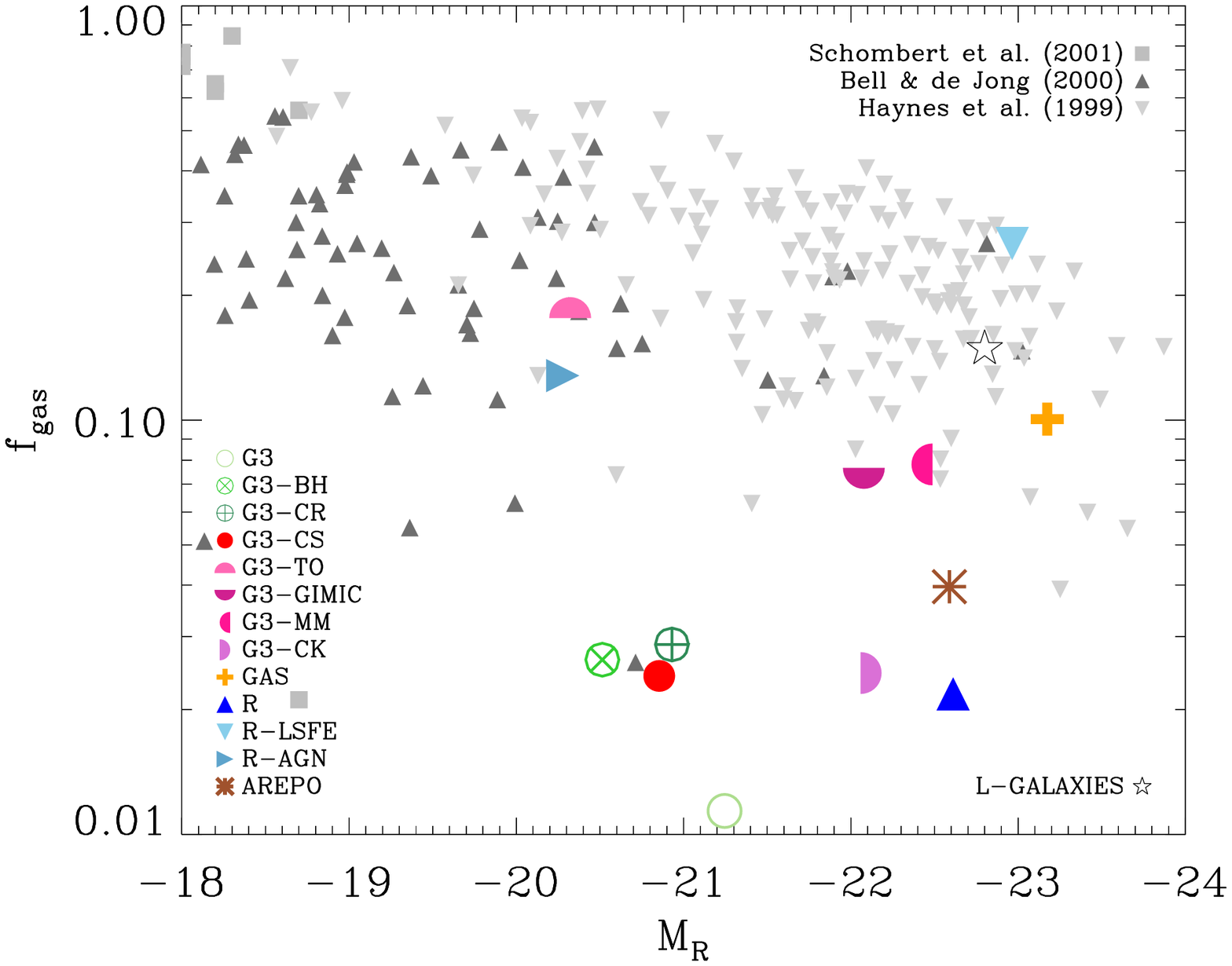}
\caption{Gas mass fraction, $f_{\rm gas}=M_{\rm gas}/(M_{\rm
    gas}+M_{\rm stellar})$, of the galaxy versus R-band absolute
  magnitude. Magnitudes have been calculated using the Bruzual \&
  Charlot (2003) population synthesis models for solar metallicity and
  a Chabrier IMF and ignoring the effects of dust extinction.  Symbols
  in grey/black show data for nearby spirals compiled from the
  references listed in the figure label. We also show the cold gas
  fraction prediction of the semi-analytic model {\sc l-galaxies}
  \citep{Guo2011} for Aq-C.  The gas fraction predicted by {\sc galform} 
\citep{Cooper2010} is close to zero and thus lies outside
  the plotted range.\label{fig:gasfraction}}
\end{center}
\end{figure*}

\subsection{Gaseous component}
\label{SecGasMass}

Fig.~\ref{fig:gasfraction} shows $f_{\rm gas}$, the fraction of the
baryonic mass of simulated galaxies in form of 
gas at $z=0$, as a
function of the R-band absolute magnitude, and compares them with data
for star forming galaxies from \citet{Schombert2001},
\citet{Bell_deJong2000} and \citet{Haynes1999}. Magnitudes for the
simulations were calculated using the dust-free \citet{BC03}
population synthesis models, for a Chabrier Initial Mass Function (IMF) and solar metallicity.

Most simulated galaxies have gas fractions below $10\%$, which puts
them at odds with observations of nearby spirals.  As for the stellar
mass, note the large code-to-code scatter in $f_{\rm gas}$, which
varies from about $1\%$ for {\small G3} to nearly $30\%$ for {\small
  R-LSFE}. As expected, galaxies with larger gas fractions are
predominantly those with morphologies that include a well-defined
stellar disk, presumably because of ongoing star formation. The
converse, however, is not always true: {\small G3-CS} and {\small R}
have low gas fractions at $z=0$ but prominent disks. 

More surprisingly perhaps, the gas fraction seems to correlate only
weakly with the present-day star formation rate (which we discuss more
thoroughly in \S~\ref{SecSFH}). For example, {\small R-AGN} has the
third largest gas fraction and by far the lowest star formation rate
at $z=0$. The same applies to {\small G3-TO}, which, despite its large
$f_{\rm gas}$, forms stars at rates well below what would be expected
for an average spiral (see Fig.~\ref{fig:sfr}).

Overall we see no obvious dependence of the gas fraction on the
numerical method: of the four galaxies with highest $f_{\rm gas}$, two
are SPH-based ({\small G3-TO} and {\small G3-MM}) and two are
AMR-based ({\small R-AGN} and {\small R-LSFE}). However, we note that
{\sc arepo} has a much higher gas fraction (and stellar mass) than
{\small G3}, despite sharing the same sub-grid physics. This supports
the conclusion that standard SPH-based methods may underestimate the
total amount of gas that cools and becomes available for star
formation, especially when feedback is as weak as implemented in the
{\small G3} and {\sc arepo} runs (see also \citealt{Agertz07,Vogelsberger2011}).

\begin{figure*}
\begin{center}
{\includegraphics[scale=0.335]{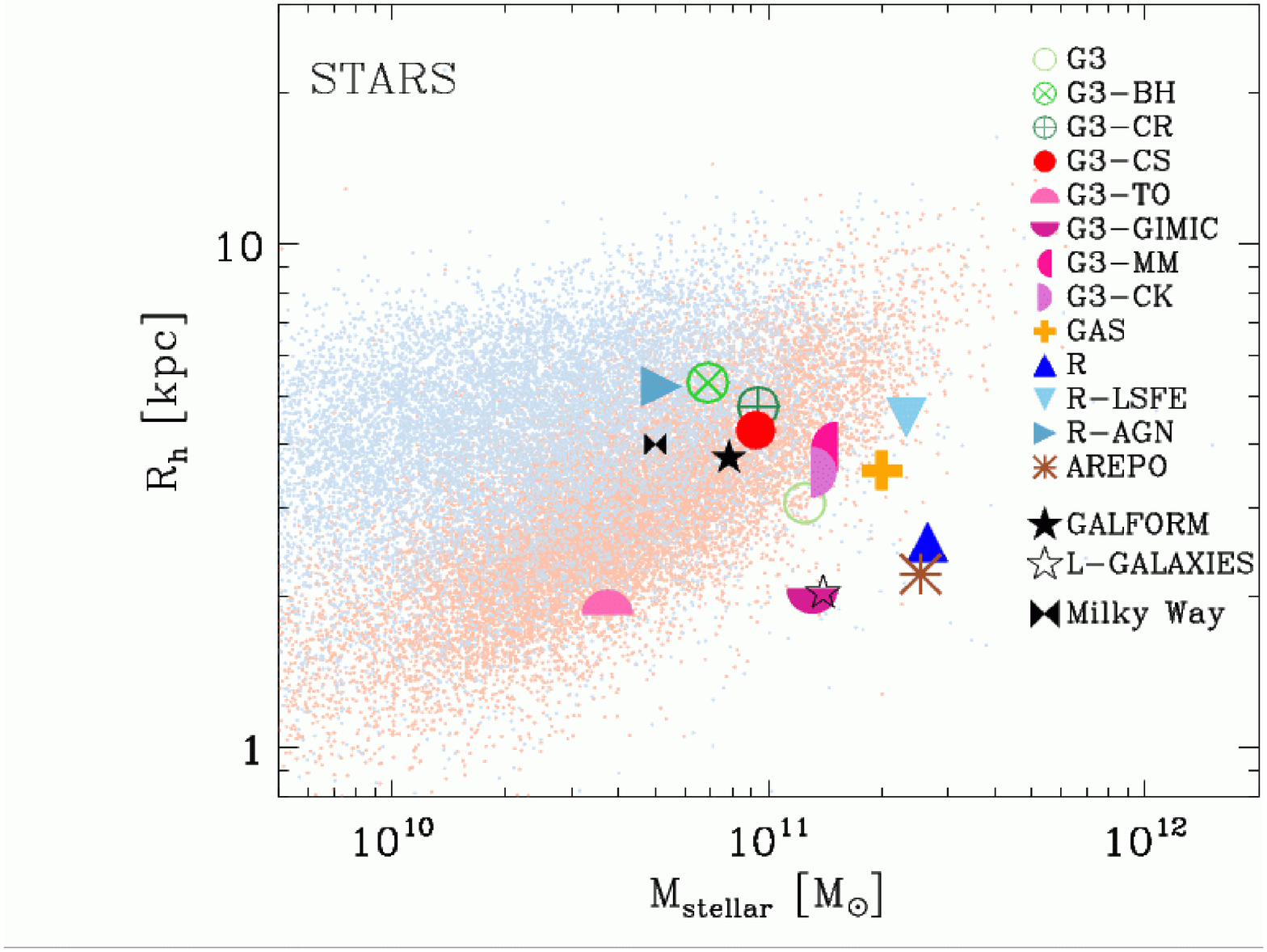}\includegraphics[scale=0.45]{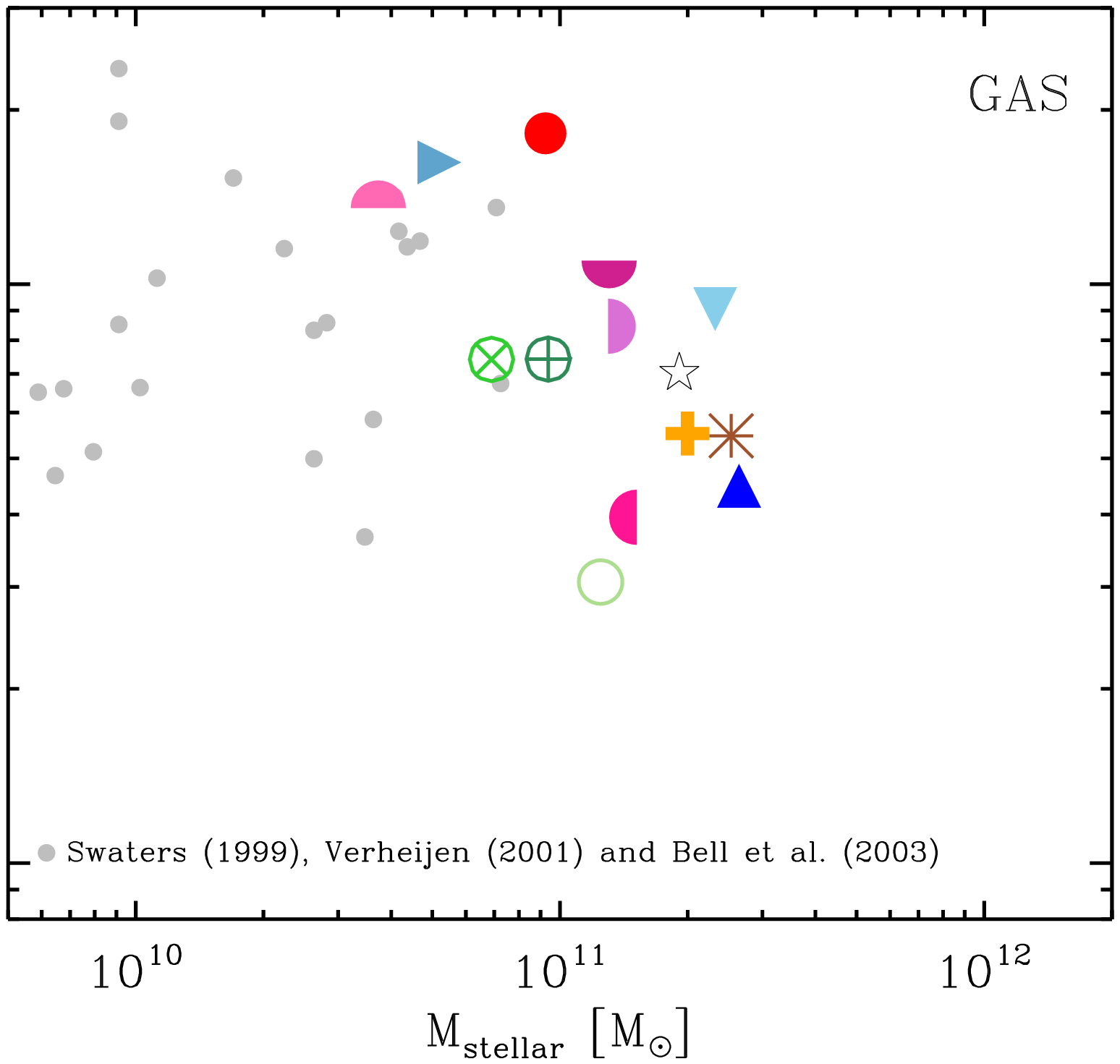}}
\caption{{\it Left:} Stellar half-mass radius as a function of the
  stellar mass of the galaxy. Blue and red dots show the Petrosian
  $r$-band half light radius for a sample of nearby ($z<0.1$) SDSS
  galaxies taken from the MPA-JHU DR7 release. The sample is split
  into ``blue cloud'' and ``red sequence'' galaxies depending on their
  colors, according to the condition $(g-r)=0.59+0.052
  \log_{10}(M_{\rm stellar}/M_\odot)-10.0$ (only $5$ per cent of
  randomly selected data points are shown).  We also show the
  predictions of the semi-analytic models {\sc galform}
  \citep{Cooper2010} and {\sc l-galaxies} \citep{Guo2011} for Aq-C,
  and the approximate location of the Milky Way in this plot.  {\it
    Right:} The projected half-mass radius of cold ($T<10^5K$) gas in
  the galaxy as a function of stellar mass. Grey circles indicate the
  half-mass radii of HI disks compiled by \citet{Dutton2011} from \citet{Swaters1999}
  and \citet{Verheijen2001}, and
  the open star symbol indicates the prediction of {\sc l-galaxies}.
  The prediction of {\sc galform} is not shown here, since it predicts
  a present-day gas mass close to zero.
\label{fig:sizes}}
\end{center}
\end{figure*}

The interpretation of these results is not straightforward. The gas content
of a galaxy is constantly evolving; supplied by accretion, depleted by
star formation, and removed by feedback-driven winds. This leads to
large fluctuations in the instantaneous gas fraction and star
formation rate of a galaxy, which may be exacerbated by chance events
such as satellite accretion.

\subsection{Galaxy size} 
\label{sec:sizes}

The left-hand panel of Fig.~\ref{fig:sizes} compares the stellar
half-mass radii of simulated galaxies with the size of observed
galaxies of similar stellar mass\footnote{Data taken from the SDSS
  MPA-JHU DR7 release for nearby ($z<0.1$) galaxies; {\tt
    http://www.mpa-garching.mpg.de/SDSS/DR7/} }, as well as with the
predictions of the {\sc galform} and {\sc l-galaxies} 
semi-analytic models 
 and the Milky Way, for reference. The
observed sizes correspond to Petrosian half-light radii in the
$r$-band rather than stellar half-mass radii so the comparison should
only be taken as indicative. Red/blue dots correspond to galaxies
redder/bluer than $(g-r)=0.59+0.052 \log_{10}(M_{\rm
  stellar}/M_\odot)-10.0$ and are meant to outline roughly the
location of early- and late-type galaxies in this plot.

As anticipated in the previous subsection, most galaxies are too
compact to be consistent with typical spiral galaxies of comparable
stellar mass. In general, the more massive the simulated galaxy, the
smaller its size, a trend that runs contrary to observation. In
particular, the most massive simulated galaxies ({\small R}, {\small
  R-LSFE}, {\sc arepo}) are even smaller than most early-type galaxies
in the nearby Universe, highlighting again the shortcomings of
simulations where cooling and star formation proceed largely unimpeded by
feedback.

Simulations where feeedback is more effective at curtailing the
formation of stars give rise to galaxies with sizes in better
agreement with observation. For example, the half-mass radii of
{\small R-AGN} and {\small G3-BH} reach $R_{\rm h}\sim 5$ kpc for $\sim 7
\times 10^{10} \, M_\odot$, at the lower end of the
distribution of spiral sizes. However, as discussed in
\S\ref{SecMorph}, neither of these galaxies has a disk-like morphology. 

The simulated galaxy with lowest $M_{\rm stellar}$ and a well-defined
disk is {\small G3-TO}, but, as seen from Fig.~\ref{fig:sizes}, it has a
half-mass radius of less than $2$ kpc, well below what would be
expected for a spiral of that mass. This is because most of the
stellar mass in {\small G3-TO} is in the form of a highly-concentrated
spheroid rather than in an extended disk. Therefore, even in this
case, feedback has apparently allowed too many low angular momentum
baryons into the galaxy. 

These results support our earlier conclusion: feedback must not only
limit how many baryons settle into the galaxy, but must also
selectively allow high angular momentum material to be accreted and
retained in order to form a realistic disk.

The right-hand panel of Fig.~\ref{fig:sizes} shows the projected
half-mass radius of cold gas in the simulated galaxies (at $z=0$) as a
function of stellar mass and compares them with HI observations
compiled by \citet{Dutton2011} from \citet{Swaters1999} and \citet{Verheijen2001}.
Despite the large code-to-code variation, the
simulated gaseous disks are systematically more extended than the
stellar component, in agreement with observation. They are also in
better agreement overall with the typical size of HI disks, a result
that suggests that material accreted relatively recently (and thus
still in gaseous form) has, on average, enough angular momentum to form
disks of realistic size. If feedback were able to favor the accretion
and retention of this late-accreting, high-angular momentum gas, then
simulated galaxies would have a much better chance of forming realistic disks.

\begin{figure}
\begin{center}
{\includegraphics[width=8.5cm]{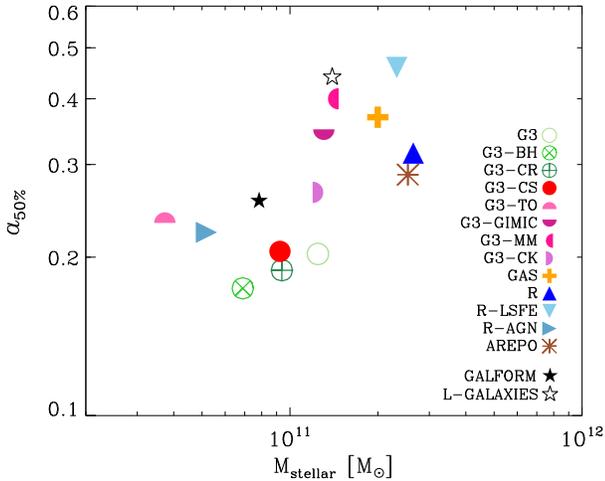}}
\caption{Median formation time of stars in the galaxy (expressed in
  terms of the expansion factor ($a_{50\%}$) as a function of total
  stellar mass at $z=0$.  We also show the
  prediction of the semi-analytic models {\sc galform} \citep{Cooper2010}
  and {\sc l-galaxies} \citep{Guo2011} for Aq-C.}
\label{FigMstarTform}
\end{center}
\end{figure}

\subsection{Star formation history}
\label{SecSFH}

\begin{figure*}
\begin{center}
\includegraphics[width=8.5cm]{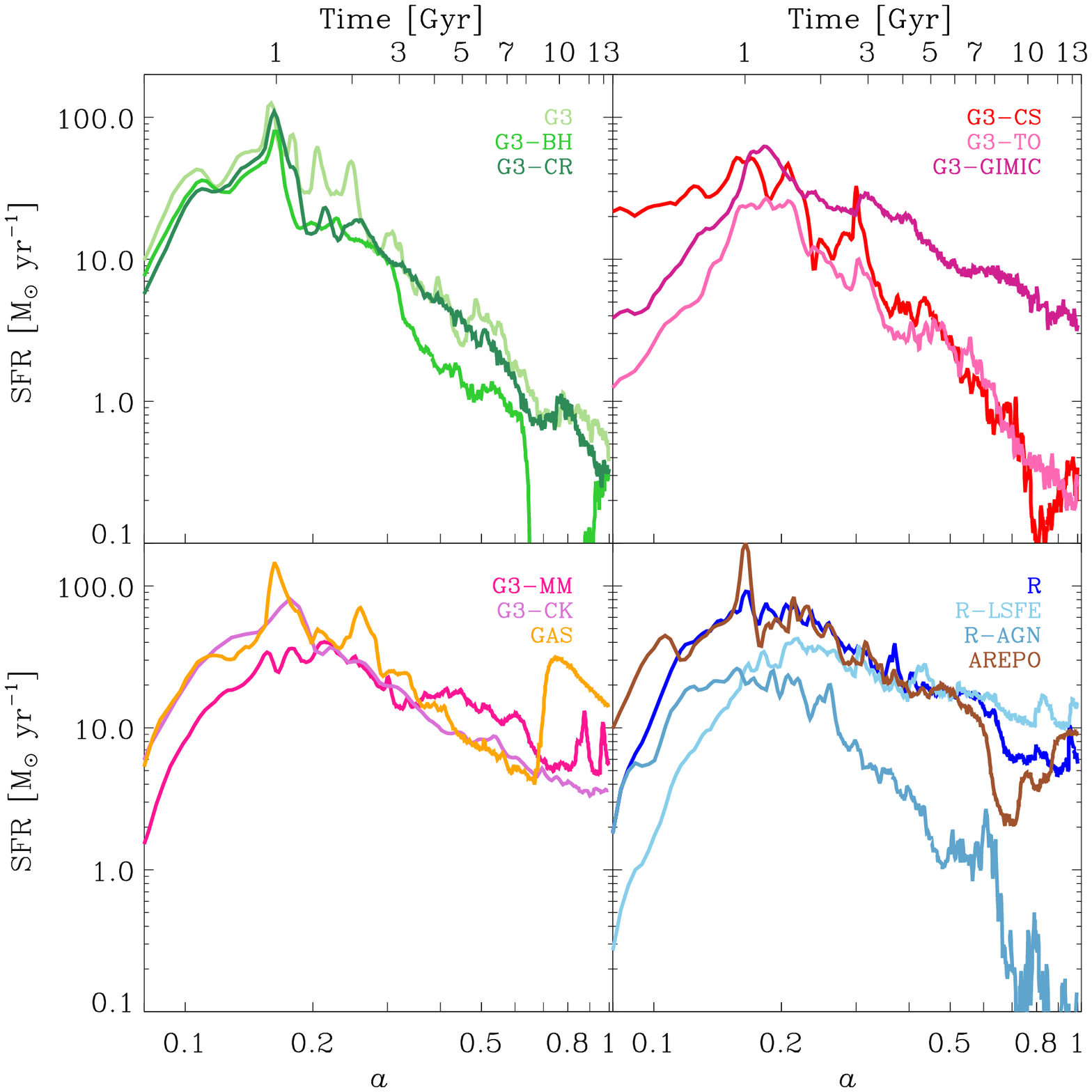}
\includegraphics[width=8.5cm]{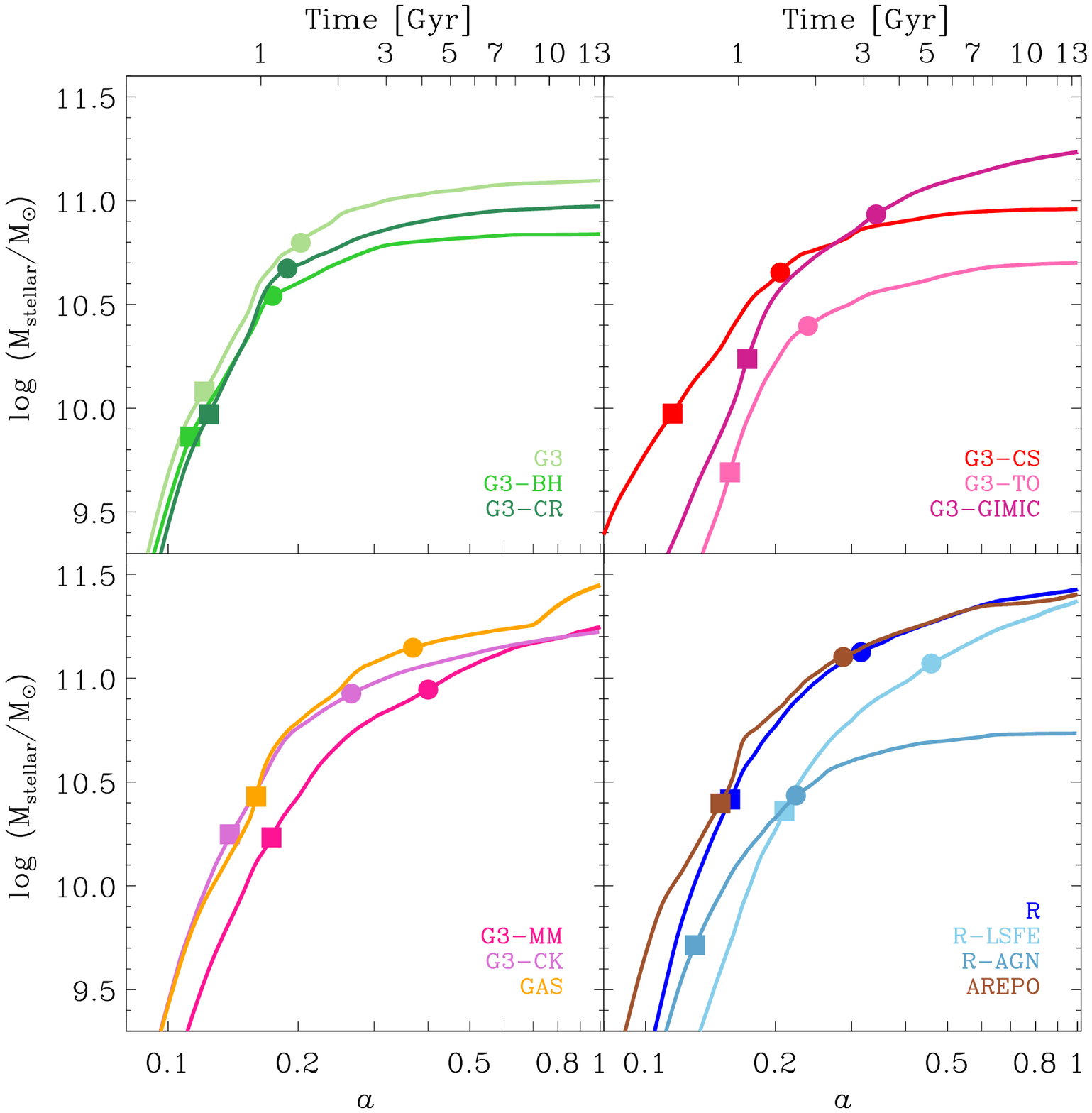}
\caption{Distribution of stellar formation times (expressed
  in terms of the expansion factor, $a$).  Left and right panels show
  the same data, in differential and cumulative form, respectively. The
  simulations are grouped according to numerical technique, as in
  Fig.~\ref{fig:vcirc}.  The squares and circles indicate  the time of
  formation of the first $10\%$ and $50\%$ of the stars.  
  \label{FigSFH}}
\end{center}
\end{figure*}

A recurring theme of our discussion so far has been the need to
prevent the assembly of an overly massive galaxy without, at the same
time, preventing  high-angular momentum, late-accreting material
from reaching the galaxy and forming a disk. This requires feedback
to act especially strongly at high redshift, when a large
fraction of baryons first become cold and dense enough to start
forming stars. None of the Aquila runs seems able to meet these
requirements satisfactorily, as shown by the star formation history of
the simulated galaxies.

Fig.~\ref{FigMstarTform} shows the stellar mass of the galaxy versus
the median formation time of the stars (expressed in terms of expansion factor,
$a=1/(1+z)$).  Note the strong correlation between the two, which
indicates that the codes best able to limit the growth of the mass
of the galaxy do so at the expense of curtailing the incorporation of
{\it late-accreting} material. Indeed, the three galaxies with the lowest
$M_{\rm stellar}$ ({\small G3-TO}, {\small R-AGN}, {small G3-BH}) form
half of their stars in the first Gyr or so of evolution, i.e., by
$z\sim 4$. It is not surprising then that two of them lack a
discernible disk, and that the disk in {\small G3-TO} is overwhelmed
by a massive, dense spheroid composed mainly of old stars.

Further details on the star formation history of individual galaxies
are presented in Fig.~\ref{FigSFH}, where we show, in cumulative and
differential form, the distribution of stellar ages of the stars in
the galaxy at $z=0$. Note that these are {\it not} star formation
rates for the main progenitor, since stars may (and some of them, indeed, do) form in
different progenitors before being accreted into the
galaxy. Nevertheless, the data in Fig.~\ref{FigSFH} show clearly that
few codes are able to prevent the early burst of star formation
activity that accompanies the collapse of the first massive
progenitors of the galaxy. Only {\small G3-TO}, {\small G3-MM},
{\small R-AGN} and {\small R-LSFE} are successful at keeping this peak
``rate'' at less than $\sim 100 \, M_{\odot}$ yr$^{-1}$ at $z\sim 4$-$5$, but even
they see a precipitous decline in star formation afterwards. (The
exception is {\small R-LSFE}, but this is achieved by artificially
delaying star formation, see \S~\ref{SecCodes}.)

Fig.~\ref{FigSFH} also shows that the morphological appearance of simulated
galaxies is very weakly correlated with star formation history. There
are examples of galaxies with lots of recent star formation that have
well-defined disks (e.g., {\small R}) and examples which do not (e.g.,
{\sc arepo}), as well as cases of galaxies with very little recent
star formation that have well-defined disks (e.g., {\small G3-CS},
{\small G3-TO}) and cases which do not ({\small R}, {\small G3-BH}).

Aside from these general considerations, the details of the star
formation history of each galaxy reflects the particular sub-grid
physics implementation chosen for each code (see \citealt{Schaye2010}). 
For example, the feedback
scheme of {\small G3-GIMIC} (where SN-driven winds are assumed to be
launched with fixed speed) imply that it is effective at early
times, when the mass of the progenitors is small, but becomes
ineffective when the main halo reaches $\sim 10^{12}\, M_\odot$. This
curtails early star formation but allows the stellar mass to increase
substantially at low redshift. On the other hand, feedback in {\small
  G3-TO} is effective at all redshifts, since its strength scales with
the potential well of the galaxy. In {\small G3-CS} star formation is
not effectively regulated at very early times because SN energy
feedback is not injected into the ISM instantaneously but rather after
a time delay which depends on the local properties of the cold and hot
neighbouring particles.

\begin{figure}
\begin{center}
\includegraphics[scale=0.31]{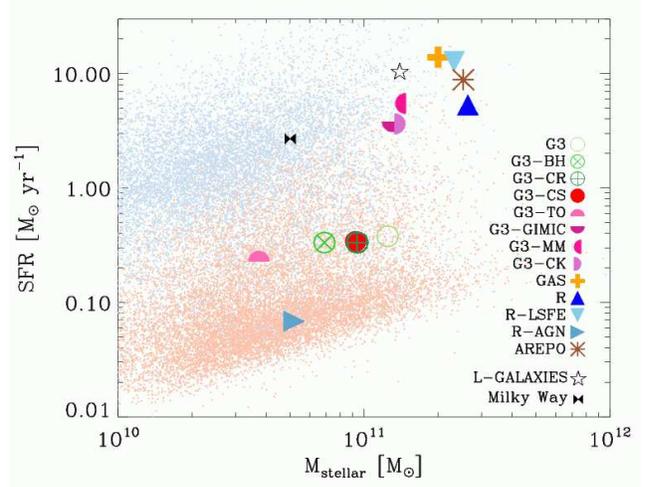}
\caption{Present-day star formation rate (averaged over the last $0.5$
  Gyr to smooth out short-term fluctuations) as a function of stellar
  mass.  Blue and red dots correspond to a sample of nearby ($z<0.1$)
  SDSS galaxies selected from the MPA-JHU DR7 catalog (only $5$
  percent of randomly selected data points are shown). The sample is
  split into ``red sequence'' and ``blue cloud'' galaxies as described
  in Fig.~\ref{fig:sizes}. We also show the prediction of the
  semi-analytic model {\sc l-galaxies }of \citet{Guo2011} for Aq-C and
  the approximate location of the Milky Way
  \citep{Oliver2010,Leitner2011}.  Note that the {\sc galform}
  semi-analytic model is not shown here since it predicts a
  present-day star formation rate close to zero. \label{fig:sfr}}
\end{center}
\end{figure}

These choices imply that at any given time there is substantial
scatter in the star forming properties of simulated galaxies,
compounded by the fact that there is a certain degree of stochasticity
in the rate at which a galaxy accretes mass (\S~\ref{SecMorph}). For
example, in the case of {\sc arepo}, the infall of a satellite at $z
\sim 0.7$ disrupts the gas disk and leads to a significant increase in
the star forming activity at $z=0$. In the {\sc gas} run, a large
burst of star formation also occurs near $z=0$, but in this case it
seems associated with enhanced gas accretion facilitated by the infall
of a satellite. Such effects are at least partially responsible for
the large code-to-code scatter in the star formation rate of the
galaxy at the present time. This is shown in Fig.~\ref{fig:sfr}, where
we compare the present-day star formation rate  (averaged over the
  past $0.5$ Gyr to smooth out short-term fluctuations) with the
stellar mass of simulated galaxies and constrast them with
observations of local SDSS galaxies. The observational sample
corresponds to nearby ($z<0.1$) SDSS galaxies selected from the
MPA-JHU DR7 catalog\footnote{\tt
  http://www.mpa-garching.mpg.de/SDSS/DR7/}. The SFR of simulated
galaxies varies from a low of $\sim 0.02\, $M$_\odot$ yr$^{-1}$
({\small R-AGN}) to nearly $20 \, $M$_{\odot}$ yr$^{-1}$ ({\sc gas}),
spanning the whole range covered by observed galaxies, from ``red and
dead'' spheroids to actively star-forming gas-rich disks. The large
scatter leads us to conclude that caution must be exercised when
analyzing the instantaneous star formation rates of simulated
galaxies, since these depend sensitively on the details of accretion
and of the implemented sub-grid physis.

\begin{figure*}
\begin{center}

\includegraphics[width=14.cm]{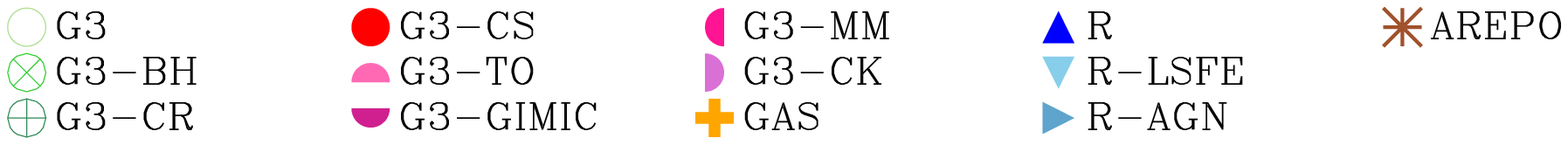}

{\includegraphics[width=7cm]{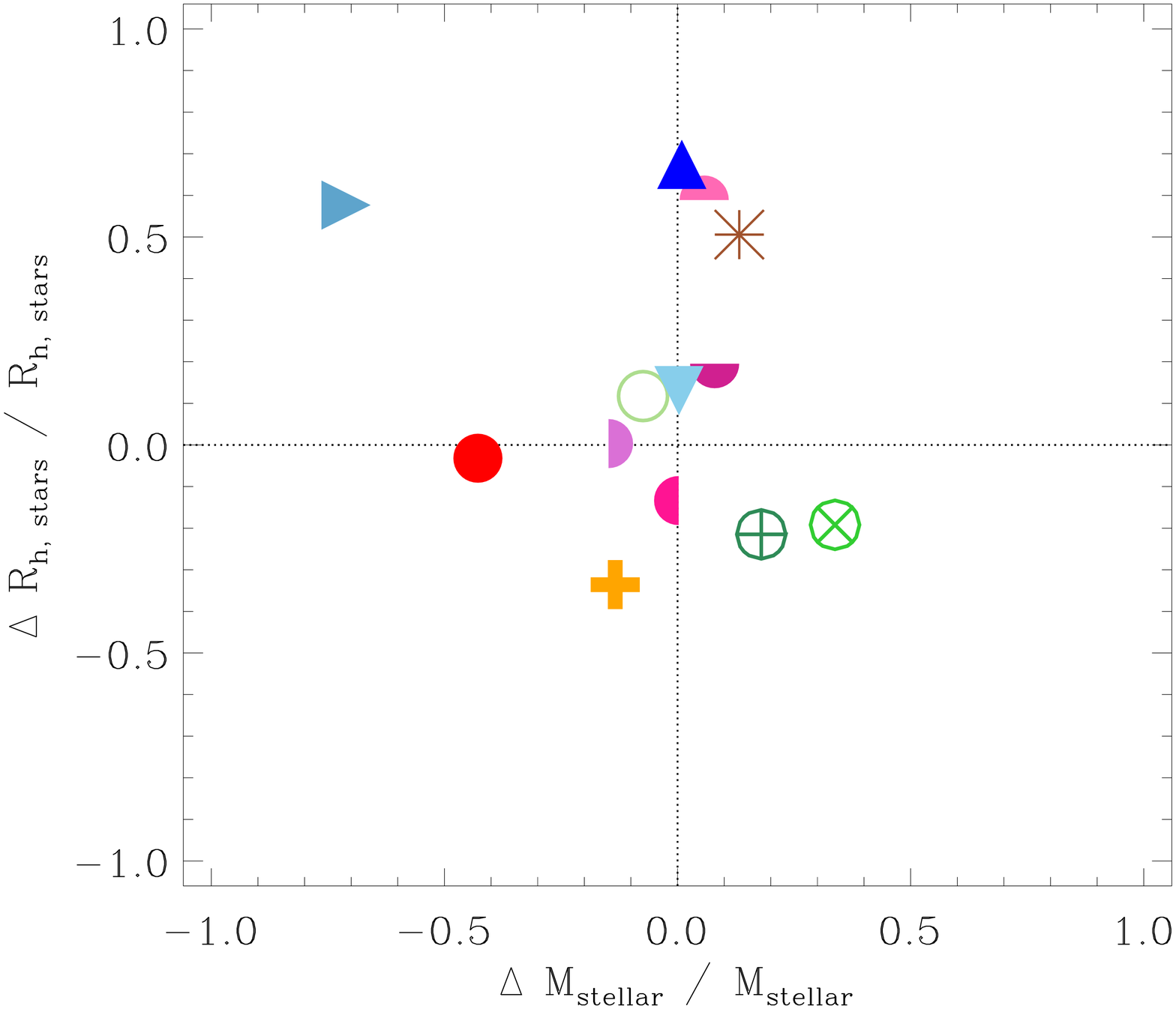}\includegraphics[width=7cm]{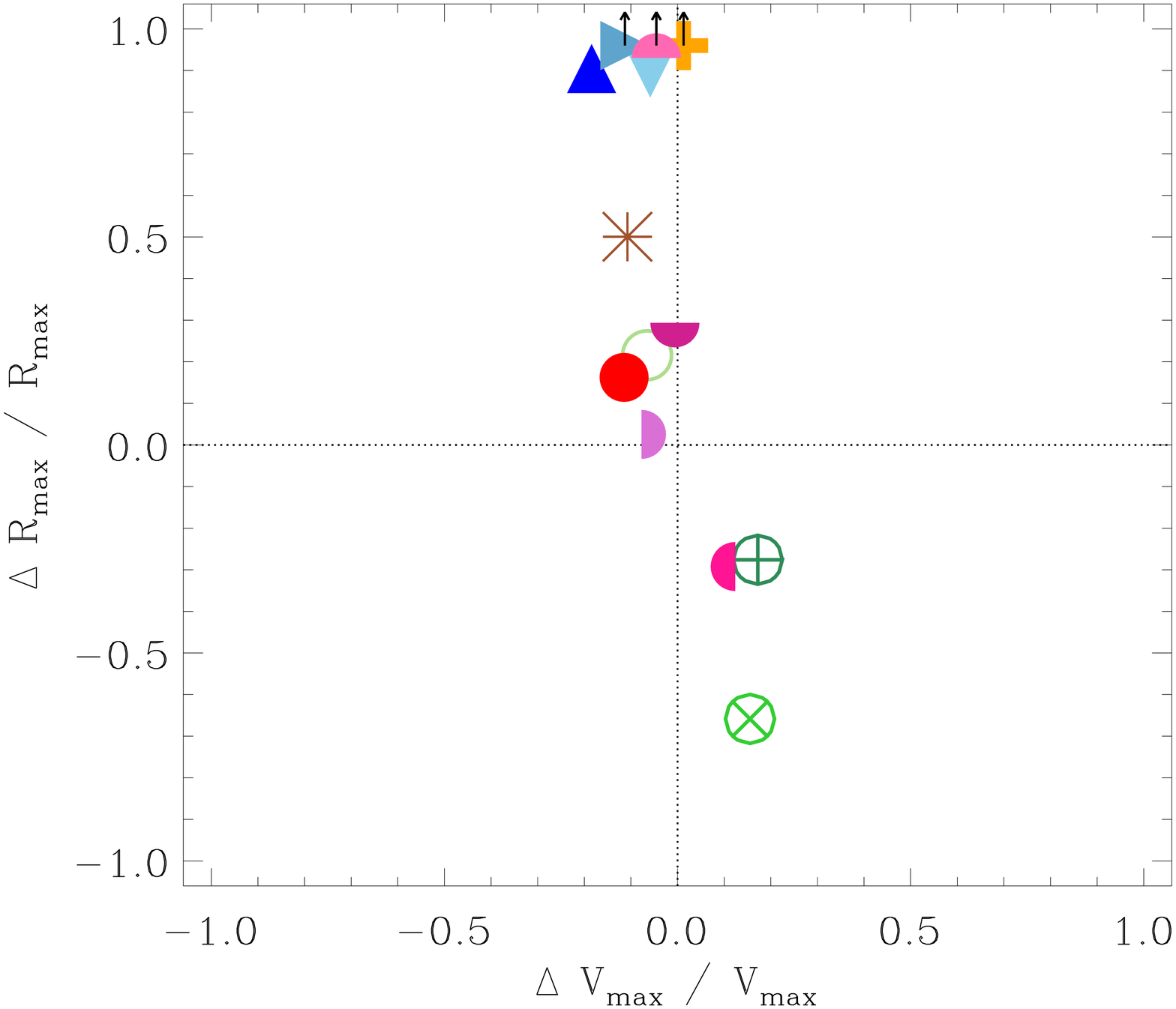}}

{\includegraphics[width=7cm]{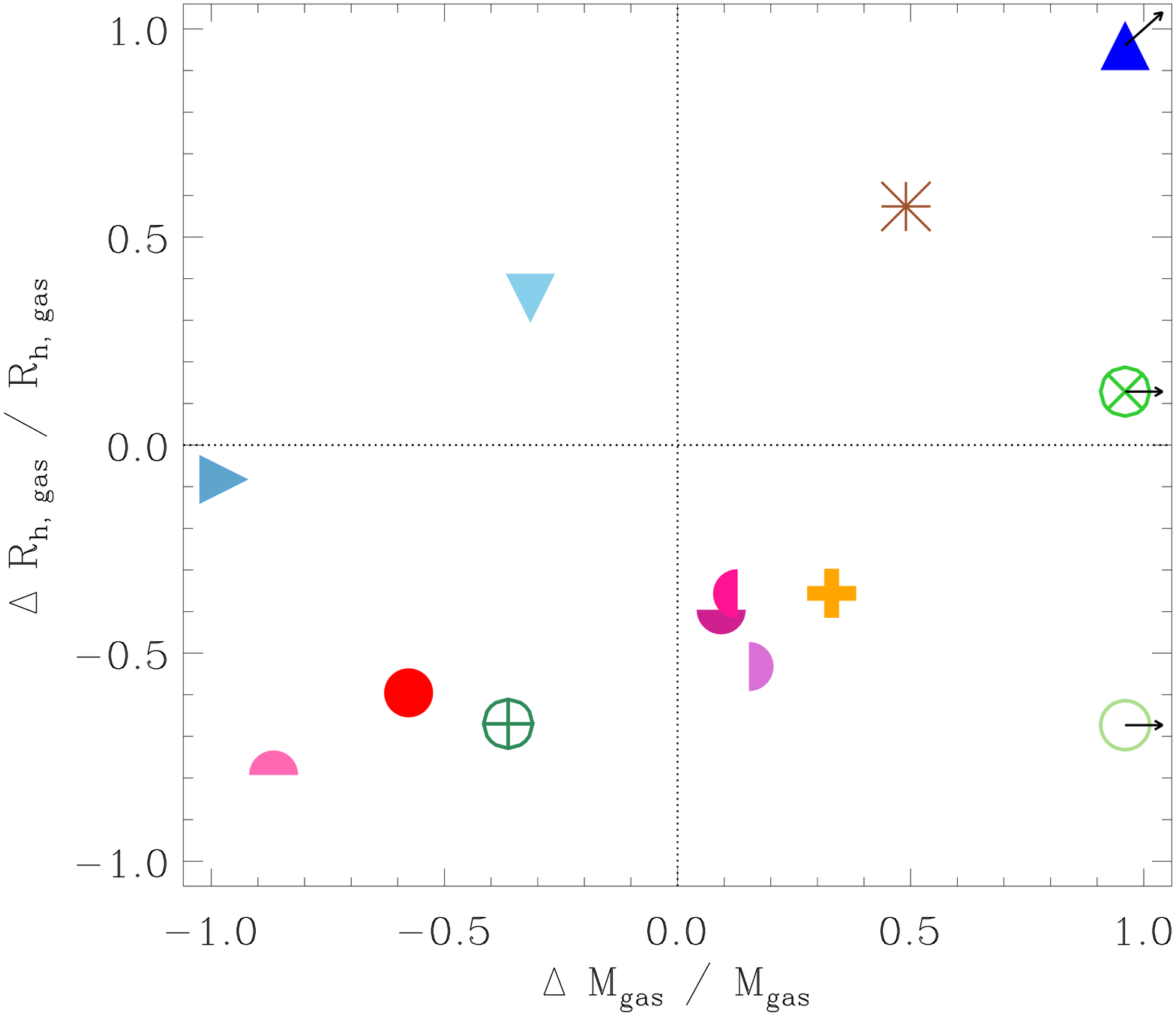}\includegraphics[width=7cm]{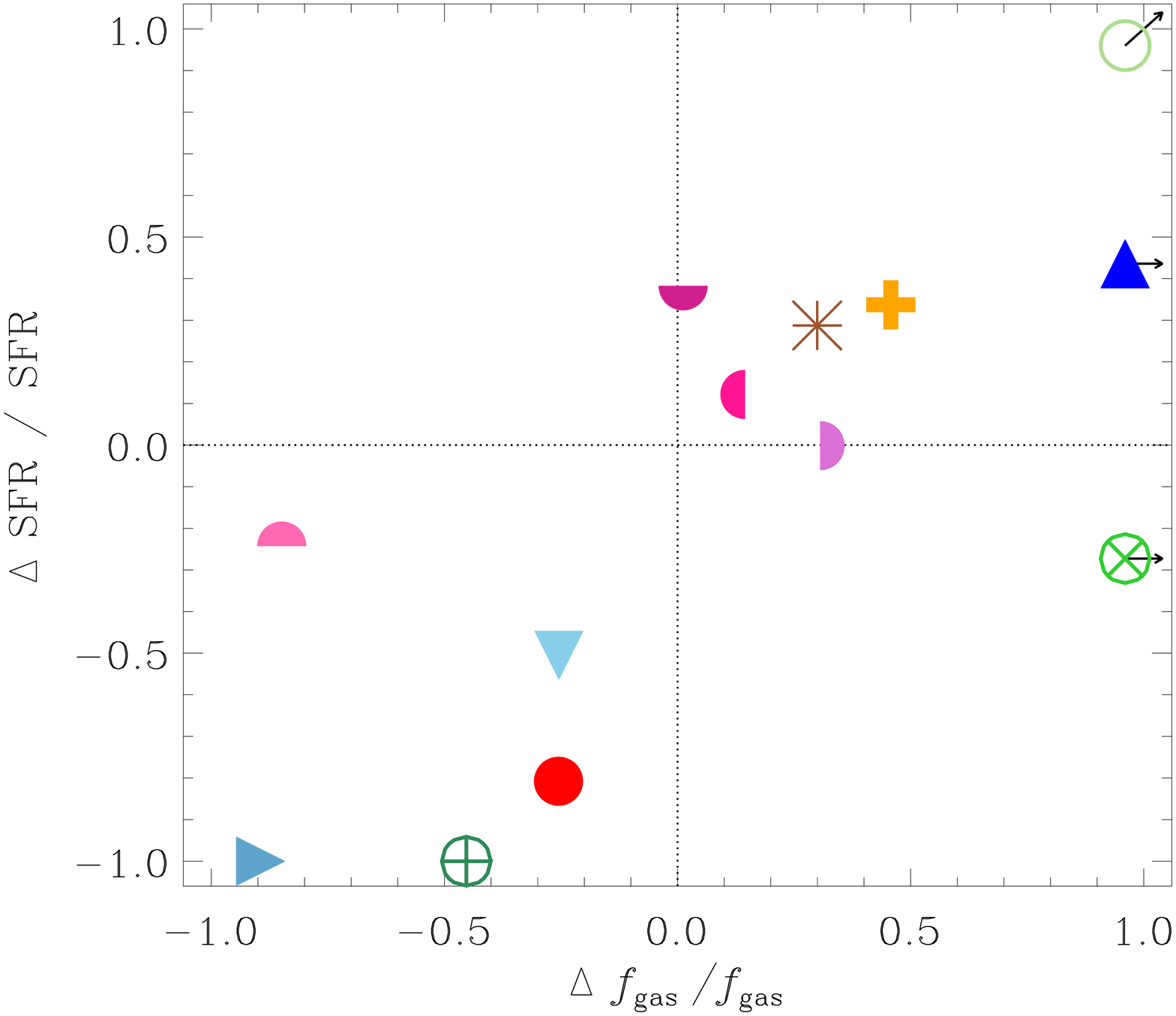}}

\includegraphics[width=7cm]{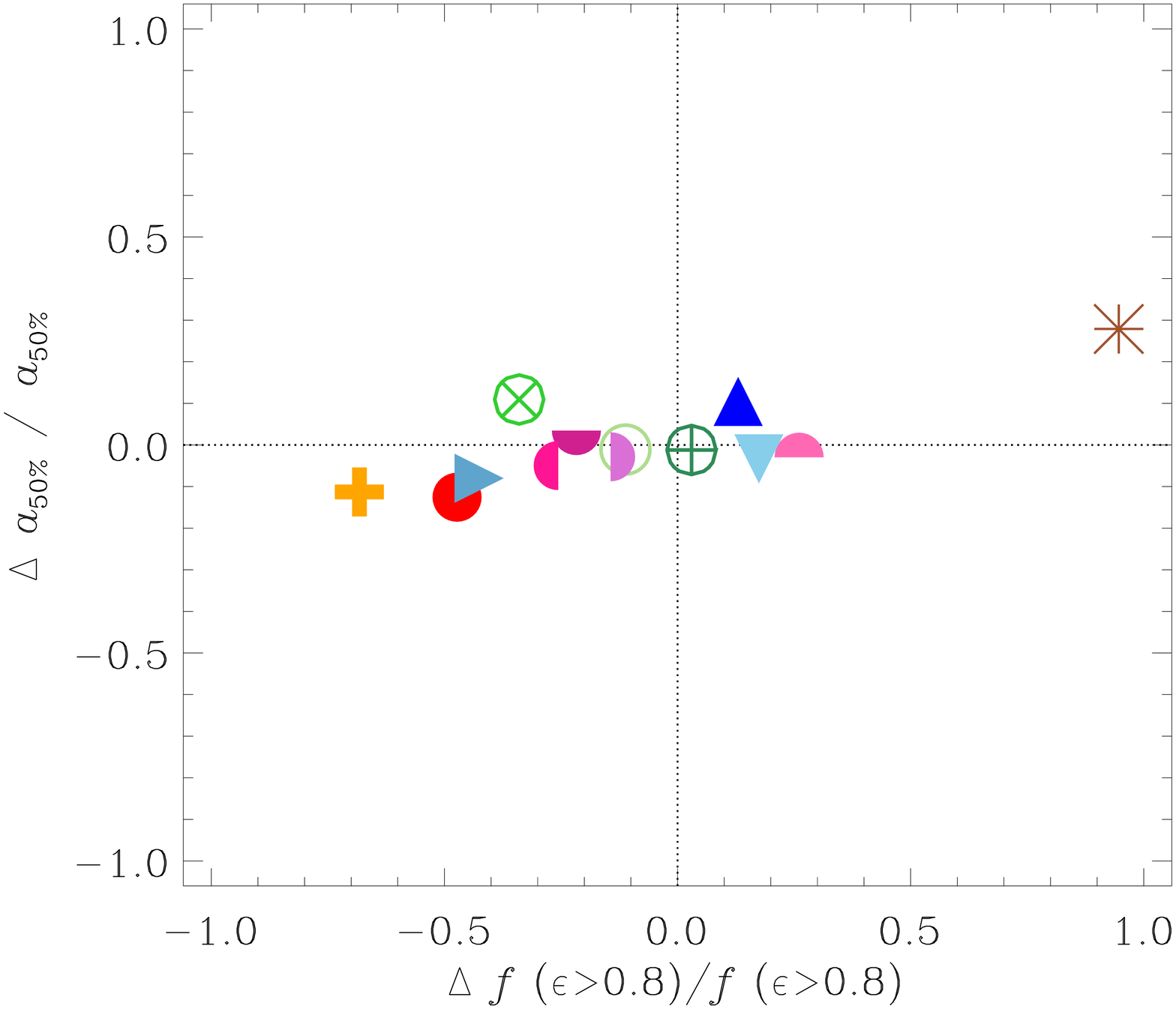}

\caption{ Numerical convergence. We show the fractional variation in
  various galaxy properties between the level-5 and level-6 resolution
  runs of each code: $\Delta Q/Q=(Q_6-Q_5)/Q_5$. Results are shown for
  the stellar mass ($M_{\rm stellar}$) and half-mass radius ($R_{\rm
    h, stars}$);  for the peak circular velocity ($V_{\rm max}$)
  and corresponding radius ($r_{\rm max}$);  for the gas mass 
($M_{\rm gas}$) and half-mass radius
  ($R_{\rm h, gas}$); for the gas fraction ($f_{\rm gas}$) and 
  present-day star formation rate (SFR); and 
for the median
  star formation time ($a_{50\%}$) and the fraction of stars with
  circularities larger than $0.8$, $f (\epsilon> 0.8)$). Arrows
  indicate that results lie outside the plotted range. The actual
  values of the quantities used for the plot are listed in
  Table~\ref{TabFigParams}.  \label{fig:resolution}}
\end{center}
\end{figure*}

\subsection{Numerical Convergence}
\label{sec:resolution}

Convergence is a necessary (but not sufficient) criterion to assess
the robustness of numerical simulation results. We discuss here the effects
of resolution by comparing the results of level-5 and level-6
simulations for each code (see also Table~\ref{TabFigParams}).  As a
quick reminder, at level 5 the parent halo of the Aquila Project,
Aq-C, has $\sim 700,000$ dark matter particles within the virial
radius; this number is reduced to $\sim 90,000$ at level 6.  

Also, when considering convergence one must note that
resolution-dependent behaviour is inevitable to some degree, since the
Jeans length and Jeans mass of star forming gas are not well resolved
either at level 6 or at level 5.  For instance, in several codes
stars are allowed to form only above a density threshold of $n_{\rm th}\sim
0.1$ cm$^{-3}$ and at temperatures of order $\sim 10^{4}$ K, which
correspond to a Jeans length of $\sim 1.5$ kpc, only slightly larger
than the gravitational softening at level 6.

 Although all codes start from initial conditions of the same mass
  and spatial resolutions, each code then adopts its own refinement
  strategy, which may result in significant effective runtime
  resolutions. For example, some codes of similar type choose
  different gravitational softening lengths, and differ as well in
  their choice of when and whether to keep it fixed in comoving or
  physical coordinates (Fig.~\ref{fig:softenings}). This makes it even
  harder to disentangle numerical resolution effects from code type
  effects.  The enhanced cooling in grid-based codes discussed in
  Sec.~\ref{sec:mstar_mhalo}, for example, seems driven by differences
  inherent to the hydrodynamical treatment, and not by resolution
  \citep[see][for details]{Vogelsberger2011}, but the distinction is
  less clear for other quantities.

We quantify the effects of resolution on a given quantity, $Q$, by the
fractional variation, measured at $z=0$, between the level-5 and
level-6 runs, 
\begin{equation}
\Delta Q/Q= (Q_{6}-Q_{5})/Q_{5}.
\end{equation}
Note that with this definition a quantity that increases with
increasing resolution will have a {\it negative} vale of $\Delta
Q/Q$.

We summarize the results in Fig.~\ref{fig:resolution}, where we show
the variations: (i) in global {\it galaxy} properties such as peak
circular velocity, $V_{\rm max}$ and the radius at which it is
achieved, $r_{\rm max}$; (ii) in properties of the {\it stellar
  component}, such as the total mass, $M_{\rm stellar}$ and the
half-mass radius, $R_{\rm h,stars}$; (iii) in properties of the {\it
  gas} component, such as the mass, $M_{\rm gas}$, and half-mass
radius, $R_{\rm h,gas}$;  (iv) in properties related to {\it star
    formation}, such as the gas fraction and the present-day star
  formation rate; and (v) in the details of the galaxy {\it
  assembly/morphology}, such as the median formation time of stars at
$z=0$ (expressed in terms of the expansion factor, $a_{\rm 50\%}$) and
the fraction of stars kinematically associated with a
rotationally-supported disk ($f(\epsilon>0.8)$, see
\S~\ref{SecMorph}).  When differences exceed $100\%$ (i.e., $|\Delta
Q/Q| > 1$), we use arrows to indicate if such deviations occur along
the $x$ and/or $y$ coordinates of the plot. The actual values of all
quantities plotted in Fig.~\ref{fig:resolution} are listed in
Table~\ref{TabFigParams}.

Aside from the fact that numerical convergence is not particularly
good for any of the codes, Fig.~\ref{fig:resolution} illustrates a few
interesting points. The first concerns the properties of the stellar
component (left two upper panels).
In particular, the total mass in stars and their median age
seem to be the most reliable results, suggesting that the star
formation/feedback scheme chosen for each code is reasonably
independent of resolution. Most codes reproduce the total stellar
mass
and $a_{\rm 50\%}$ to better than $20$-$30\%$.
Interestingly, the peak circular velocity is one of the
properties least affected by resolution (see also Fig.~\ref{fig:vcirc}). 
This is encouraging, since it
implies that diagnostics such as the observed Tully-Fisher relation
may be used to assess the success of a particular model.

Convergence deteriorates when considering the detailed properties
of the stellar component, such as the fraction of stars in
high-circularity orbits: $f(\epsilon >0.8)$. Although half the codes
give results that converge to better than $\sim 30\%$ the variations
can be much larger for some codes. Interestingly, {\it increasing the
  resolution does not always leads to better-defined disks}: for
example, the fraction of ``disk'' stars increases by a large fraction
in the case of {\sc gas} but actually decreases for {\sc
  arepo}. Finally, there is some indication that the most extreme
feedback models (i.e., those that result in the lowest $M_{\rm
  stellar}$; e.g., {\small R-AGN} and {\small G3-BH}) are the most
vulnerable to resolution-induced changes.

Even larger variations are seen for the properties of the gas
component: only five of the thirteen simulations show variations in
$M_{\rm gas}$ and $R_{h, {\rm gas}}$ smaller than $50\%$, and three
simulations have differences in $M_{\rm gas}$ larger than $100$
percent.  Similar results are found for the gas fractions, 
and, consequently, for the present-day average star formation rates. 
In general, increasing the resolution leads to larger gaseous
disks, but in some cases the total mass in gas increases while it
decreases in others. These large variations are at least partly due to
the fact that some galaxies (e.g., {\small R-AGN}) have almost no gas
left at $z=0$, and therefore even small changes can lead to large
fractional variations. The properties of the gaseous component seem
the most vulnerable to numerical resolution effects and therefore
caution must be exercised in their interpretation.

\section{Summary and Conclusions}
\label{SecConc}

The Aquila Project compares the results of simulations of galaxy
formation within a Milky Way-sized $\Lambda$CDM halo. We use $9$
gas-dynamical codes developed and run independently by different groups
adopting in each case their preferred implementation of radiative cooling,
star formation, and feedback. The codes include SPH-based, grid-based,
and moving-mesh techniques; all include supernova feedback in thermal
and/or kinetic form and primordial or metal-dependent cooling
functions. Two of the codes ({\sc gadget} and {\sc ramses}) were run
three times with three different sub-grid physics for a total of $13$
different simulations. In addition, each code was run at two different
resolution levels in order to investigate numerical convergence. All
runs share the same initial conditions and are analyzed with a set of
consistent analysis tools. Our main conclusions may be summarized as
follows.

{\tt Galaxy Morphology.} At $z=0$, simulated galaxies exhibit complex
morphologies, with spheroids, disks, and bars of varying
importance. Morphology shows no obvious dependence on the
hydrodynamical method adopted or on numerical resolution, and seems to
be mostly related to how and when gas is accreted and transformed into
stars. The best indicator of the presence of a disk seems to be the
median age of the stars; the later stars form the more prevalent the
disk component is.  This suggests that, to be successful at forming
disks, codes must be able to preempt the early transformation of gas
into stars while at the same time promoting the accretion and
retention of late-accreting, high-angular momentum gas.

{\tt Stellar Mass.}  Despite the common halo-assembly history, we find
large code-to-code variations in the final mass of simulated
galaxies. The stellar mass ranges from $4\times 10^{10} \, M_\odot$ to
$\sim 3\times 10^{11} \, M_\odot$, depending largely on the adopted
strength of the feedback. There is also an indication that the
numerical method may play a role: {\sc arepo} is able to form almost
twice as many stars as {\small G3} although they both adopt the same
sub-grid physics. Most simulated galaxies are too massive compared
with theoretical expectations based on abundance-matching
considerations.  The median stellar age also correlates with galaxy
mass, indicating that models that favor late star formation (as needed
to form a disk) do so at the expense of allowing too many stars to
form overall.

{\tt Rotation Curves.} All simulated galaxies have rotation speeds in
excess of what is expected from the Tully-Fisher relation of late-type
spirals. The disagreement worsens as the stellar mass of the simulated
galaxy increases, both for galaxies with and without a well-defined
stellar disk. At the high mass end, simulated galaxies have
unrealistically sharply-peaked, strongly declining rotation
curves. Although reasonably flat rotation curves are obtained at low
$M_{\rm stellar}$, the Tully-Fisher zero-point offset persists for
these systems.

{\tt Galaxy Size.} The difficulties matching the Tully-Fisher relation
are due to the fact that most simulated galaxies have stellar
half-mass radii much smaller than expected from observation given
their stellar mass. This is especially true for galaxies where
inefficient feedback allows the formation of an overly massive galaxy;
these systems are, quite unrealistically, smaller even than dense
early-type galaxies. Galaxies where feedback is able to limit the
stellar mass to more acceptable levels are also more concentrated than
typical spirals, highlighting the difficulty that all codes face to
prevent too many low-angular momentum baryons from assembling into the
galaxy.

{\tt Star Formation History.} The excess of low-angular momentum
baryons alluded to above may be traced to the inability of feedback
schemes to prevent large bursts of star formation at early times. This
is clearly seen in the stellar age distribution, which shows that star
formation peaks at $z\sim 4$ and declines steadily afterwards. The
same trend holds for essentially all runs, albeit modulated by the
particular implementation of feedback adopted by each individual
code. Indeed, essentially all models allow more than half of the stars
to form in the first $\sim 3$ Gyrs of evolution. The relative
insignificance of star formation in recent times compared to the
intensity of the early burst seems to be at the root of many of the
difficulties that simulated galaxies have in matching observation.

{\tt Gas Component.} The properties of the gaseous component of the
galaxy at $z=0$ show even wider code-to-code variations than the
stars, since it is continuously resupplied by accretion, depleted by
star formation, and dislodged by feedback-driven winds. This results
in large short-lived fluctuations that lead to poor numerical
convergence and large code-to-code scatter.  Most simulated galaxies
have gaseous disks with  sizes that compare favorably with observation,
although their gas fractions are systematically lower than those of
star-forming spirals of comparable mass.

{\tt Numerical Convergence.} We have investigated numerical
convergence by comparing the results at two different numerical
resolutions, spanning a range of $\sim 2$ and $\sim 8$ in spatial and
mass resolution, respectively.  Although numerical convergence
is not particularly good for any of the codes, reasonably-good
convergence is found for the properties of the
stellar component, such as total mass and median age. Less
well-converged are the internal properties of the galaxy, such as the
half-mass radius, or the fraction of stars in a rotationally-supported
disk. For the same reasons cited in the previous item, the properties
of the gas are the ones that are most poorly reproduced at the two different
resolutions. There is also indication that feedback schemes that are
more effective at limiting the stellar mass of the galaxy (such as
through the inclusion of AGN-related feedback in addition to
supernovae) converge less well than other implementations. Overall,
the variations introduced by resolution are small compared to
code-to-code variations, which leads us to conclude that none of the
trends highlighted above are driven primarily by resolution.

{\tt SPH vs AMR vs Moving-Mesh.} Our results suggest that numerical
hydrodynamics techniques have some influence on the outcome of a
simulation. This is most clearly demonstrated by comparing the results
of {\small G3} and {\sc arepo} which, despite adopting the same
sub-grid physics modules, yield galaxies that differ by almost a
factor of $2$ in stellar mass. The AMR technique also yields large
stellar masses (when similarly inefficient feedback is adopted, as in
run {\small R}), lending support to the view that standard SPH-based
codes may underestimate the total amount of gas that can cool and
become available for star formation at least in the weak feedback
regime. On the other hand, the galaxies formed by {\small R}, {\small
  G3}, and {\sc arepo} are unrealistically massive and concentrated,
so large modifications to the feedback implementation of these codes
are needed in order to bring them into agreement with
observation. These changes may overwhelm the method-induced
differences; for example, {\small R} makes a prominent disk while
{\small R-AGN} has $5$ times fewer stars and no disk.  It is therefore
unclear at this point whether the shortcomings of SPH are actually
significant compared with the uncertainties involved in designing a
star formation/feedback scheme that can yield realistic galaxies,
although it is obviously desirable to avoid numerical inaccuracies as
far as possible. Further investigation of this question is needed to
clarify this issue.

Aside from these considerations, perhaps the main result of the Aquila
Project is that, despite the large spread in properties spanned by the
simulated galaxies, none of them has properties fully consistent with
theoretical expectations or observational constraints in terms of
mass, size, gas content, and morphology. Despite this apparent
failure, we believe that the Aquila Project nevertheless yields
interesting clues to guide how codes might be modified to yield
realistic galaxies. For example, the need (i) to control effectively
the overcooling of baryons; (ii) to curtail the early burst in star
forming activity; and (iii) to promote the accretion and retention of
the late-accreting, high-angular momentum baryons needed to form disks
similar to those of normal spirals are of general applicability to all
codes. 

There seems to be little predictive power at this point in
state-of-the-art simulations of galaxy formation; these seem best
suited to the identification of the role and importance of various
mechanisms rather than to the detailed modeling of individual systems.
It may be argued that the strength of this conclusion depends on
whether the parent halo of the Aquila runs (Aq-C) is truly destined to
harbor a disk galaxy and that there is no hard proof for this.
Further, the possibility that Aq-C might be an unrepresentative
outlier should also be considered, as suggested by the {\sc l-galaxies}
semi-analytic model (see, e.g., Fig.~\ref{fig:mstar_mhalo}).

These objections may be addressed when simulations are able to
follow a statistically-significant number of galaxies in a
cosmologically-representative volume. We might not know what kind of
galaxy inhabits an individual halo, but we do know what the {\it
  population} of galaxies looks like. Evolving a region large enough
to contain at least a few dozen Milky Way-sized galaxies at the resolution
achieved here seems like a natural next step, and one that should be
achievable in the not too-distant future.

Finally, the complexity of the problem suggests that the best approach
to improving galaxy formation simulations may be one where multiple
numerical alternatives are developed and explored simultaneously and
independently, provided that they are periodically contrasted in
controlled experiments such as the one presented here. Given the
intricacy of the task and in the absence of a clear front-runner, the
diversity of numerical techniques presently available might actually
turn out to be a strength rather than a shortcoming.

\section{Acknowledgments}
We thank the referee  for a thorough reading of this work and for his/her
helpful comments and suggestions.
We are grateful to Aaron Dutton and Marla Geha for
providing the compilations of observational data used in this work. 
We thank Andrew Cooper for providing the predictions of the {\sc
  galform} semi-analytic model and for stimulating discussions and
suggestions.  We also thank the referee for a prompt and useful report.
The  G3, G3-BH, G3-CR, G3-CS and {\sc arepo} simulations were 
carried out at the Computing Center of the
Max-Planck-Society in Garching.
M.W. acknowledges support by the DFG cluster of excellence 
'Origin and Structure of the Universe' and 
V.S. by the DFG Research Center SFB 881 'The Milky Way System'.
The RAMSES simulations were performed thanks 
to the HPC resources of CINES under the allocation 2010-GEN2192 made by GENCI, France.
P.M. and G.M. acknowledge a {\it standard 2010 grant} 
from CASPUR, and support from PRIN-INAF 2009 titled 
``Towards an italian network for computational cosmology''.
This work was supported by Marie Curie Initial 
Training Network CosmoComp (PITN-GA-2009-238356). 
The {\sc Gasoline} simulations used computers graciously 
provided by {\sc sharcnet}, Compute Canada and the HPCAVF 
at the University of Central Lancashire in Preston, UK.
The G3-TO simulations were performed 
with T2K-Tsukuba at the Center for Computational
Sciences at the University of Tsukuba.
TO acknowledges financial support from Grant-in-Aid for Young Scientists
(start-up: 21840015) and from MEXT HPCI STRATEGIC PROGRAM.
Some of the calculations for this paper were performed on the Green
Machine at Swinburne University of Technology and on the ICC 
Cosmology Machine, which is part of the DiRAC Facility, jointly funded 
by STFC, the Large Facilities Capital Fund of BIS, and Durham
University.  We thank the DEISA Consortium (www.deisa.eu), co-funded
through the EU FP6 project RI-031513 and the FP7 project RI-222919,
for support within the DEISA Extreme Computing Initiative.
TQ and EC acknowledge support from NSF grant AST-0908499.

\begin{landscape}
\begin{table}
\caption{Properties of simulated
galaxies at $z=0$, for the level 5 (upper row) and level 6 (lower row)
resolution simulations. We also list the prediction of the semi-analytic models 
{\sc galform} \citep{Cooper2010} and {\sc l-galaxies} \citep{Guo2011}, and
of the dark matter-only simulation Aq-C-4 of the Aquarius Project. }
\label{TabFigParams}
{\small
\begin{tabular}{lcccccccccccccccc}
\hline\hline
Code &  $M_{200}$ & $r_{200}$ &  $V_{200}$ &  $M_{\rm stellar}$ &  $M_{\rm cold\, gas}$  &  $f(\epsilon > 0.8)$ &  $a_{\rm 50\%}$ &  $R_{\rm h, stars}$ &  $R_{\rm h, gas}$ &  $V_{1/2}$ &  SFR &  $M_{\rm R}$ &  $f_{\rm gas}$ &  $V_{\rm DM(R_{h})}/V_{\rm C(R_{h})}$ \\
& [$10^{10}$M$_\odot$] &  [kpc] &  [km s$^{-1}$] &  [$10^{10}$M$_\odot$]  &  [$10^{10}$M$_\odot$]  & & &  [kpc] &  [kpc]&  [km s$^{-1}$] &  [M$_\odot$ yr$^{-1}$]  & & &&\\
\hline

G3 &     164.94 &     239.12 &     172.24 &      12.47 &      0.063 &       0.13 &       0.20 &       3.06 &       3.06 &     348.19 &      0.378 &     -21.25 &      0.011 &      0.521\\
 &     172.39 &     242.23 &     174.95 &      11.54 &      0.003 &       0.11 &       0.20 &       3.43 &       1.00 &     325.44 &      0.906 &     -21.17 &      0.046 &      0.552\\

\\
G3-BH &     157.61 &     233.48 &     170.39 &       6.89 &      0.161 &       0.15 &       0.17 &       5.31 &       7.42 &     224.88 &      0.333 &     -20.51 &      0.026 &      0.695\\
 &     167.38 &     242.35 &     172.34 &      11.07 &      0.027 &       0.10 &       0.19 &       3.43 &       8.36 &     304.63 &      0.242 &     -20.99 &      0.070 &      0.557\\

\\
G3-CR &     154.32 &     234.29 &     168.31 &       9.38 &      0.247 &       0.09 &       0.19 &       4.75 &       7.41 &     262.93 &      0.333 &     -20.93 &      0.029 &      0.639\\
 &     172.85 &     239.25 &     176.27 &       9.22 &      0.001 &       0.09 &       0.19 &       4.28 &       2.45 &     272.73 &      0.000 &     -20.77 &      0.016 &      0.652\\

\\
G3-CS &     164.11 &     237.45 &     172.41 &       9.24 &      0.200 &       0.24 &       0.21 &       4.27 &      18.22 &     270.82 &      0.341 &     -20.85 &      0.024 &      0.619\\
 &     149.46 &     230.01 &     167.17 &       5.28 &      0.061 &       0.13 &       0.18 &       3.79 &       7.37 &     236.04 &      0.065 &     -20.19 &      0.018 &      0.706\\

\\
G3-TO &     147.32 &     228.40 &     166.56 &       3.73 &      0.850 &       0.28 &       0.24 &       1.94 &      14.30 &     213.98 &      0.253 &     -20.32 &      0.187 &      0.533\\
 &     147.85 &     228.70 &     166.74 &       3.94 &      0.084 &       0.35 &       0.24 &       3.39 &       3.39 &     216.65 &      0.199 &     -20.43 &      0.028 &      0.677\\

\\
G3-GIMIC &     161.84 &     235.76 &     171.82 &      13.06 &      0.982 &       0.39 &       0.34 &       1.95 &      10.39 &     398.37 &     3.425 &     -22.08 &      0.072 &      0.353\\
 &     167.57 &     238.41 &     173.86 &      14.10 &      1.082 &       0.31 &       0.34 &       2.44 &       5.97 &     388.78 &       4.635 &     -22.18 &      0.073 &      0.355\\

\\
G3-MM &     176.11 &     245.02 &     175.82 &      14.07 &      0.205 &       0.16 &       0.31 &       3.96 &       3.96 &     335.49 &      5.471 &     -22.43 &      0.078 &      0.542\\
 &     176.69 &     242.57 &     177.00 &      13.74 &      0.318 &       0.11 &       0.34 &       2.28 &       2.54 &     362.90 &       6.138 &     -22.33 &      0.087 &      0.374\\

\\
G3-CK &     166.45 &     237.61 &     173.57 &      14.00 &      0.324 &       0.20 &       0.27 &       3.52 &       8.46 &     373.97 &      3.637 &     -22.12 &      0.025 &      0.444\\
 &     177.52 &     243.66 &     177.01 &      12.30 &      0.263 &       0.18 &       0.26 &       3.18 &       3.96 &     346.24 &       3.633 &     -22.15 &      0.033 &      0.421\\

\\
GAS &     183.18 &     246.12 &     178.91 &      19.98 &      0.749 &       0.39 &       0.37 &       3.55 &       5.52 &     440.25 &     13.892 &     -23.17 &      0.101 &      0.427\\
 &     183.31 &     246.23 &     178.93 &      17.31 &      1.090 &       0.12 &       0.33 &       2.85 &       3.55 &     505.07 &     18.574 &     -22.90 &      0.147 &      0.385\\

\\
R &     202.25 &     256.20 &     184.26 &      26.43 &      1.084 &       0.45 &       0.32 &       2.56 &       4.48 &     580.44 &      5.327 &     -22.61 &      0.022 &      0.387\\
 &     178.74 &     244.00 &     177.49 &      26.67 &      2.063 &       0.51 &       0.35 &       3.66 &      10.37 &     504.81 &       7.650 &     -22.66 &      0.054 &      0.457\\

\\
R-LSFE &     210.27 &     258.70 &     186.97 &      23.21 &      3.101 &       0.53 &       0.46 &       4.53 &       9.06 &     444.55 &     12.704 &     -22.96 &      0.267 &      0.464\\
 &     180.34 &     245.10 &     177.89 &      23.28 &      3.450 &       0.62 &       0.44 &       5.51 &      12.25 &     428.59 &       6.284 &     -22.73 &      0.199 &      0.554\\

\\
R-AGN &     150.63 &     231.80 &     167.17 &       5.19 &      0.512 &       0.20 &       0.22 &       5.22 &      16.23 &     222.24 &      0.068 &     -20.28 &      0.128 &      0.677\\
 &     147.61 &     229.10 &     166.46 &       1.50 &      0.319 &       0.11 &       0.21 &       6.87 &      14.89 &     169.58 &      0.000 &     -18.86 &      0.014 &      0.834\\

\\
AREPO &     204.54 &     257.45 &     184.85 &      25.33 &      0.382 &       0.19 &       0.29 &       2.21 &       5.47 &     498.77 &      8.827 &     -22.59 &      0.040 &      0.343\\
 &     206.21 &     257.54 &     185.57 &      28.68 &      0.951 &       0.36 &       0.37 &       3.48 &       8.61 &     464.29 &       11.364 &     -22.88 &      0.051 &      0.416\\

\\
{\sc galform} &     203.27 &     261.01 &     183.01 &       7.84 &      0.004 &  & 0.26 &       3.77 &      10.43 &  &      0.004 &     -20.99 &      0.001 & \\
\\
{\sc l-galaxies} & 178.01 & 243.10 & 177.46 & 13.95 & 2.44 &  & 0.44 & 2.03 & 5.13 &  & 10.328 & -22.80 &      0.149 & \\
\\

{Aq-C-4} & 179.30 & 243.68 & 177.92 &  &  &  &  &  &  &  &  &  &   & \\

\hline
\end{tabular}
}
\end{table}
\end{landscape}

\appendix

\section{Description of the codes}
\label{code_description}

\begin{figure*}
\begin{center}
\includegraphics[width=12cm]{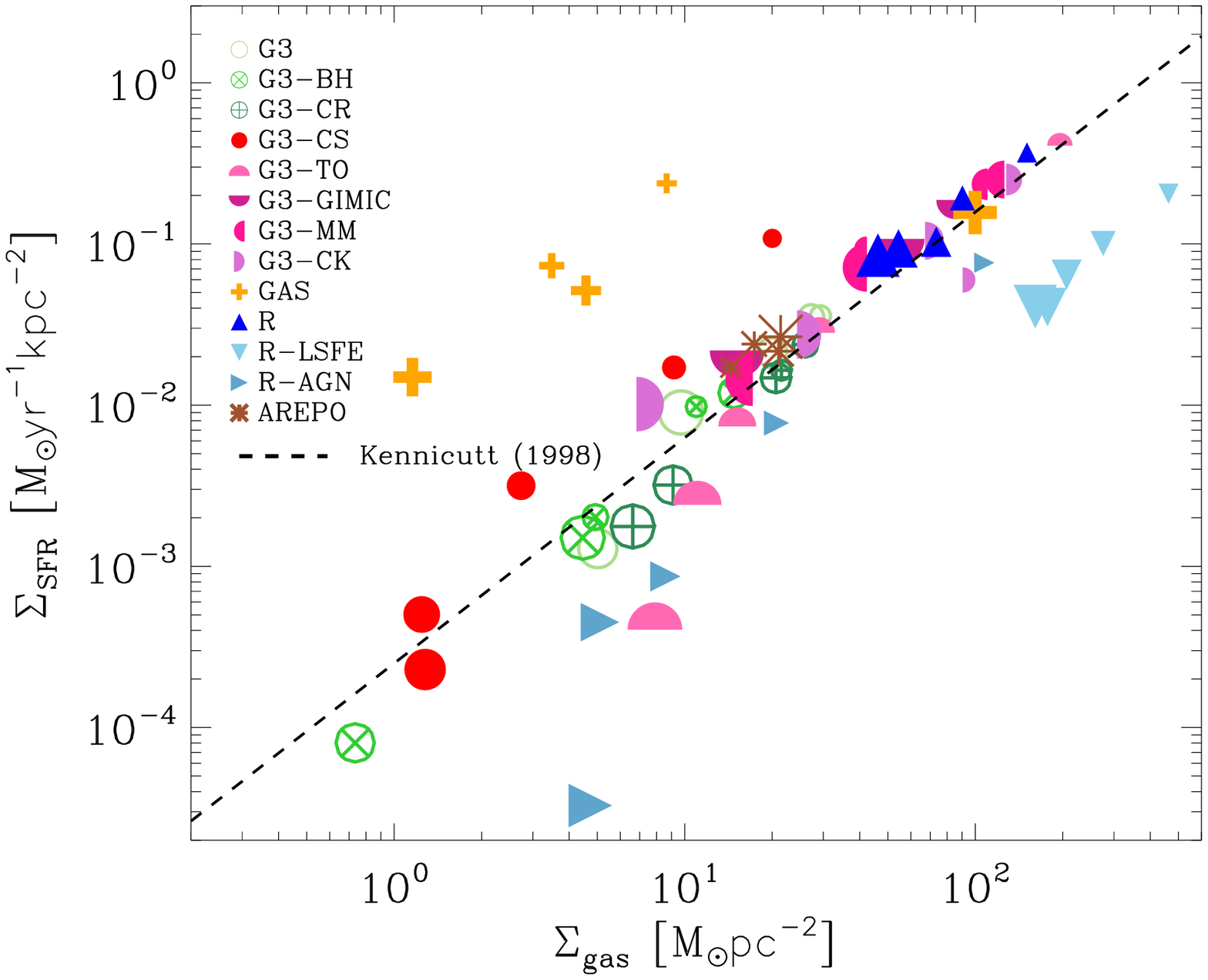}
\caption{Cold ($T<10^5K$) gas surface density versus star formation
  rate per unit area in the simulated galaxies, measured face-on and
  averaged within a cylinder of radius equal to the (projected)
  half-mass radius of the cold gas (see the right panel of
  Fig.~\ref{fig:sizes}). Each set of symbols indicate results for
  level-5 simulations at $z=2$,$1.5$,$1$,$0.5$ and $0$, with the size
  of the symbols increasing with decreasing redshift.  The dashed line
  indicates, for reference, the Schmidt-Kennicutt law for nearby
  ``normal'' and ``star-bursting'' disks, as given by
  \citet{Kennicutt98}.
  \label{fig:Kennicutt}}
\end{center}
\end{figure*}

In this Appendix we summarize the main properties of the various
simulation codes and implemented physics. These succinct descriptions
have been provided by each of the individual groups participating in
the Aquila Project. As explained below, the different models use a
variety of implementations of the processes of star formation and
feedback, resulting in the star formation rate surface densities
shown in Fig.~\ref{fig:Kennicutt}.

\vspace{0.3cm}
\centerline{\it  Gadget-3 models (G3, G3-BH, G3-CR)}
\vspace{0.1cm}

The simulations {\small G3}, {\small G3-BH} and {\small G3-CR} are
based on {\sc gadget3}. This is a significantly updated version of
{\small GADGET2} \citep{Springel2005}, a fully cosmological code
based on smoothed particle hydrodynamics (SPH) and on the Tree-PM
method to evaluate gravitational forces.  All three of these
simulations use the star formation model introduced by
\citet{SpringelHernquist2003}. {\small G3-BH} includes AGN feedback
following \citet{SpringelDiMatteo2005}, while the {\small G3-CR} model
includes, in addition, energy deposition by cosmic rays 
\citep[see][and references therein]{Jubelgas2008}.

The \citet{SpringelHernquist2003} star formation model uses a
primordial cooling function following \citet{Katz1996} and includes a
uniform UV background, based on an updated version of
\citet{HaardtMadau1996}, that is switched on at $z=6$. Stars are
formed stochastically in dense regions assuming a Salpeter IMF. This
model pictures each gas element as a two-phase mixture of hot and cold
gas in approximate pressure equilibrium.  Cold gas is converted into
``star'' particles on a characteristic timescale $t_{\star}$. Assuming
that a fraction $\beta$ of the stellar population represented by each
star particle is so short-lived that they explode as supernovae instantly, 
their (metal-enriched) mass is fed back to the interstellar medium. The star
formation rate is given by
\begin{equation}
 \frac{\rm{d}\rho_{\star}}{\rm{d}t} =
\frac{\rho_c}{t_{\star}} -
\beta\frac{\rho_c}{t_{\star}}=(1-\beta)\frac{\rho_c}{t_\star},
\end{equation}
where $\rho_c$ is the mean mass density of gas in the cold phase, 
and $t_\star$ is a density-dependent star formation timescale. The
energy feedback of supernovae explosions is used to heat the ambient
hot gas as well at to evaporate cold clouds, leading to a
self-regulating cycle for star-forming gas. The net effect is that the
dynamics of the multi-phase medium is governed by an effective
equation-of-state \citep[see][]{SpringelHernquist2003}.

The simulations {\small G3-BH} and {\small G3-CR} both include thermal AGN feedback
associated with a Bondi-Hoyle-Lyttelton parameterization of the
(spatially unresolved) gas accretion onto a central supermassive black
hole \citep[see][for a detailed
description]{SpringelDiMatteo2005}. The thermal feedback energy is
parameterized by $\dot{E}_{\rm feed} = \epsilon_f \epsilon_r
\dot{M}_{\rm BH}c^2$, where $\epsilon_r$ is the radiative efficiency
of the black hole and $\epsilon_f$ encodes the feedback efficiency.

In {\small G3-CR} the distribution function of cosmic rays (CR)  is modeled as a power law,
\begin{equation}
\frac{\rm{d}^2N}{\rm{d}p\;\rm{d}V} = Cp^{-\alpha}\theta(p-q),
\end{equation}
in momentum space.  The non-thermal pressure of this cosmic ray population is
then given by
\begin{equation}
P_{\rm{CR}} = \frac{Cm_pC2}{6} \mathcal B_{\frac{1}{1+q2}}\left(\frac{\alpha-2}{2},\frac{3-\alpha}{2}\right),
\end{equation}
which is added to the ordinary gas pressure \citep{Jubelgas2008}.
Here $\mathcal B_n(a,b)$ denotes incomplete Beta functions
and $\alpha$ is the assumed power-law slope of the CR particles.  The
energy injected by supernova explosions into the CR population is
given as $\Delta \tilde{\varepsilon}_{\rm SN} = \zeta_{\rm
  SN}\epsilon_{\rm SN} \dot{m}_{\star} \Delta t$ where $\zeta_{\rm
  SN}$ is the is the fraction of the supernova energy that first
appears in cosmic rays.

The numerical parameters used for the Aquila Project simulations are
as follows: a star formation timescale of $t_{\star} = 2.1$ Gyr and a
supernova energy of $E_{\rm SN} = 10^{51}$ erg. The {\small G3-BH}
model adopts $\epsilon_r=0.1$ and $\epsilon_f=0.05$. The {\small
  G3-CR} run uses a cosmic ray generation efficiency of $\zeta_{\rm
  SN} = 0.3$ and a spectral index $\alpha = 2.5$

\vspace{0.3cm}
\centerline{\it  The CS model (G3-CS)}
\vspace{0.1cm}

The CS model is a {\sc gadget3}-based code that includes stochastic
star formation, chemical enrichment and supernova (thermal) feedback
from Type II and Type Ia SN explosions, a multi-phase model for the
gas component and metal-dependent cooling. Full details of the
implementation are given in \citet{CS2005} and \citet{CS2006} and
previous applications to cosmological galaxy formation and disk
formation can be found in \citet{CS2008, CS2009, CS2010, CS2011}.

The model includes a UV background field which turns on at $z=6$ and
follows the formulation of \citet{HaardtMadau1996} and
metal-dependent cooling functions from \citet{SutherlandDopita1993}
are used.  Star formation occurs stochastically in gas particles with
high density ($n > n_{\rm th}$) and in a convergent flow
\citep{SpringelHernquist2003}, following the Schmidt-Kennicutt law:
\begin{equation}\label{eq_kenn_law}
{d\rho_*\over{dt}} = c_*\ {\rho_{\rm gas}\over{t_{\rm dyn}}}
\end{equation}
where $\rho_*$ and $\rho_{\rm gas}$ are the stellar and gas densities,
$c_*$ is a star formation efficiency and $t_{\rm dyn}$ is the
dynamical time of the gas.  Stars return metals and energy during
supernova Type II and Type Ia explosions.  Both types of SNe eject the
same amount of energy $E_{\rm SN}$ to the interstellar medium, but
they have different chemical yields, explosion times and rates.  SN
energy is dumped in given fractions to the cold ($\epsilon_{\rm c}$)
and hot ($1-\epsilon_{\rm c}$) neighbors of exploding stars. Energy
deposition into hot neighbors occurs at the time of explosion;
however, for cold neighbors energy from successive explosions is
instead accumulated and deposited only after a time-delay which
depends on the local conditions of the cold and hot gas phases.  In
this way, artificial loss of SN energy in high-density regions is
prevented.  Star particles are treated as single stellar populations
with a Salpeter initial mass function.

We introduce a multiphase gas model which allows gas in both dense and
diffuse phases to coexist in the same spatial region.  In our model, SPH
particles of a given physical state (density, entropy) ignore
neighboring particles that have a much lower (a factor of $50$)
entropy. This scheme also makes the deposition of SN energy more
efficient.

The input parameters used for the simulations analyzed in this paper
are: $n_{\rm th}=0.03$  cm$^{-3}$, $c_*=0.1$, $E_{\rm
  SN}=0.7\times 10^{51}$ erg and $\epsilon_c=0.5$. Finally, the input
parameters for the chemical model are identical to those used in
\citet{CS2009}.

\vspace{0.3cm}
\centerline{\it   The TO model (G3-TO)}
\vspace{0.1cm}

The TO model is described in \citet{Okamoto2010} and is based on an
early version of the {\sc gadget3} code.  It incorporates
metal-dependent cooling, star formation, thermal and kinetic feedback
from supernovae (SNe) and enrichment by AGB stars and SNe.  Examples
of its application in cosmological galaxy formation simulations
include \citet{OkamotoFrenk2009}, \citet{Okamoto2010} and
\cite{Parry2011}.

Photo-heating and cooling are implemented as described in
\citet{Wiersma2009a}, including contributions from eleven elements in
the presence of a spatially uniform, time evolving UV background of
the form calculated by \citet{HaardtMadau2001}.  Gas above a density $n_{th}$ forms stars at a rate
normalized to reproduce the Schmidt-Kennicutt law.  Star particles
represent simple stellar populations with a \citet{Chabrier2003}
initial mass function.  Following \citet{Wiersma2009b}, energy, mass
and metals are returned to the ISM by AGB stars, type Ia and type II
SNe on timescales appropriate for the age and metallicity of the
stellar population, with yields and stellar lifetimes taken from
\citet{Portinari1998} and \citet{Marigo2001}.

The unresolved interstellar medium  is modeled using a sub-grid
prescription.  Each gas particle with $n > n_{th}$ is treated as
a series of cold clouds, with an empirically motivated mass function,
embedded in a hot ambient medium.  The two phases exchange mass
through thermal instability and cloud evaporation.  Each cloud has a
star formation rate that is inversely proportional to its dynamical
time.  The hot phase is supported by imposing a minimum pressure of
$P_{\rm{min}}\propto\rho^{1.4}$.  Type Ia SNe increase the thermal
energy of the gas around star forming regions, whilst type II SNe are
assumed to drive large-scale winds.  The wind speed is chosen to be
proportional to the halo circular velocity, using the local dark
matter velocity dispersion as a proxy ($v=\alpha\sigma_{DM}$).  The
mass loading then follows by requiring conservation of all of the
available SNe energy.  Wind particles are decoupled from hydrodynamic
forces for a short time, to allow them to escape the star forming
region and ensure that the specified mass loading and wind velocity
are not modified by viscous drag from the ISM.

The density required for star formation is set at $n_{th} =$$\rm
0.1$ cm$^{-3}$ in the level 6 simulation and $\rm 0.4$
cm$^{-3}$ in the level 5 simulation.  For the wind speed parameter,
$\alpha=5$ is used, which has been shown to produce a good match to
the Milky Way satellite luminosity function
\citep{Okamoto2010,Parry2011}.

\vspace{0.3cm}
\centerline{\it   The GIMIC model (G3-GIMIC)}
\vspace{0.1cm}

The GIMIC model is a {\sc gadget3}-based code that includes metal
dependent cooling on an element-by-element basis in the presence
of a UV background, star formation and supernovae (SNe) driven winds,
as well as mass and metal recycling by AGB stars, type Ia and type II
SNe. GIMIC is identical to model MILL of the OWLS suite of simulations
\citep{Schaye2010}. A full description of the model can be found in
\citet{Crain2009} and \citet{Schaye2010} and further applications of this model can be found
in \citet{Crain2010}, \citet{Cui2011}, \citet{Deason2011} and
\citet{Font2011,McCarthy11}.

The model includes a spatially uniform, time evolving UV background
following the formulation of \citet{HaardtMadau2001}. Hydrogen
reionises at $z=9$, Helium II at $z=3.5$. Radiative cooling 
and heating processes
are implemented on an element-by-element basis, using interpolation
tables computed with {\sc cloudy} \citep{Ferland1998}, as described by
\citet{Wiersma2009a}.  The star formation rate (SFR) prescription,
described by \citet{SchayeDallaVecchia2008}, is pressure-dependent
and enforces a local
Schmidt-Kennicutt law and requires only that the slope and
normalization of the observed relation are specified as input
parameters. Gas particles are converted to star particles
stochastically, with a probability that depends on their associated
SFR. Each star particle represents a simple stellar population (SSP)
with a \citet{Chabrier2003} initial mass function and inherits the
elemental abundances of its parent gas particle. The chemodynamical
evolution of SSPs and the associated recycling of heavy elements into
surrounding gas, is followed on an element-by-element basis and
includes contributions from AGB stars and both Type Ia and Type II
supernovae, as described by \citet{Wiersma2009b}.

GIMIC appeals to a phenomenological treatment to model the energetic
feedback resulting from stellar evolution and supernovae. 
As the simulation lacks both the physics and the resolution to model
the multiphase ISM, an effective equation of state is imposed onto gas particles
that are sufficiently dense ($n_{\rm H}> 0.1$ cm$^{-3}$) and cold ($T<10^5$ K)
to be subject to gravitational instability \citep{Schaye2004}. The effective
equation of state, $P = \kappa \rho^{4/3}$, 
is chosen to ensure that the Jeans mass and the ratio between
the Jeans length and the SPH smoothing length are independent of the
gas density, thus preventing spurious fragmentation due to a lack of
numerical resolution \citep{SchayeDallaVecchia2008}. SN energy is 
assumed to drive galaxy-wide outflows, therefore gas particles
neighboring star-forming regions are given a randomly orientated
velocity kick of 600 kms$^{-1}$. A probabilistic scheme ensures that,
on average, the mass put into the wind is a factor of four times the
amount of stars formed. Unlike many similar schemes, these particles are
not temporarily decoupled from hydrodynamic forces \citep[for further
  discussion, see][]{DallaVecchiaSchaye2008}.

Values for all model parameters can be found in \citet{Crain2009}, but
to summarize: the initial wind mass loading and velocity are set to 4 and 600
kms$^{-1}$ respectively (with $10^{51}$ ergs per SN), the star
formation threshold to $n > 0.1$ cm$^{-3}$ and the input star
formation law to have slope 1.4 and normalization coefficient
$\sim1.5\times10^{-4}{\rm M_{\odot}yr^{-1}kpc^{-2}}$.

\vspace{0.3cm}
\centerline{\it  The MM model (G3-MM)}
\vspace{0.1cm}

The MM model (or MUlti Phase Particle Integrator, MUPPI), fully
described in \cite{M10}, is also implemented within {\sc gadget3}.  We
include radiative cooling of a gas with primordial composition;
heating from a uniform UV background of the form given by
\cite{HaardtMadau1996}, turned on at a redshift $z=6$; star formation
and stellar feedback with the multi-phase model described below.  We
assume a Salpeter IMF and the Instantaneous Recycling Approximation to
treat gas restoration and SN energy feedback.  No chemical enrichment
is included in the present version of the code.  No treatment of
accreting black holes or cosmic ray feedback is implemented.

In the MUPPI sub-resolution model of star formation and stellar
feedback, gas particles at relatively high density ($n > n_{\rm th}$)
and low temperature ($T<T_{\rm th}$) are treated as a multi-phase
system, made up by cold and hot gas phases coexisting in pressure
equilibrium, and a stellar component.  Hot phase has low mass fraction
but high filling factor, so its cooling time is much longer than that
computed for the average particle density, and thus thermal energy is
{\it not} quickly dissipated as soon as it is injected.  
Regarding the cold phase, a part of it is assumed to be in molecular
form.  Blitz \& Rosolowsky \citep{Blitz2006} found a phenomenological
correlation between the ratio of molecular and HI gas surface
densities and the external disk pressure. Inspired by this result, we
compute the fraction of cold gas in molecular form as a function of
hydrodynamical pressure $P$:
\begin{equation}
f_{\rm mol} = \frac{1}{1+ \frac{P_0}{P}}\, .
\end{equation}

Mass flows among the components as follows.  Cooling of the hot phase
feeds the cold gas phase.  Stars form from the {\it molecular} cold
gas:
\begin{equation}
\dot{M}_\star = f_\star f_{\rm mol} \frac{M_{\rm cold}}{t_{\rm dyn}}
\end{equation}
where $t_{\rm dyn}$ is the dynamical time of the cold phase at the
onset of a multi-phase cycle (see \citealt{M10} for details) and
$f_\star$ is the fraction of molecular cloud transformed into stars
per dynamical time.  Our prescription for star formation is not based
on imposing a Schmidt-Kennicutt relation, which is instead naturally
produced by our model (see Monaco et al. 2012).  A fraction $f_{\rm
  ev}$ of the star-forming gas is evaporated back to the hot phase,
while a fraction $f_{\rm re}$ of stellar mass is instantaneously
restored to the hot phase.  The code also tracks the hot phase thermal
energy. This gained or lost through hydrodynamics, lost by cooling and
gained from type II SN explosions: a small fraction $f_{\rm in}$ of
their energy is deposited into the hot phase of the same star-forming
gas particle, while a more significant fraction $f_{\rm out}$ is
distributed to neighboring particles, preferentially along the ``least
resistance path'' defined as a cone of semi-aperture $\theta$ and
anti-aligned with the density gradient.  Finally, to mimic the
destruction of molecular clouds a multi-phase particle returns
single-phase after a time $t_{\rm clock}$, scaled with the dynamical
time.

A system of ordinary differential equations evolves the mass and
energy flows described above.  This is solved on-the-fly within each
SPH time-step with a Runge-Kutta integrator with adaptive time-step.
This dynamical system has an intrinsically runaway behavior: energy
from SNe increases gas pressure, which in turn increases SFR through
the higher molecular fraction.  This runaway is stabilized by the
hydrodynamic response of gas: when a particle receives enough energy,
it expands thereby decreasing its pressure. The equations MM model
solves are similar to those used in the star formation model by
Springel \& Hernquist (2003); the main differences are that in MM
model no equilibrium solution is assumed, and the effect of
hydrodynamics on the multi-phase gas is explicitly taken into account.

MUPPI produces reservoirs of ``virtual'' stars, that are transformed
into star particles using the stochastic algorithm of Springel \&
Hernquist (2003). Each gas particle produce up to 4 generations of
star particles.

We used here the same ``standard'' set of parameters described in
\cite{M10}, namely: {\it(i) star formation efficiency $f_\star=0.02$},
amount of molecular gas which is converted into stars; {\it
(ii)} $P_0=35000$ K cm$^{-3}$, pressure normalization for the Blitz \&
Rosolowsky relation; {\it(iii)} $T_c=1000$K, temperature of the cold
phase; {\it (iv) $f_{\rm out}=0.3$}, fraction of SNe energy given to
neighboring gas particles; {\it (v) $f_{\rm in}=0.02$}, fraction given
to the hot gas of the particle itself; {\it (vi) $f_{\rm ev}=0.1$},
fraction of cold gas mass evaporated by SNe; {\it (vii)
$E_{SN}=10^{51}$} erg, SN energy; {\it (vii) $\theta=60$},
semi-aperture of the cone determining the neighboring gas particles
which receive energy; {\it (viii)} $n_{\rm th}=0.01$  cm$^{-3}$,
density threshold for entering the multi-phase stage; {\it (ix)
$T_{\rm th}=50000$}K, temperature threshold for entering the
multi-phase stage; $t_{\rm clock}=2t_{\rm dyn}$, time after which a
particle {\it exits} the multi-phase stage.

\vspace{0.3cm}
\centerline{\it  The CK model (G3-CK)}
\vspace{0.1cm}

The CK model is a {\sc gadget3}-based code that includes star formation, chemical enrichment, and (thermal) feedback from stellar winds, core-collapse supernovae, Type Ia supernovae, and asymptotic giant branch (AGB) stars.
The details are described in \citet{kob04}, \citet{kob07}, and \citet{kob11}.

The UV background radiation is included with \citet{HaardtMadau1996} from $z=6$.
Radiative cooling is computed with the metal-dependent cooling functions \citep{SutherlandDopita1993} as a function of [Fe/H].
[O/Fe] is fixed with the observed [O/Fe]-[Fe/H] relation in the solar neighborhood.
The star formation criteria are 
1) converging flow, 2) rapid cooling, and 3) Jeans unstable.
The star formation rate (SFR) is determined from the Schmidt-Kennicutt law (Eq.~\ref{eq_kenn_law}).
If a gas particle satisfies the star formation criteria, a part of the mass of the gas particle turns into a new star particle.
We then treat the star particle as a simple stellar population and calculate the evolution of the stellar population every time-step.
The masses of the stars associated with each star particle are distributed according to an initial mass function.
We adopt the Salpeter IMF that is invariant to time and metallicity with a slope $x=1.35$ for $0.1-120M_\odot$, to be consistent with the Galactic chemical evolution \citep{kob06}.

For the feedback of energy and heavy elements, we do not adopt the instantaneous recycling approximation.
Instead, via stellar winds, core-collapse supernovae, Type Ia supernovae, and AGB stars, thermal energy and heavy elements are ejected from an evolved star particle as a function of time, and distributed to a constant number $N_{\rm FB}$ of surrounding gas particles.
Among core-collapse supernovae, we include the effect of hypernovae, which are observationally known to produce more than ten times larger explosion energy and iron than normal Type II supernovae.
We adopt the metal-dependent efficiency of hypernovae ($\epsilon_{\rm HN}=0.5, 0.5, 0.4, 0.01$, and $0.01$ for $Z=0, 0.001, 0.004, 0.02$, and $0.05$) for the initial masses of $M \gtsim 20M_\odot$, to be consistent with the observed [Zn/Fe] ratios in the Milky-Way Galaxy \citep{kob11}.
The ejected energy ($1-40 \times 10^{51}$ erg) and nucleosynthesis yields are taken from  \citet{kob06} as a function of progenitor mass and metallicity.
With hypernovae, cosmological simulations give a better agreement with observed  cosmic SFRs \citep{kob07}.
For Type Ia Supernovae, we use our single-degenerate model with the metallicity effect \citep{kob98,kob09}.
The lifetimes span a range of $0.1-20$ Gyr as a function of progenitor metallicity, which is consistent with the observed supernova rates and with the observed [$\alpha$/Fe] relations in the Milky-Way Galaxy.
The ejected energy is $1.3 \times 10^{51}$ erg per explosion.
>From stellar winds, $\sim 0.2 \times 10^{51}$ erg is ejected depending on metallicity for the stars with $\ge 8M_\odot$.
The adopted nucleosynthesis yields of supernovae and AGB stars are summarized in \citet{kob11b}.

The input parameters used for the simulations analyzed in this paper are: the star formation timescale of $c_*=0.02$ and the feedback number of $N_{\rm FB}=64$.

\vspace{0.3cm}
\centerline{\it  The Gasoline model (GAS)}
\vspace{0.1cm}

The GAS model uses the SPH code \textsc{gasoline}, 
which is described in detail in \citet{Wadsley2004}.
It includes a UV background heating, metal cooling, star formation, 
thermal stellar feedback and chemical enrichment from SNII, SNIa, 
and mass return from stellar winds.

\textsc{gasoline} contains metal
cooling based on radiative transfer found in \textsc{cloudy} as
described in \citet{Shen2010}.  The \textsc{cloudy} cooling was
calculated using an external UV radiation field starting at $z=8.9$.
Star formation is calculated and supernova feedback is implemented
using the blastwave formalism as described in \citet{Stinson2006}.
Recent examples of simulations that use similar physics include
\citet{Governato2007, Governato2009, Governato2010},
\citet{Stinson2010}, \citet{Brooks2011} and \citet{Guedes2011}.

Star formation is calculated using the commonly used recipe described
in \citet{Katz1992}.  Stars form from gas below a maximum temperature
of 15,000 K and above a density of 1 cm$^{-3}$ with an efficiency
of 5\%.  
However, the high resolution runs presented in \citet{Governato2010} and 
\citet{Guedes2011}  used a higher threshold for star formation (100 and 5 cm$^{-3}$
respectively) that leads to more efficient gas outflows than the SF
recipe adopted for the Aquila simulation.
The star particles are treated as single stellar populations
using the IMF described in \citet{Kroupa1993}.  In this context,
ejecta from both Type II and type Ia supernovae are considered.  These
supernovae feed both energy and metals back into the interstellar
medium gas surrounding the region where they formed.  Type II
supernovae deposit $7\times10^{50}$ erg of energy into the surrounding
interstellar medium.  Since this gas is dense, it would be quickly
radiated away due to the efficient cooling.  For this reason, cooling
is disabled for particles inside the blast region $r_{CSO} =
10^{1.74}E_{\rm 51}^{0.32}n_0^{-0.16}\tilde{P}_{\rm 04}^{-0.20} {\rm
  pc}$ for the length of time $t_{CSO} = 10^{6.85}E_{\rm
  51}^{0.32}n_0^{0.34}\tilde{P}_{\rm 04}^{-0.70} {\rm yr}$ given in
\citet{McKee1977}.

In summary, the input parameters used for the simulations analyzed in this paper are: 
$n_{\rm th} = 1.0$ cm$^{-3}$, $c_* = 0.05$, and $E_{\rm SN} = 0.7\times 10^{51}$ erg. 

\vspace{0.3cm}
\centerline{\it  The Ramses  models (R, R-LSFE, R-AGN)}
\vspace{0.1cm}

The {\sc ramses} code is an Eulerian Adaptive Mesh Refinement code that uses
the Particle Mesh techniques for the N body part (stars and dark
matter) and a shock-capturing, unsplit second-order MUSCL scheme for
the fluid component. For the latter, we use the MinMod slope limiter
and the HLLC Riemann solver, ensuring both stability and proper
capturing of discontinuities \citep{Teyssier2002}.

Star formation is implemented stochastically using a Schmidt law, with
density threshold for star formation held fixed at $n_{\rm th} =
0.1 $  cm$^{-3}$ and an efficiency parameter chosen between $1\%$
(R-LSFE) and $5\%$ (R, R-AGN). Stellar feedback is modeled using a
thermal dump of $10^{51}$ erg per supernova, assuming a Salpeter
IMF. Cooling is performed using a tabulated cooling function depending
on gas metallicity, the latter being modeled self-consistently as a
additional scalar hydro variable and initiated during supernovae
explosions with a yield y=$10\%$ \citep{Rasera2006, Dubois2008}.

The grid is refined following a quasi-Lagrangian strategy, each cell
being refined if the number of dark matter particles exceed 8, or if
the baryonic mass (star + gas) exceeds 8 times the initial mass
resolution set by the Aquila initial conditions. In order to avoid
catastrophic refinement at early times, we carefully trigger new
levels so that the maximum level {\it levelmax} is opened only at
expansion factor $a=0.8$, the previous one at $a=0.4$, {\it
  levelmax$-2$} at $a=0.2$ etc.  This ensures that the resolution of
the grid in physical units remains roughly constant (although in a
discrete, stepwise sense).  For the level 6 simulation, we have set
levelmax=17 (cell size is 1 kpc physical) and for the level 5
simulation, levelmax=19 (cell size is 261 pc physical).  This
corresponds also to the maximum levels reached by a pure dark matter
simulation with the same number of dark matter particles for each
case, minimizing spurious effects due to two-body relaxation.  For the
R-AGN simulation only, AGN feedback has been implemented using the
model of \citet{Booth2010}; see \citet{Teyssier2011} for
details.  A sphere of size 4 cells is defined around each sink
particle defining the SMBH. This spherical region is used to determine
the accretion rate on the SMBH, but also to spread on the grid the AGN
feedback energy, using only a pure thermal dump.

To summarize, the {\sc ramses} simulations use a star formation density
threshold of $n_{\rm th} = 0.1 $  cm$^{-3}$, an energy per SN of
$10^{51}$ ergs and a star formation efficiency of $5\%$ (R and R-AGN)
or $1\%$ (R-LSFE).

\vspace{0.3cm}
\centerline{\it  The Arepo model (AREPO)}
\vspace{0.1cm}

The {\small AREPO} code \citep{Springel2010} is a novel
pseudo-Lagrangian hydrodynamical code that works with an unstructured,
fully dynamic and adaptive mesh. The mesh is defined as the Voronoi
tessellation \citep[e.g.][]{Okabe2000} of the simulation volume
generated by a set of mesh-generating points. The hydrodynamics is
calculated with a second-order accurate finite volume approach on this
mesh, based on the MUSCL-Hancock scheme that is also widely employed
in ordinary Eulerian mesh codes. This involves a spatial
reconstruction step and flux estimates at all cell faces by solving
Riemann problems at the interfaces. The very important new ingredient
in the scheme is that the mesh-generating points are allowed to move
freely during the calculation. In particular, they can be moved with
the local fluid velocity such that the mesh dynamically follows the
fluid motion without showing any problematic mesh twisting effects. In
this mode, {\small AREPO} minimizes advection errors in the
hydrodynamics and produces Galilean-invariant results that help to
avoid accuracy problems with high velocity cold flows that can occur
in ordinary mesh codes.  A more detailed description and an investigation
of numerous test problems that demonstrate the high accuracy of the
method can be found in \citet{Springel2010}.

Even though the hydrodynamics is solved completely differently in
{\small AREPO} than in {\sc gadget3}, the Lagrangian character of
both codes makes it possible to implement the physics of star
formation and feedback in very similar ways in both codes. In fact,
the {\small AREPO} simulations analyzed here implement exactly the
same sub-grid model for dense gas as well as the same supernova
feedback recipe as our default {\sc gadget} run G3. Since also the
gravitational solver for both codes is identical, any differences
found in the results hence reflect changes due to the numerical
treatment of hydrodynamics alone.

\section{Gravitational softening}\label{app:softening}

\begin{figure}
\begin{center}
\includegraphics[width=8cm]{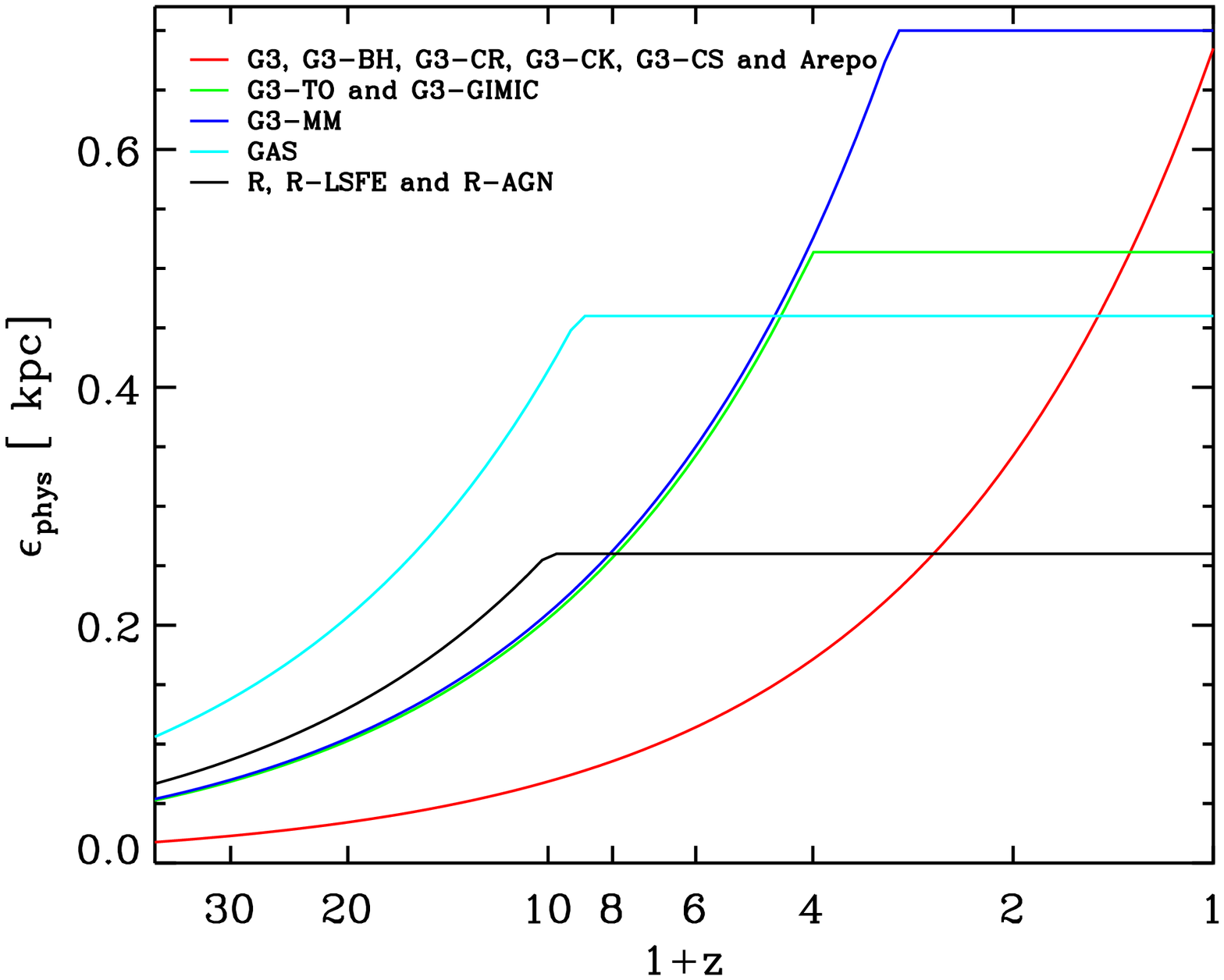}

\caption{Evolution of the {\it physical} gravitational softening
in the different models.  \label{fig:softenings}}
\end{center}
\end{figure}

 Our 13 simulations have  used a variety of  choices for the evolution 
 of the gravitational softening. This evolution is
governed by two parameters: $z_{\rm fix}$ and $\epsilon_{\rm
  g}^{z=0}$. The former divides the period where the softening
changes from being fixed in comoving coordinates ($z>z_{\rm fix}$) to
being fixed in physical coordinates ($z\le z_{\rm fix}$).
 The different simulations have various choices for
$z_{\rm fix}$, as shown in shown in Table~\ref{table_params}.
The second parameter, $\epsilon_{\rm
  g}^{z=0}$, is the value of the gravitational softening
at the present time, and  also varies slightly for our 13 simulation
(Table~\ref{table_params}).
Fig.~\ref{fig:softenings} shows the evolution of the gravitational softening
in physical coordinates for our simulations.

\section{Disk orientation}\label{app:disk_orientation}

\begin{figure*}
\begin{center}
{\includegraphics[width=90mm]{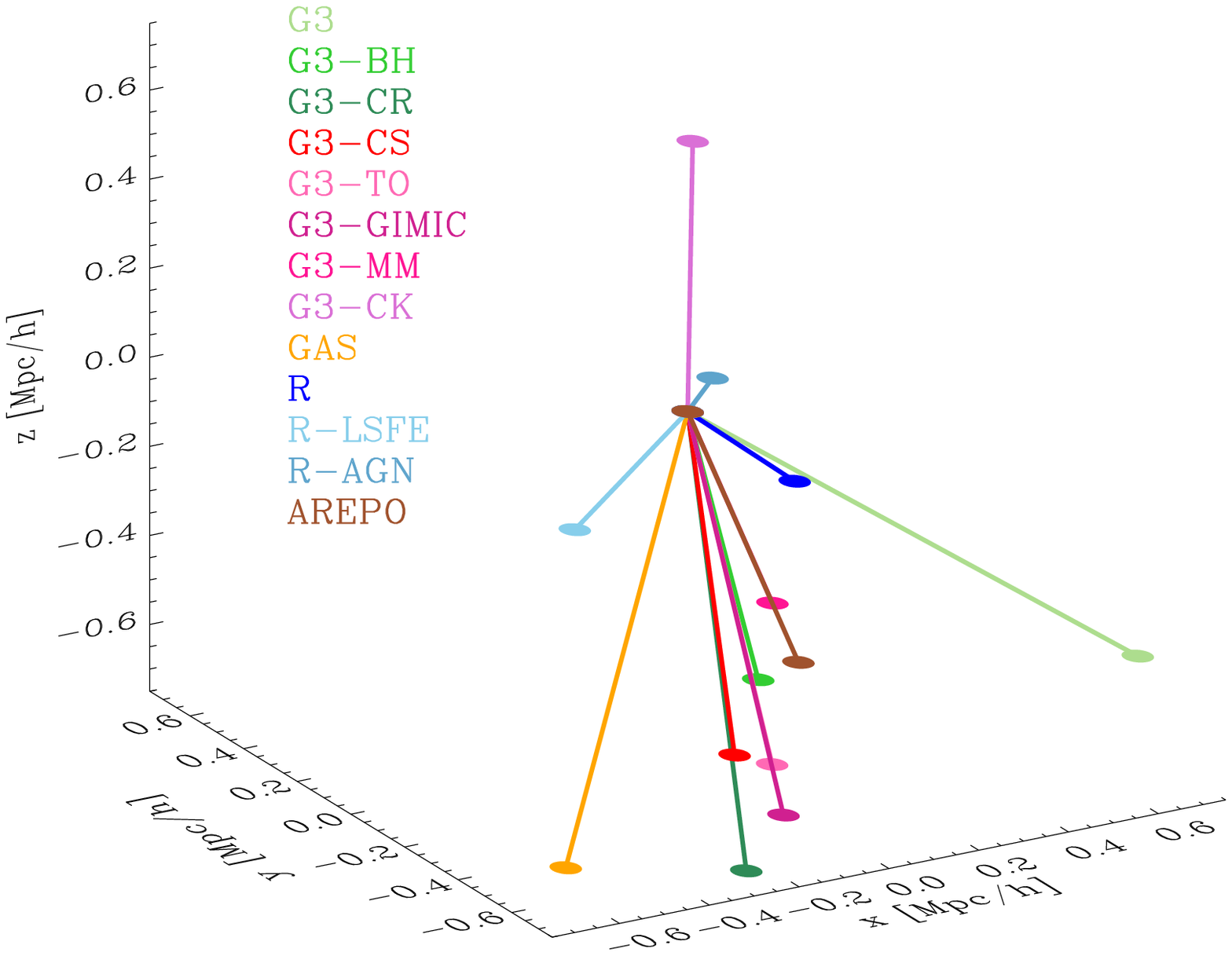}\includegraphics[width=90mm]{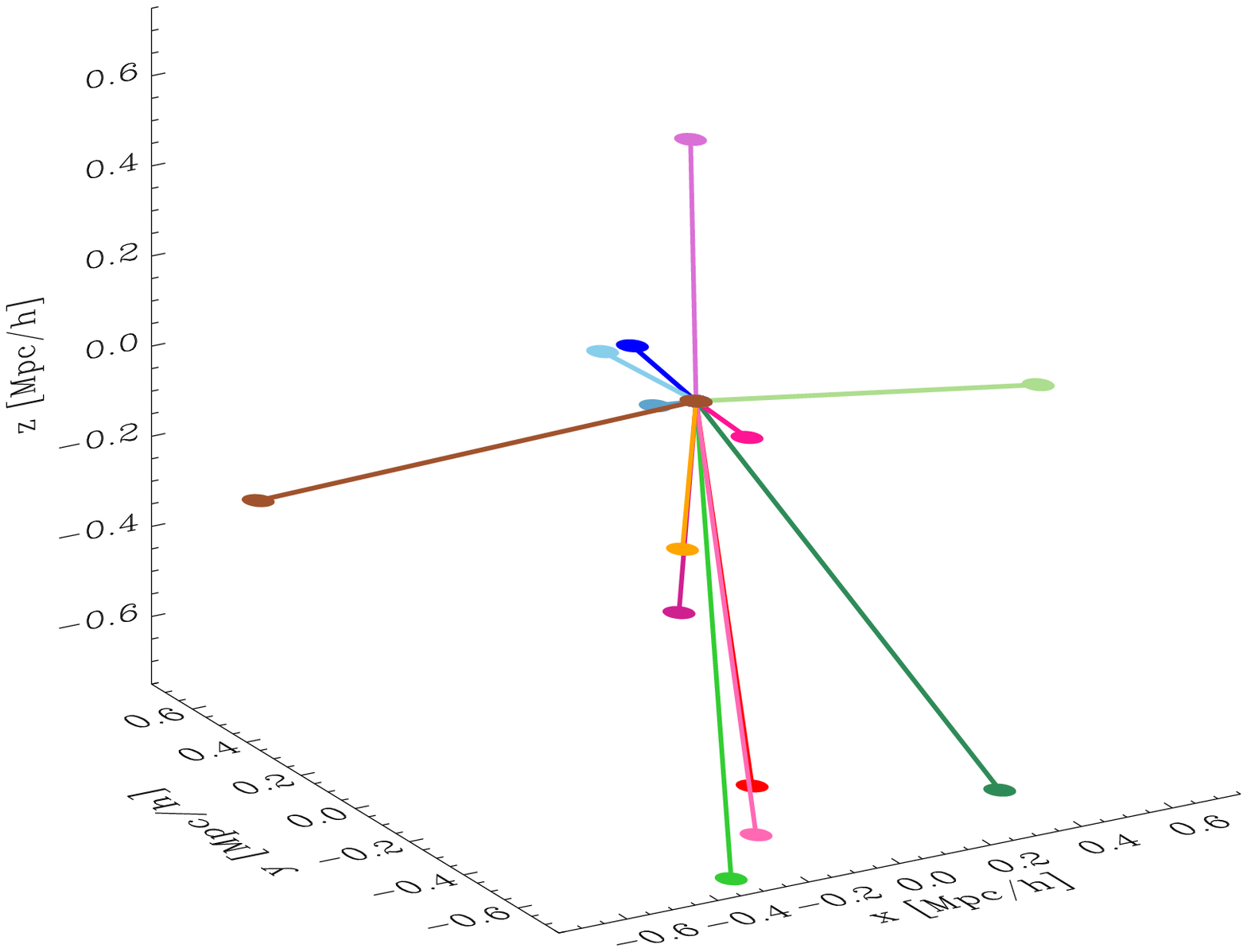}}
\end{center}
\caption{Disk orientation of the level 5 (left-hand panel) and level 6 
(right-hand panel) resolution simulations. The
vectors indicate the orientation of the angular momentum of galactic stars, and are
normalized to the maximum among all simulations. }
\label{fig:disk_orientation}
\end{figure*}

 Fig.~\ref{fig:disk_orientation} illustrates, for our level 5 and
  level 6 runs, the spin of the stellar component of the simulated
  galaxies at $z=0$.  The two panels show, in a 3D rendering, the
  specific angular momentum vector of all stars in each galaxy,
  normalized to the maximum among all simulations.  Clearly, disks are
  not all aligned in the same direction, nor is the specific angular
  momentum the same. This is not surprising, given the wide range of
  stellar masses spanned by the different simulations.  In spite of
  this, the orientation is actually similar for some simulations. As
  explained in Section~\ref{SecRuns}, for orientation-dependent
  diagnostics we have rotated each simulated galaxy to a new
  coordinate system where the angular momentum vector of its stellar
  component coincides with the $z$ direction. 

\bibliographystyle{mn2e}

\bibliography{bibliography.bib}

\end{document}